\documentclass[acmsmall]{acmart}

\newif\ifarxiv
\arxivtrue   

\newcommand{\appref}[2]{%
  \ifarxiv
    \Cref{#1}\apprefcheck{#1}{#2}%
  \else
    the extended paper~\cite[App.~#2]{cse4-extended}%
  \fi}

\newcommand{\apprefcheck}[2]{%
  \ifcsname r@#1\endcsname
    \edef\apprefactual{\getrefnumber{#1}}%
    \edef\apprefgiven{#2}%
    \ifx\apprefactual\apprefgiven\else
      \ClassError{cse4}
        {Appendix `#1' is numbered \apprefactual, but \string\appref\space says #2}
        {Change the second argument of that \string\appref\space to \apprefactual.}%
      \batchmode \read -1 to \apprefactual  
    \fi
  \fi}

\setcopyright{cc}
\setcctype{by}
\acmDOI{10.1145/3871709}
\acmYear{2027}
\acmJournal{PACMPL}
\acmVolume{11}
\acmNumber{POPL}
\acmArticle{7}
\acmMonth{1}
\acmSubmissionID{popl27main-p99-p}
\ifarxiv\else 
  \received{2026-07-09}
  \received[accepted]{2026-10-05}
\fi

\usepackage{cse}

\usepackage{subcaption}
\usepackage{tabularray}
\usepackage{tabularx}
\UseTblrLibrary{counter, varwidth}
\usepackage{wrapfig}
\usepackage{threeparttable}
\usepackage{multicol}

\usepackage{tikz}
\usetikzlibrary{calc, positioning, shapes.symbols}
\usepackage{pgfplots}
\pgfplotsset{compat=1.18}

\usepackage[frozencache]{minted}

\setminted{fontsize=\footnotesize}
\setmintedinline{fontsize=\normalsize}
\setminted[coq]{
  linenos,
  breakautoindent=true,
  breakindent=1.5em,
  breaklines=true,
  frame=lines,
  framerule=0.3pt,
  framesep=3pt,
  numbersep=6pt,
  rulecolor=\color{RocqCodeRule},
  tabsize=2,
  xleftmargin=1.5em
}
\setminted[c]{
  breakautoindent=true,
  breakindent=1.5em,
  breaklines=true,
  tabsize=4
}

\usepackage{cleveref}
\crefname{figure}{Fig.}{Figs.}
\crefname{section}{Sec.}{Secs.}
\crefname{appendix}{App.}{Apps.}
\crefname{subappendix}{App.}{Apps.}

\ifarxiv
  \title[An End-to-End Theory for CSE (Extended Version)]
        {An End-to-End Theory for Compositional Symbolic Execution: From Expressive Specifications to SMT\texorpdfstring{\\}{ }(Extended Version)}
\else
  \title[An End-to-End Theory for CSE]
        {An End-to-End Theory for Compositional Symbolic Execution: From Expressive Specifications to SMT}
\fi

\author{Shivanandan Tamil Kumaran}
\correspondingauthor
\orcid{0009-0009-8126-3868}
\affiliation{%
 \institution{Imperial College London}
 \city{London}
 \country{United Kingdom}}
\email{shiva.tamil-kumaran21@imperial.ac.uk}

\author{Andreas Lööw}
\orcid{0000-0002-9564-4663}
\affiliation{%
 \institution{Royal Holloway, University of London}
 \city{Egham}
 \country{United Kingdom}}
\email{andreas.loow@rhul.ac.uk}

\author{Philippa Gardner}
\orcid{0000-0002-4187-0585}
\affiliation{%
 \institution{Imperial College London}
 \city{London}
 \country{United Kingdom}}
\email{p.gardner@imperial.ac.uk}



\begin{abstract}
Compositional symbolic execution (CSE) is a powerful technique for program verification and bug-finding, yet existing formal theories of CSE do not sufficiently bridge the gap between idealised models and real-world tools. We present progress towards a foundational, realistic, end-to-end theory of CSE in the Rocq proof assistant. Our theory advances realism at both ends of the CSE pipeline: it supports an expressive frontend with algebraic data types, logical functions, and lemmas; it also provides a verified connection to SMT-LIB compliant SMT solvers. Our contributions include: (1) a CSE theory that captures practical verification patterns; (2) a standalone, reusable Rocq mechanisation of the SMT-LIB standard; (3) a proven-correct encoding from CSE backend queries to SMT-LIB queries that accounts for semantic discrepancies and optimises using type information; and (4) empirical evidence for the backend by extracting our verified SMT-LIB encoding and running it alongside the CSE tool Gillian's own encoding on a majority of its solver queries.
\end{abstract}

\begin{document}

\maketitle

\section{Introduction}%
\label{sect:introduction}

Compositional symbolic execution (CSE) provides symbolic execution~\cite{Baldoni18} for creating and using function specifications grounded in separation logic (SL)~\cite{OHearn2001,Reynolds2002} and its derivatives, such as incorrectness SL (ISL)~\cite{Raad2020} and implicit dynamic frames~\cite{implicit-dynamic-frames-smans-2012}. SL reasoning enables CSE to use reusable function specifications, as well as verify and synthesise its own. CSE is applicable to both over-approximate (OX) correctness analysis for semi-automatic verification and under-approximate (UX) incorrectness analysis for automatic true bug-finding; modern CSE tools include VeriFast~\cite{Jacobs2011}, Viper~\cite{Muller2016}, and CN~\cite{Pulte2023} for OX, Infer-Pulse~\cite{Le2022} for UX, and Gillian~\cite{Maksimovic2021,Loow2024} for both.

\Cref{fig:interop} illustrates the architecture of modern CSE tools. The CSE engine symbolically executes the supported programming language(s). Its user-facing \emph{frontend} consists of user-specified annotations such as logical data types (\(\pname{List}\)), logical functions (the mathematical list reverse function $^\dag$), representation predicates (\(\pname{list}\), definition in \Cref{fig:example-spec}), and function specifications (an SL specification for \texttt{reverse}\footnote{As usual, logical variable $x$ records the old value of $\pvar x$, and the distinguished program variable $\pvar{ret}$ records the return value.}). The additional annotation required depends on the analysis: e.g., in verification, function specifications are manually provided; in bug-finding with bi-abduction~\cite{Calcagno2011}, they are automatically synthesised. At the other end, the \emph{backend} handles first-order reasoning, offloading it to \emph{satisfiability modulo theories} (SMT) solvers adhering to the SMT-LIB standard~\cite{Barrett2026}, such as Z3~\cite{DeMoura2008} or cvc5~\cite{Barbosa2022}; spatial reasoning (about ownership of heap memory) remains inside~the~CSE~engine.

\begin{wrapfigure}[15]{r}{0.39\textwidth}
\centering
\begin{tikzpicture}[>=stealth,thick,node distance=0.42cm]

  \node[draw, dashed, rounded corners, align=center, inner sep=3pt,
        text width=0.375\textwidth]
    (cloud)
    {
      \small \textbf{Definitions} \\
      \small $\pname{List} ::= [] \mid \vals :: \pname{List}$;
      \small $(\alpha : \pname{List})^\dag : \pname{List} = \cdots$; \text{ etc. } \\
      \small \textbf{Function specifications} \\
      \small $\triple{\pvar{x} = x \lstar \\
      \llist{x, \alpha}}{\pfuncspec{reverse}{\pvar x}}{\llist{ x, \alpha} \lstar \\ \llist{\pvar{ret}, \alpha^\dag}}$ \\
    };

  \node[draw, minimum width=0.20\textwidth, minimum height=0.6 cm, align=center,
        below=0.58cm of cloud]
    (box) {\small CSE Engine};

  \node[draw, minimum width=0.24\textwidth, minimum height=0.6cm, align=center,
        below=0.58cm of box]
    (solver) {\small SMT-LIB Compliant Solver};

  \draw[->] ([xshift=-4pt]cloud.south) -- ([xshift=-4pt]box.north);
  \draw[->] ([xshift=4pt]box.north) -- ([xshift=4pt]cloud.south);

  \draw[->] ([xshift=-4pt]box.south) -- ([xshift=-4pt]solver.north);
  \draw[->] ([xshift=4pt]solver.north) -- ([xshift=4pt]box.south);

  \draw[dotted]
    ($(cloud.south)!0.7!(box.north) + (-0.18\textwidth, 0)$) --
    ($(cloud.south)!0.7!(box.north) + ( 0.18\textwidth, 0)$)
    node[pos=0.02, above=1pt, anchor=south west, inner sep=1pt] {\scriptsize Frontend};

  \draw[dotted]
    ($(box.south)!0.3!(solver.north) + (-0.18\textwidth, 0)$) --
    ($(box.south)!0.3!(solver.north) + ( 0.18\textwidth, 0)$)
    node[pos=0.02, below=1pt, anchor=north west, inner sep=1pt] {\scriptsize Backend};

\end{tikzpicture}
\caption{Architecture of modern CSE tools.}%
\label{fig:interop}
\end{wrapfigure}

Recent work has introduced several \emph{CSE theories}: mathematical models of CSE engines with correctness proofs. The most complete are Featherweight VeriFast (FVF)~\cite{Jacobs2015}, a simplified, tool-dependent model of VeriFast; and the theories of \citet{Dardinier2025} and \citet{Loow2024,Loow2025}, loosely inspired by Viper and Gillian respectively, which aspire to tool independence.

No existing CSE theory has achieved the \emph{gold standard} of formal methods: being simultaneously foundational, realistic, and end-to-end. Here \emph{foundational} means mechanised inside an interactive theorem prover (e.g., Rocq, Isabelle/HOL, or Lean); \emph{realistic} means capturing enough features of corresponding unverified tools for real-world tasks; and \emph{end-to-end} means using the same standard, well-specified interfaces and/or languages as corresponding unverified tools. The best-known project meeting this standard is CompCert~\cite{Leroy2009}: mechanised in Rocq; realistic enough to include many real-world optimisation passes and generate efficient code; and end-to-end because it takes (a large subset of) standard C and outputs standard assembly.

We make substantial progress towards this gold standard for CSE. Previous CSE theories~\cite{Jacobs2015,Dardinier2025,Loow2025} are foundational, but fall short of being realistic and end-to-end: their frontends are too limited for real-world tasks, and their backends have no formal connection to SMT-LIB compliant SMT solvers, relying instead on, e.g., a toy solver or an assumed SL solver. The theory we contribute is mechanised in Rocq, covers many of the CSE features used in real-world verification case studies, and comes with an end-to-end guarantee down to standard SMT solvers through a new Rocq mechanisation of the SMT-LIB standard and a verified encoding of expressions in our theory into SMT-LIB. The solver is the one component we axiomatise, and we assume of it only what the standard prescribes. A tool following our theory therefore trusts its solver for SMT-LIB compliance and nothing more. Otherwise, our results rely only on the Rocq kernel and classical axioms.\footnote{Excluded middle, constructive indefinite description, functional extensionality, proof irrelevance, and an axiom of the real numbers, all from Rocq's standard library.}

We detail the frontend and backend gaps of previous work (\Cref{sect:introduction:frontend,sect:introduction:backend}) and summarise our contributions (\Cref{sect:introduction:contributions}). Our main contribution is closing the backend gap, but both gaps must be closed together, since everything the frontend adds to expressions must be encoded into SMT-LIB.

\subsection{CSE Frontend Gap (Realism Gap)}%
\label{sect:introduction:frontend}

Modern CSE tools, verification tools in particular, support a wide range of expressive features. A realistic CSE theory must not only define these features but also, critically, prove them sound. To illustrate the frontend gap of existing CSE theories, we return to the list-reversal function from \Cref{fig:interop}, which we implement as $\texttt{reverse}(\pvar{x})~\{ \pfuncall{\pvar r}{\mathtt{reverse'}}{\pvar x, \nil}; \preturn{\pvar r} \}$ using a helper function $\mathtt{reverse'}$ that reverses $\pvar x$ onto an accumulator. \Cref{fig:example-spec} gives $\mathtt{reverse'}$ and its full verification setup, highlighting the \grayboxed{\text{expressive features}} missing from previous CSE theories. To verify \(\mathtt{reverse'}\) (\Cref{fig:example-spec:spec}), a proof developer defines an algebraic data type (ADT) for list contents (\Cref{fig:example-spec:adt}) and logical functions over it, such as concatenation (\(\concat\)) and reversal (\(^\dag\)) (\Cref{fig:example-spec:lfunc}). The \(\pname{list}\) predicate (\Cref{fig:example-spec:predicate}) is an inductively defined spatial predicate asserting that \(\lvar{x}\) points to a heap-allocated linked list with contents \(\alpha\). The specifications for \(\mathtt{reverse'}\) and \(\mathtt{prepend}\) use these definitions. During the verification of \(\mathtt{reverse'}\) by symbolic execution, calls to \(\mathtt{prepend}\) and \(\mathtt{reverse'}\) update the symbolic state via their specifications: given the list at pointer $x$ representing logical list $\alpha = a :: \alpha'$ and the list at pointer $y$ representing $\beta$, prepending $a$ to $\beta$ then reversing $\alpha'$ into the prepended list yields \(\alpha'^\dag \concat (a :: \beta)\).
\begin{figure}[t]
    \centering
    \footnotesize
    \begin{tblr}{
      width=\linewidth,
      colspec = {X[c] X[c] X[c]},
      measure = vbox,
      rowsep=0pt}
      \begin{minipage}{0.3\textwidth}
        \centering
        $
        \begin{array}{l}
          \specline{\pvar{x} = x \lstar \pvar{y} = y \lstar \\
          \llist{ x, \grayboxed{\asrtcolor{\alpha}}} \lstar \llist{{y}, \grayboxed{\asrtcolor{\beta}}}} \\
          \pfuncspec{reverse'}{\pvar{x}, \pvar{y}}~\{ \\
          \quad \pifelses{\pvar{x} = \nil}{\passign{\pvar r}{\pvar y}} \pifelsem \\
          \quad \quad \pderef{\pvar h}{\pvar x}; \\
          \quad \quad \pderef{\pvar t}{\pvar x + 1}; \\
          \quad \quad \pfuncall{\pvar{y}}{\mathtt{prepend}}{\pvar h, \pvar y}; \\
          \quad \quad \pfuncall{\pvar r}{\mathtt{reverse'}}{\pvar t, \pvar y}; \\
          \quad \quad \grayboxed{\ghostcolor{
          \begin{array}{l}
            \pmatchs{\alpha}{\pname{List}} \\
            \quad \pcase{[]}{\pskip} \\
            \quad a :: \alpha' \rightarrow \\
            \quad \papply{concatr}{\alpha'^\dag, a, \beta} \\
            \pmatche
          \end{array}}
          } \\   
          \quad \pifelsee; \\
          \quad \preturn{\pvar r} \\
          \} \\
          \specline{\llist{x, \grayboxed{\asrtcolor{\alpha}}} \lstar \llist{\pvar{ret}, \grayboxed{\asrtcolor{\alpha^\dag \concat \beta}}}}
        \end{array}
        $
        \subcaption{Function and its specification}\label{fig:example-spec:spec}
      \end{minipage}
      &
      \begin{minipage}{0.3\textwidth}
        \begin{minipage}{\linewidth}
        \[
            \grayboxed{\pname{List} ::= [] \mid \vals :: \pname{List}}
         \]
        \subcaption{ADT definition}\label{fig:example-spec:adt}
        \end{minipage}

        \vspace{1em}

        \begin{minipage}{\linewidth}
        \[
            \grayboxed{
            \begin{array}{l}
                (\alpha : \pname{List}) \concat (\beta : \pname{List}) : \pname{List} = \\
                \quad \match{\alpha}{\pname{List}}{}\\
                \;\;\;\; \begin{array}{l l}
                                []           &\rightarrow \beta \\
                                a :: \alpha' &\rightarrow a :: (\alpha' \concat \beta)
                            \end{array} \\
                \mathrm{\mathbf{when}}\ \true
            \end{array}
            }
        \]
        \vspace{0.5em}
        \[
            \grayboxed{
            \begin{array}{l}
                (\alpha : \pname{List})^\dag : \pname{List} = \\
                \quad \match{\alpha}{\pname{List}}{}\\
                \quad \begin{array}{l l}
                          []           &\rightarrow [] \\
                          a :: \alpha' &\rightarrow \alpha'^\dag \concat [a]
                      \end{array} \\
                \mathrm{\mathbf{when}}\ \true
            \end{array}
            }
        \]
        \subcaption{Logical function definition}\label{fig:example-spec:lfunc}
        \end{minipage}
      \end{minipage}
      &
      \begin{minipage}{0.3\textwidth}
        \begin{minipage}{\linewidth}
          \centering
          $
          \grayboxed{
          \begin{array}{l}
            \mathtt{concatr} : \forall \alpha, a, \beta.\\
            \quad \asrtcolor{\alpha \in \pname{List} \lstar \beta \in \pname{List}} \Rightarrow \\
            \quad \asrtcolor{\alpha \concat (a :: \beta) =} \\
            \quad \asrtcolor{(\alpha \concat [a]) \concat \beta} \\
            \{ \\
            \quad \pmatchs{\alpha}{\pname{List}} \\
            \quad \quad \pcase{[]}{\pskip} \\
            \quad \quad \ghostcolor{a' :: \alpha' \rightarrow} \\
            \quad \quad \quad \ghostcolor{\papply{concatr}{\alpha', a, \beta}} \\
            \quad \pmatche \\
            \} \\
            (|\alpha|\ \mathrm{\mathbf{decreases}})
          \end{array}
          }
          $
          \setcounter{subfigure}{4}
          \subcaption{Lemma statement and proof}\label{fig:example-spec:lemma}
        \end{minipage}

        \vspace{1em}

        \begin{minipage}{\linewidth}
          \centering
          $
          \begin{array}{c}
            \specline{\pvar{a} = a \lstar \pvar{x} = x \lstar \llist{x, \grayboxed{\asrtcolor{\alpha}}}} \\
            \pfuncspec{prepend}{\pvar a, \pvar x} \\
            \specline{\llist{\pvar{ret},  \grayboxed{\asrtcolor{a :: \alpha}}}}
          \end{array}
          $
          \subcaption{Prepend's specification}\label{fig:example-spec:append}
        \end{minipage}
      \end{minipage}
    \end{tblr}

    \begin{subfigure}{\textwidth}
      \[
        \llist{x, \alpha} =
          \asrtcolor{(x = \nil \lstar \alpha = \grayboxed{[]}) \lor (\exists a, \alpha', y. \,\twocell{x}{a}{y} \lstar \alpha = \grayboxed{a :: \alpha'} \lstar \llist{y, \alpha'})}
      \]
      \setcounter{subfigure}{3}
      \caption{List predicate definition}\label{fig:example-spec:predicate}
    \end{subfigure}
    \caption{An example verification of a specification in our CSE theory, using \grayboxed{\text{expressive features}} missing in previous theories. We omit ghost commands for folding and unfolding user-defined predicates (to avoid clutter) and the definition of the list-length function $|\cdot|$ (as it is similar to the logical functions shown).}%
    \label{fig:example-spec}
\end{figure}
Establishing the final postcondition of \(\mathtt{reverse'}\) requires the inductive fact \(\lemstmt{\alpha, a, \beta}{\alpha \in \pname{List} \lstar \beta \in \pname{List}}{\Rightarrow}{\alpha \concat (a :: \beta) = (\alpha \concat [a]) \concat \beta}\). As SMT solvers lack inductive reasoning, CSE tools support lemmas (\Cref{fig:example-spec:lemma}). The developer provides a lemma and a ghost command $\pvar{apply}$ to trigger its application.

The difficulty lies in proving these features sound, and in doing this in one formalism. For example, \(\mathtt{reverse'}\) is verified using its own specification at the recursive call, before that specification is established, so the soundness proof must break this circularity. Similarly, the inductive proof of \(\mathtt{concatr}\) applies \(\mathtt{concatr}\) itself, so the theory must show that all proofs it accepts are well-founded, e.g.\ by checking that a user-provided measure, here \(|\alpha|\), decreases at each application.

\paragraph{Previous work.} Prior CSE theories have inexpressive frontend languages, lacking user-defined data types, logical functions, and lemmas: FVF supports only basic arithmetic and boolean operations; \citet{Dardinier2025} and \citet{Loow2025} extend this only modestly. No prior theory can verify recursive functions against specifications whose meaning is independent of the engine. These theories therefore cannot model realistic verification tasks.

\subsection{CSE Backend Gap (End-to-End Gap)}%
\label{sect:introduction:backend}

Frontend expressivity increases backend encoding complexity: as the frontend becomes richer, the semantic gap to SMT-LIB widens.
In particular, user-defined functions expose a mismatch between CSE's operational semantics and SMT-LIB's declarative semantics. Operationally, in CSE, a function might not terminate but still have meaning in SMT-LIB: for example, the definition \(f(x) = f(x)\) does not terminate, but in SMT-LIB it corresponds to the axiom \(\forall x.\ f(x) = f(x)\), which places no constraint on \(f\). As a result, the solver may interpret \(f\) arbitrarily, allowing it to be ``evaluated'' despite its intended non-termination. Partial operations exacerbate this mismatch. CSE typically treats division or data type destructors as erroring on certain inputs, whereas SMT-LIB assumes total functions and leaves such cases underspecified. Solvers may therefore assign arbitrary values to erroneous cases (e.g.\ interpreting \(1 / 0 = 42\)), yielding spurious satisfiability results.

Typing adds another mismatch. SMT-LIB is many-sorted and statically typed, so SMT interaction must coerce CSE values into SMT-LIB sorts. The gap is widest for dynamically typed CSE tools: Gillian's frontend, for example, is dynamically typed to support dynamically typed languages, tracing its ancestry to the JavaScript verifier JaVerT~\cite{Fragososantos2018}. But even statically typed CSE tools need coercions, as their types need not coincide with SMT-LIB sorts. In either case, SMT interaction must rule out SMT models containing values not \emph{representable} in the CSE logic: if 256-bit unsigned integers are encoded using SMT-LIB's unbounded integer type, solver values such as \(-1\) or \(2^{256}\) must be excluded despite being valid SMT-LIB integers.

\paragraph{Previous work.} Existing theories do not adequately account for SMT interaction. FVF implements a toy decision procedure for its assertion language, bypassing SMT-LIB. Dardinier et al.\ connect their formalisation to \citet{Parthasarathy2024}, establishing an end-to-end connection via Boogie but no connection to SMT interaction. L\"{o}\"{o}w et al.\ assume a separation-logic solver for their assertion language. Thus, no prior CSE theory provides a verified connection to off-the-shelf SMT solvers.

\subsection{Contributions}%
\label{sect:introduction:contributions}

We contribute a mechanised CSE theory that substantially narrows our stated realism and end-to-end gaps, building on~\citet{Loow2025}.  L\"{o}\"{o}w et al.'s theory is the only one that creates and uses specifications of functions whose meaning is given by separation-logic satisfaction, independently of the engine, an independence essential for our end-to-end agenda. L\"{o}\"{o}w et al.'s theory also uniquely covers both SL correctness analysis and ISL incorrectness analysis; our extension preserves both.

\smallskip
\noindent
\textbf{Frontend extensions (\Cref{sect:frontend}).} Our CSE theory has an expressive frontend language with user-defined algebraic data types (\(\pname{List}\), \Cref{fig:example-spec:adt}), with their associated match expressions (\(\match{\alpha}{\pname{List}}{\cdots}\), \Cref{fig:example-spec:lfunc}) and command-level match (\(\pmatch{\alpha}{\pname{List}}{\cdots}\), \Cref{fig:example-spec:spec}); logical functions (\(\concat\) and \(^\dag\), \Cref{fig:example-spec:lfunc}); definitions and proofs of lemmas (\(\mathtt{concatr}\), \Cref{fig:example-spec:lemma}); and soundness proofs for recursive function verification and measure-based total-correctness arguments.

\smallskip
\noindent
\textbf{SMT-LIB mechanisation (\Cref{sect:smt}).} To state correctness of our backend extensions, we contribute the first line-by-line mechanisation of SMT-LIB 2.7, provided as a reusable Rocq package. It covers:
\begin{enumerate}
\item the syntax and semantics of the SMT-LIB language;
\item the definition of SMT-LIB theories and the composition of SMT-LIB theories;
\item the SMT-LIB theories most commonly used by CSE engines: the standard theories \texttt{Core}, \texttt{HO-Core}, \texttt{Reals\_Ints}, and (the core part of) \texttt{Strings}, as well as the common but nonstandard theory \texttt{Seq} (i.e., the theory of sequences).
\end{enumerate}
\smallskip

\noindent
\textbf{Backend extensions (\Cref{sect:backend}).} Built on our SMT-LIB mechanisation, our CSE theory contains a verified encoding of CSE state satisfaction and implication queries into SMT-LIB queries, covering:
\begin{enumerate}
\item all our expressive specification features such as user-defined data types and functions;
\item partial expression evaluation of CSE into SMT-LIB's total expression evaluation;
\item dynamically typed expressions of CSE into SMT-LIB's statically typed expressions;
\item representability conditions ensuring that satisfying SMT-LIB models correspond to CSE values and partial logical interpretations.
\end{enumerate}
We also verify two type-based encoding optimisations:
\begin{enumerate}
\item inspired by gradual type systems~\cite{SiekTaha2006} with incomplete type information, and compilers exploiting type information for efficient code generation~\cite{Gal2009,Leroy1992}, we map subexpressions with known types to native SMT-LIB sorts;
\item inspired by work on ill-typing~\cite{Ramsay2024}, we use known types to rule out unsatisfiable queries early, without consulting the underlying SMT solver.
\end{enumerate}

\smallskip
\noindent
\textbf{Empirical evaluation (\Cref{sect:eval}).} We evaluate the frontend against examples from VeriFast and Gillian, and the backend by extracting the verified encoder to OCaml and connecting it to Gillian's solver loop. A bridge translates 75\% of Gillian's queries into our theory; the rest mostly contain negative integers, which our theory lacks. On the queries that both backends answer, the verified encoder agrees with Gillian's except on 14 queries where the expression languages differ in meaning, and solver times are comparable except on a small tail (1.5\%) with a known cause.


\section{Overview}%
\label{sect:overview}

We summarise the theory of \citet{Loow2025}, then describe our frontend and backend extensions.

\subsection{Background: A Crash Course in Compositional Symbolic Execution}

\newcommand{\csesimp}[3]{#1, #2 \baction #3}
\newcommand{\csesimpgamma}[4]{#1, #2 \baction_{#3} #4}
\newcommand{\csesimpgammamode}[5]{#1, #2 \baction_{#3}^{#4} #5}

L\"{o}\"{o}w et al.'s theory uses a small demonstrator language with a big-step semantics over concrete states \(\st\) consisting of a variable store and heap.  The judgement 
\(\csesimpgamma{\st}{\cmd}{\gamma}{o: \st'}\) reads: evaluating command \(\cmd\) in state \(\st\), with function definitions $\gamma$, produces outcome $o$ and final state \(\st'\).

\begin{figure}[t]
    \centering
    \begin{minipage}[t]{0.48\textwidth}
        \centering
        \usetikzlibrary{positioning}
        \begin{tikzpicture}[scale=0.9, transform shape]
            \node (init) {$\sst, \; \pifelse{\overbrace{\pvar x = \nil}^{\pexp}}{\cmd_1}{\cmd_2}$};
            \node (err) [below=of init] {(1) $\sst[\sstupdate{pc}{(\pexp \notin \bools \wedge \access{\sst}{pc})}], -$};
            \node (true) [below=of init, yshift=-0.75cm, xshift=-1cm] {(2) $\sst[\sstupdate{pc}{(\pexp = \true \wedge \access{\sst}{pc})}], \; \cmd_1$};
            \node (false) [below=of init, yshift=-1.5cm, xshift=0.5cm] {(3) $\sst[\sstupdate{pc}{(\pexp = \false \wedge \access{\sst}{pc})}], \; \cmd_2$};
            \draw[->]
                (init) to
                (err);

            \draw[->]
                (init) to[bend right]
                ($(true.north)+(-1.75,0)$);

            \draw[->]
                ($(init.south)+(2,0)$) to[bend left]
                ($(false.north)+(1.75,0)$);
        \end{tikzpicture}
        \caption{Part of the symbolic execution tree of $\mathtt{reverse'}$. Symbolic state is treated like a record and \texttt{pc} refers to ``path condition.''}%
        \label{fig:branching}
    \end{minipage}%
    \hfill
    \begin{minipage}[t]{0.48\textwidth}
        \centering
        \usetikzlibrary{positioning,decorations.pathreplacing}
        \begin{tikzpicture}[
            scale=0.9,
            transform shape,
            >=stealth,
            state/.style={draw, rectangle, minimum width=0.75cm,
                          minimum height=0.45cm, inner sep=1.5pt},
            font=\small]
            \node[state] (preF) at (0.7,0) {\(\sst_F\)};
            \node[state, right=-0.4pt of preF] (preP) {\(\asrtcolor{P}\)};

            \node[state, below=2.0cm of preF] (postF) {\(\sst_F\)};
            \node[state, right=-0.4pt of postF] (postQ) {\(\asrtcolor{Q}\)};

            \draw[decorate, decoration={brace, amplitude=4pt, raise=2pt}]
              (preF.north west) -- node[above=7pt] {\(\sst\)} (preP.north east);
            \draw[decorate, decoration={brace, mirror, amplitude=4pt, raise=2pt}]
              (postF.south west) -- node[below=7pt] {\(\sst'\)} (postQ.south east);

            \draw[->] (preF.south) -- node[right, align=left, fill=white, inner sep=3pt] (cmd) {%
              \(\pfuncall{\pvar y}{\mathtt{prepend}}{\pvar h, \pvar y}\)} (postF.north);

            \node[align=center, right=-0.45pt of cmd] (spec) {
              $
              \begin{array}{c}
                \asrtcolor{\{ \overbrace{\pvar a = a \lstar \pvar x = x \lstar \llist{x, \alpha}}^{P} \}} \\
                \pfuncspec{prepend}{\pvar a, \pvar x} \\
                \asrtcolor{\{ \underbrace{\llist{\pvar{ret}, a :: \alpha}}_{Q} \}}
              \end{array}
              $
            };

            \draw[->] (spec.north) to[out=155, in = 0]
              node[above, align=center, inner sep=3pt, xshift=4pt] {%
                \(\consume(\asrtcolor{P},\hat{\theta}, \sst) \rightsquigarrow (\hat{\theta}', \sst_F)\)}
              (preP.east);
            \draw[->] (spec.south) to[out=205, in=0]
              node[pos=0.48, below, align=center, inner sep=3pt] {%
                \(\produce(\asrtcolor{Q}, \hat{\theta}' , \sst_F) \rightsquigarrow
                \sst'
                \)}
              (postQ.east);
        \end{tikzpicture}
        \caption{Functional compositionality, underpinned by \(\consume\) and \(\produce\) operations.}%
        \label{fig:consume-produce}
    \end{minipage}
\end{figure}

\paragraph{CSE engine judgement}

The CSE engine is defined over symbolic states \(\sst\) with first-order path expressions recording branch constraints. The symbolic semantics is defined by the judgement $\csesimpgammamode{\sst}{\cmd}{\Gamma}{m}{o: \sst'}$, which is interpreted analogously to the concrete semantics, with two differences: the semantics is parameterised by a mode ($m \in \{\mOX, \mUX\}$);  and the context $\Gamma$ provides callable function specifications rather than the function bodies supplied by $\gamma$. \Cref{fig:branching} illustrates branching and path-condition updates for \(\pifelse{\pvar x = \nil}{\cdots}{\cdots}\) in \(\mathtt{reverse'}\) from \Cref{fig:example-spec:spec}: evaluation of the condition either (1) fails or yields a non-boolean, with error outcome $\oxerr$, (2) yields \(\true\) with success outcome $\oxok$, or (3) yields \(\false\) with $\oxok$.

\paragraph{OX and UX soundness}
The key CSE-engine properties are OX and UX soundness; we give a simplified account here and a precise one in \Cref{sect:background}. Let $\st \models \sst$ mean concrete state $\st$ models symbolic state $\sst$. Then OX soundness is: when $o = \oxok$ or $\oxerr$,
\[
\csesimpgamma{\st}{\cmd}{\gamma}{o: \st'} \land \st \models \sst \Rightarrow \exists \sst'.\; \csesimpgammamode{\sst}{\cmd}{\Gamma}{\mOX}{o: \sst'} \land \st' \models \sst',
\]
i.e., each concrete execution has a corresponding symbolic execution. Concrete execution never gets stuck, as all errors are explicit outcomes. UX soundness guarantees the opposite relation between concrete and symbolic execution: when $o = \oxok$ or $\oxerr$, 
\[
\csesimpgammamode{\sst}{\cmd}{\Gamma}{\mUX}{o: \sst'} \land \st' \models \sst' \Rightarrow \exists \st.\;  \csesimpgamma{\st}{\cmd}{\gamma}{o: \st'} \land \st \models \sst,
\]
i.e., each concrete state modelling the result of a symbolic execution is reached by a corresponding concrete execution.
OX soundness supports verification: concrete behaviour is contained in symbolic behaviour, so concrete bugs are reflected symbolically and all-successful symbolic paths rule out concrete bugs. UX soundness supports bug finding: a symbolic bug corresponds to a concrete bug.

\paragraph{Consume and produce operations}

Functional compositionality in CSE comes from the connection between symbolic state and separation-logic assertions used in function specifications. L\"{o}\"{o}w et al. capture this interface using mathematical descriptions of the \(\consume\) and \(\produce\) operations, characterising the consume and produce implementations of, for example,  VeriFast, Viper, and Gillian: \(\consume\) \emph{matches and removes} symbolic state corresponding to an assertion; $\produce$ dually \emph{adds} it. These operations implement function-specification verification, function-call reasoning, and predicate folding/unfolding. \Cref{fig:consume-produce} illustrates their use in symbolically executing the call $\pfuncall{\pvar y}{\mathtt{prepend}}{\pvar h, \pvar y}$ of \Cref{fig:example-spec:spec}. Under substitution \(\hat{\theta}\), which maps the specification's logical variables \(a\) and \(x\) to the symbolic values of the arguments \(\pvar h\) and \(\pvar y\), consuming \(\asrtcolor{P}\) from \(\sst\) removes the list at \(\pvar y\). It yields the frame \(\sst_F\) and the substitution \(\hat{\theta}'\), which extends \(\hat{\theta}\) by mapping \(\alpha\) to the contents of the list at \(\pvar y\). Producing \(\asrtcolor{Q}\) with \(\hat{\theta}'\) on \(\sst_F\) then yields \(\sst'\) by adding the returned list, now at \(\pvar y\), with contents \(\hat{\theta}'(a :: \alpha)\), which is the value of \(\pvar h\) prepended to the old contents. Function-specification verification is the dual: (1)~produce $\asrtcolor{P}$ into an empty symbolic state; (2)~execute the function body symbolically; and (3)~consume $\asrtcolor{Q}$ from the final state. Success in all three steps verifies the specification.

\subsection{Frontend Extensions}

\begin{table}[t]
\centering
\begin{threeparttable}
\caption{Frontend features in CSE tools and theories. For tools, entries (e.g., \cmark{} and \xmark{}) indicate whether a feature is supported; for theories, whether it is justified by the theory. The feature columns cover whether logical functions are total or partial (logical funs.), whether the logic is statically or dynamically typed (typing), user-defined predicates (UDP), user-defined logical types (UDLT), user-defined logical functions (UDLF), function specification application and verification (spec.), and lemma application and verification (lem.).}%
\label{table:frontend}
\begin{tabularx}{\textwidth}{Xlllllll}
\toprule
\textbf{CSE tool} & \textbf{Logical funs.} & \textbf{Typing} & \textbf{UDP} & \textbf{UDLT} & \textbf{UDLF} & \textbf{Spec.} & \textbf{Lem.} \\ \midrule
VeriFast~\cite{Jacobs2011} & Total & Static & \cmark & ADTs & \cmark & \cmark & \cmark \\
CN~\cite{Pulte2023} & Total & Static & \cmark & ADTs & \cmark & \cmark & \almostmark\tnote{1} \\
Viper~\cite{Muller2016} & Partial & Static & \cmark & Domains\tnote{2} & \cmark & \cmark & \xmark\tnote{3} \\
Gillian~\cite{Maksimovic2021} & Total\tnote{4} & Dynamic & \cmark & \xmark & \xmark & \cmark & \cmark \\ \midrule
\textbf{CSE theory} & ~ & ~ & ~ & ~ & ~ & ~ & ~ \\ \midrule
FVF~\cite{Jacobs2015} & Total & Static & \cmark & \xmark & \xmark & \almostmark\tnote{5} & \xmark \\
\citet{Dardinier2025} & Partial & Static & \xmark & \xmark & \xmark & \xmark\tnote{6} & \xmark \\
\citet{Loow2025} & Partial & Dynamic & \cmark & \xmark & \xmark & \almostmark\tnote{7} & \xmark \\
This paper & Partial & Dynamic & \cmark & ADTs & \cmark & \cmark & \cmark \\ \bottomrule
\end{tabularx}

\begin{tablenotes}
\footnotesize
    \item[1] Lemmas must be encoded as C functions with an empty body, which can be exported to Rocq and proved there.
    \item[2] We do not discuss domains further. There is an ADT plugin for Viper that automatically translates ADTs to domains.
    \item[3] No special lemma support is provided; lemmas must be encoded as functions.
    \item[4] This is not a principled design decision, and the tool may deviate from it.
    \item[5] The theory does not assign meaning to function specifications that are independent of the engine, meaning that specifications cannot be transferred in and out of the engine. On top of this, there is no support for~verifying~termination.
    \item[6] The theory has no functions; inhale and exhale apply and verify only loop and parallel-branch specifications.
    \item[7] The theory does not justify the verification of recursive functions.
\end{tablenotes}
\end{threeparttable}
\end{table}

We discuss our frontend extensions against existing tools and theories; see \Cref{table:frontend}.

\paragraph{Non-extensions: total vs. partial logical functions; statically typed vs. dynamically typed logic; and user-defined predicates}
For the first three columns, we inherit L\"{o}\"{o}w et al.'s design choices and Rocq implementation. As discussed in the introduction, a dynamically typed logic with partial operations creates a relatively large gap to SMT-LIB.
Despite the popularity of dynamically typed languages such as JavaScript and Python, most analysis tools based on separation logic target only statically typed languages. The only exception we know is Gillian~\cite{Maksimovic2021}, and its predecessor JaVerT~\cite{Fragososantos2018}, which can analyse JavaScript. Nagini~\cite{Eilers18}, built on Viper, analyses Python but only for typed programs.

\paragraph{User-defined logical data types and functions}
Previous CSE theories lack user-defined logical ADTs, e.g. the ADT $\pname{List}$ in \Cref{fig:example-spec:adt}, and functions, e.g. $\pname{++}$ in \Cref{fig:example-spec:lfunc}. We extend L\"{o}\"{o}w et al.\ with ADT constructor application (e.g.\ \(a :: \alpha\)), logical function application (e.g.\ \(\alpha^\dag \concat \beta\)), ADT match expressions (e.g.\ \(\match{\alpha}{\pname{List}}{\cdots}\)), and command-level ADT match (e.g. \(\pmatch{\alpha}{\pname{List}}{\dots}\)). This raises Rocq challenges (\Cref{sect:frontend}), but the main work is in the SMT-LIB encoding (\Cref{sect:backend}).

\paragraph{Function specification application and verification}
L\"{o}\"{o}w et al.'s account of function calls and verification is the most complete among CSE theories: function specifications are used at function calls (\Cref{fig:consume-produce}), and the results of verification are valid function specifications in an SL. It lacks a soundness proof for recursive function verification, which we provide by adapting \emph{depth-indexing} and induction on depth from Nipkow's work for Hoare logic~\cite{Nipkow2002}.

\paragraph{Lemma application and verification} 
User-defined ADTs require inductive reasoning, but no previous CSE theory formalises \emph{lemmas} (e.g.\ \Cref{fig:example-spec:lemma}) and their ghost commands. Lemma application and verification mirror function calls and verification: lemmas are functions whose bodies consist solely of ghost commands representing proofs. \Cref{fig:example-spec:lemma} gives the measure needed to check that the inductive proof of \(\mathtt{concatr}\) is well-founded, analogous to checking function termination. We again adapt~\cite{Nipkow2002}, using a \emph{measure-based} approach: the user provides a measure over lemma arguments, here \(|\alpha|\), which must decrease on each inductive application. Soundness then follows by induction over the measure. We also apply this argument to function termination, i.e., \emph{total correctness}.

\subsection{Backend Extensions}

The backend handles symbolic state satisfaction, $\satpred(\sst)$ meaning $\exists \st, \st \models \sst$, and implication queries, $\sst \models \sexp$ meaning $\sexp$ is true in all models of $\sst$. Its main tasks are (1)~cutting branches with state $\sst$ when \(\neg \satpred(\sst)\), and (2)~checking entailment $\sst \models \sexp$ during consume.

Branch cutting differs between OX verification and UX true bug-finding. OX soundness requires exploring \emph{all} satisfiable branches, so cutting unsatisfiable branches is sound but optional; it improves performance by controlling exponential branching and prevents false-negative verification failures from unsatisfiable error branches. UX soundness requires exploring \emph{only} satisfiable branches, so cutting unsatisfiable branches is mandatory. In \Cref{fig:branching}, the engine branches on $\pvar{x} = \nil$: the error branch is unrealisable because $\pvar x$ is always present as a function parameter, so OX cutting avoids spurious verification failure and UX cutting avoids a false bug report.

No previous CSE theory accounts for encoding such queries into SMT-LIB: for example, L\"{o}\"{o}w et al.\ assume an independent oracle for \(\satpred(\sst)\) and \(\sst \models \sexp\), although $\sst$ and $\sexp$ are engine-specific constructions. We address this with a Rocq mechanisation of SMT-LIB and a verified encoding of symbolic expressions into SMT-LIB queries. We extend symbolic states with a typing environment \(\ste\), which records types of logical variables already entailed by the state. The encoding is a judgement \(\ste \vdash \sexp \leadsto t : \tau \mid \Phi\), which encodes the symbolic expression \(\sexp\) under \(\ste\) as an SMT-LIB term \(t\) of sort \(\tau\) that faithfully represents \(\sexp\) whenever the side conditions \(\Phi\) hold. We use the encoding to define expression satisfiability checks \(\issat(\sexp)\), OX and UX state satisfiability checks \(\issat(\mOX / \mUX, \sst)\), and entailment checks \(\isentails(\sst, \sexp)\). OX and UX soundness reduce to the following properties of the checks, where \(\esem{\sexp}{\subst}\) evaluates \(\sexp\) under a logical interpretation \(\subst\), mapping logical variables to values:
\[
\begin{aligned}
(\text{OX})~& \quad \esem{\sexp}{\subst} = \true \Rightarrow \issat(\sexp) \in \{\sat, \unknown\}
\\
(\text{UX})~& \quad \issat(\sexp) = \sat \Rightarrow \exists \subst \ldotp \esem{\sexp}{\subst} = \true .
\end{aligned}
\]

Calling symbolic expressions, their values, and logical interpretations the \emph{source} side, the encoding solves four problems. First, it reflects user-defined ADTs and logical functions into the SMT-LIB signature. Second, source evaluation may error, whereas SMT-LIB functions are total and underspecified on the corresponding arguments, so the side conditions rule out these underspecified SMT-LIB evaluations. Third, source expressions are dynamically typed, whereas every SMT-LIB term must have a single, statically known sort. The encoder is therefore a type-directed translation based on a gradual type system~\cite{SiekTaha2006}, which assigns each expression a sort. It wraps values of unknown type in an SMT-LIB ADT with one constructor per type and coerces between it and native sorts where needed, while \(\ste\) lets variables of known type use native sorts and lets the check detect some type errors before calling the solver. Fourth, the queries restrict SMT-LIB models to values representable as source values, because some target sorts over-approximate source values: e.g., integers encoding naturals can be negative.


\section{Background: CSE Theory}\label{sect:background}

We introduce the parts of \citet{Loow2025}'s CSE theory needed for our frontend (\Cref{sect:frontend}) and backend (\Cref{sect:backend}) extensions; further details are in~\appref{app:background-defs}{A}. This CSE theory is uniquely parametric on the memory model, expressed algebraically using a partial commutative monoid with memory actions satisfying frame and soundness properties familiar to separation logics. Our Rocq mechanisation keeps this parametricity, but here we use the standard linear memory, as parametricity matters only for symbolic state satisfiability and entailment (\Cref{sect:backend:satisfiability}).

\subsection{Syntax}

\newcommand{\prg}{\mathsf{PF}}
\newcommand{\prgs}{\mathsf{PrgFrg}}

\newcommand{\specthing}[1]{{\asrtcolor{\left\langle\!\! \left\langle #1 \right\rangle\!\! \right\rangle}}}

\begin{figure}[t]
\centering
\begin{minipage}[t]{0.47\textwidth}
\centering
{\small
\[
\begin{array}{r@{~}c@{~}l}
\gv \in \vals &::=& \nil \mid \bv \in \bools \mid \nv \in \nats \\
&\mid& q \in \rationals \mid \sv \in \strings \mid [\lst{v}] \\
\type \in \types &::=& \vals \mid \nils \mid \bools \mid \nats \\
&\mid& \rationals \mid \strings \mid \listsof{\type} \\
\multicolumn{3}{l}{\pvar x, \pvar y, \pvar z, \ldots \in \pvars \quad \odot \in \{ \neg, \dots \} \quad \oplus \in \{ +, /, \dots \}} \\
\pexp \in \pexps &::=& \gv \mid \pvar x \mid \odot \pexp \mid \pexp \oplus \pexp \quad\quad f \in \strings \\
\cmd \in \pcmds & ::= &\pskip \mid \psequence{\cmd}{\cmd} \mid \pifelse{\pexp}{\cmd}{\cmd} \\
&\mid& \passign{\pvar{x}}{\pexp} \mid \pfuncall{\pvar{x}}{f}{\lst{\pexp}} \mid \pderef{\pvar{x}}{\pexp} \\
&\mid& \pmutate{\pexp}{\pexp} \mid \palloc{\pvar{x}}{\pexp} \mid \pdealloc{\pexp}
\end{array}
\]
}
\end{minipage}%
\hfill%
\begin{minipage}[t]{0.52\textwidth}
\centering
{\small
\[
\begin{array}{r@{~}c@{~}l}
x, y, z, \ldots &\in& \lvars \quad\quad p \in \strings \\
\lexp \in \lexps &::=& \gv \mid \pvar x \mid x \mid \odot \lexp \mid \lexp \oplus \lexp \mid [\lst{\lexp}] \mid \lexp \in \type \\
\gcmd \in \gcmds & ::= & \ghostcolor{\pgskip} \mid \ghostcolor{\pgseq{\gcmd}{\gcmd}} \mid \ghostcolor{\pgifelse{\lexp}{\gcmd}{\gcmd}} \\
& \mid & \ghostcolor{\pfold{\pred}{\vec{\lexp}}} \mid \ghostcolor{\punfold{\pred}{\vec{\lexp}}} \\
\cmd \in \cmds &::=& \ldots \mid \ghostcolor{\gcmd} \\
\asrtcolor{P, Q} \in \Assert &::=& \asrtcolor{\lexp} \mid \asrtcolor{\cell{\lexp}{\lexp}} \mid \asrtcolor{\pred(\vec{\lexp})} \mid \asrtcolor{\emp} \mid \asrtcolor{\AssTrue} \\
&\mid& \asrtcolor{P \lstar Q} \mid \asrtcolor{P \vee Q} \mid \asrtcolor{P \Rightarrow Q} \mid \asrtcolor{\exists x \ldotp P} \mid \ldots \\
\prg \in \prgs &::=& \prg; \prg \mid \pred(\vec{x}) = \asrtcolor{P} \\
&\mid& \triple{P}{f(\lst{\pvar{x}})~\{~\psequence{\cmd}{\preturn{\pexp}}~\}}{Q}
\end{array}
\]
}
\end{minipage}
\caption{The syntax of the programming language without annotations (left) and with annotations (right). Arrows denote lists, e.g., \(\lst{v}\) is a list of values and \(\lst{\pexp}\) a list of (program) expressions.}%
\label{fig:syntax}
\end{figure}

We use the standard demonstrator language of \cref{fig:syntax} (left), with values $\vals$, types $\types$,\footnote{Like \(\bools\) and \(\nats\), \(\listsof{\type}\) is a built-in type, distinct from the user-defined ADT \(\pname{List}\) of \Cref{fig:example-spec:adt}.} (program) variables $\pvars$, (program) expressions $\pexps$, and (program) commands $\pcmds$ with memory actions: lookup $\pderef{\pvar{x}}{\pexp}$, mutation $\pmutate{\pexp}{\pexp}$, allocation $\palloc{\pvar{x}}{\pexp}$, and deallocation $\pdealloc{\pexp}$.

Which program annotations are relevant depends on the analysis; UX bug-finding, for example, requires none. We illustrate those needed for verifying OX function specifications in \cref{fig:syntax} (right): logical variables $\lvars$ distinct from program variables; logical expressions $\lexps$, built like $\pexps$ but including logical variables and type membership; ghost commands $\gcmds$ for folding and unfolding user-defined predicates; commands extended with ghost commands $\cmds$; assertions $\Assert$, of which we give a representative fragment that, as is common in tools, omits universal quantification, which SMT solvers handle poorly, while existentially quantified variables are instantiated during consumption by matching plans~\cite{Loow2024Matching}; and program fragments $\prgs$ comprising predicate\footnote{Our presentation of user-defined predicates is simplified. In the full theory, predicate parameters are split into in- and out-parameters. The distinction is used for frame inference through matching plans~\cite{Loow2024Matching}, which requires no discussion here.} and function definitions. We work with program fragments rather than whole programs because SL analyses can analyse functions separately. Throughout, \(\pv{\cdot}\) and \(\lv{\cdot}\) denote the program and free logical variables of an expression, command, assertion, or, later, symbolic state.

\subsection{Concrete Semantics}

As usual in separation logic work, our concrete operational semantics uses {\em partial} concrete memories. Concrete states are pairs \(\st = (\sto, \mem)\) of a {concrete variable store} \(\sto : \pvars \pfmap \vals\) (a partial finite map from program variables to values) and a partial {concrete memory} \(\mem\). Memories differ between instantiations: the theory requires them to form a partial commutative monoid with memory actions satisfying the OX and/or UX frame properties. For this paper, we work with the standard linear memory model of SL, \(\mem : \nats \pfmap \vals\), which is compatible with OX analysis.\footnote{The standard ISL model~\cite{Raad2020}, \(\mem : \nats \pfmap (\vals \uplus \{\cfreed\})\), is similar, where the symbol \(\cfreed\) records that a memory cell has been freed, so that the UX frame property holds; it can be used for UX analyses such as bi-abduction.}

Program expressions $\pexp \in \pexps$ are evaluated by a standard evaluation function \(\esem{\pexp}{\sto} = v\) over the store only, i.e., program-expression evaluation does not access or modify the heap. If evaluation errors, e.g. division by zero, we write \(\esem{\pexp}{\sto} = \undefd\); we call evaluations that may error \emph{partial}.

The big-step semantics for commands $\cmd \in \cmds$ has judgement \(\csesemtransabstract{\st}{\cmd}{\st'}{\fictx}{\outcome}\), where outcome \(o ::= \oxok \mid \oxerr \mid \oxm \) denotes successful execution, language error, or missing-resource error (discussed below), and \(\fictx\) is a \emph{function implementation context}: a partial function providing definitions for function calls, i.e., \(\fictx(\pvar{f}) = (\lst{\pvar{x}}, \cmd, \pexp)\), often written 
\(\pfunction{f}{\lst{\pvar{x}}}{\cmd; \preturn{\pexp}} \in \gamma\). The rules extend normal operational semantics with expected language- and missing-error rules (erroneous executions record an error description in the distinguished program variable \(\pvar{err}\)), and ignore user annotations, so ghost commands $\gcmds$ are no-ops. The semantics never gets stuck, since errors are explicit outcomes, which UX bug finding needs to report them. As a representative case, we give the most interesting rules of lookup, illustrating each type of outcome:%
\begin{mathpar}
\small
\inferrule{
    \esem{\pexp}{\sto} = n \\
    \mem(n) = v
  }{
    \csesemtrans{\sto, \mem}{\pderef{\pvar{x}}{\pexp}}{\sto[\pvar{x} \storearrow v], \mem}{\fictx}{\oxok}
  }
\and
\inferrule{
    \esem{\pexp}{\sto} = v \\
    v \not \in \nats
  }{
    \csesemtrans{\sto, \mem}{\pderef{\pvar{x}}{\pexp}}{\sto[\pvar{err} \storearrow [\text{``Type''}]], \mem}{\fictx}{\oxerr}
  }
\and
\inferrule{
    \esem{\pexp}{\sto} = n \\ n \in \nats \\
    n \notin \dom(\mem)
  }{
    \csesemtrans{\sto, \mem}{\pderef{\pvar{x}}{\pexp}}{\sto [\pvar{err} \storearrow [\text{``MissingCell''}, n]], \mem}{\fictx}{\oxm}
  }
\end{mathpar}
The lookup rules for the $\oxok$ and $\oxerr$ judgements should come as no surprise. The lookup rule for the $\oxm$ judgement is perhaps less familiar. It arises due to a missing resource such as $n \not \in \dom(\mem)$. Missing-resource errors must be singled out from other errors because they do not satisfy the frame properties: for example, framing on a cell can turn a missed lookup into a successful lookup. They are also useful for bi-abduction, used in automatic UX bug-finding, which fixes a missing-resource error by adding the missing resource, here a cell at $n$.  

\subsection{Semantics of Assertions and Program Logics}

The theory builds on a series of definitions for the assertions and program logic, provided in detail in \appref{app:background-defs}{A}. We recall the most important definitions to establish notation.  
The satisfaction relation for assertions, denoted $\subst, (\sto, \mem) \models \asrtcolor{P}$, where $\subst : \lvars \pfmap \vals$ is a logical interpretation giving semantics to logical variables, is standard. Logical expressions are evaluated by \(\esem{\lexp}{\subst, \sto}\), which extends program-expression evaluation with \(\subst\) for logical variables; for logical expressions without program variables, we write \(\esem{\lexp}{\subst}\).  Given $\csesemtransabstract{\st}{\cmd}{\st'}{\fictx}{\outcome}$ and $\subst, (\sto, \mem) \models \asrtcolor{P}$, we can define  OX and UX judgements using quadruples in SL and ISL respectively of the form:
\[
\fictx \models \quadruple{P}{\cmd}{\Qok}{\Qerr}
\qquad\text{and}\qquad \fictx \models
\islquadruple{P}{\cmd}{\Qok}{\Qerr},
\]
with precondition \(\asrtcolor{P}\) and postconditions \(\asrtcolor{\Qok}, \asrtcolor{\Qerr} \in \Assert\) for successful and erroneous termination. The quadruples are required in the unifying approach to  OX and UX reasoning given by CSE. The details are given in \appref{app:background-defs}{A}.  For this paper, it is enough to  work with the traditional SL triple \(\triple{P}{\cmd}{Q}\) which is  syntactic sugar for an SL quadruple where \(\asrtcolor{\Qok} = \asrtcolor{Q}\) and \(\asrtcolor{\Qerr} = \asrtcolor{\false}\), asserting that an erroneous execution does not occur.
The symbol \(\fsctx \in 2^{\specs}\) denotes a function specification context, where $\specs$ is the set of function quadruples. Lastly, the judgement $\models (\fictx, \fsctx)$ holds when all specifications of $\fsctx$ are valid under function implementation context $\fictx$.

\subsection{Symbolic Engine}%
\label{sect:background:symbolic-semantics}

We now define the symbolic execution engine of the language, also called its symbolic semantics.

\paragraph{Symbolic states}
The symbolic state of the engine, denoted \(\sst\), is a structure built on symbolic  expressions \(\sexp \in
\sexps \subset \lexps\) which are logical expressions that do not contain program variables. A symbolic state \(\sst\) is a tuple \((\ssto, \smem, \spred, \spc)\), with symbolic variable store \(\ssto : \pvars \pfmap \sexps\); symbolic memory \(\smem : \sexps \pfmap \sexps\); symbolic predicates \(\spred\), a multiset of symbolic user-defined predicates, where a symbolic predicate has the form \(p(\vec\sexp)\) and \(p \in \strings\) is a predicate name; and the path condition \(\spc \in \sexps\), which captures constraints imposed during symbolic execution. We treat a symbolic state as a record, writing \(\access{\sst}{sto}\) for the symbolic store of \(\sst\) and \(\sst[\sstupdate{pc}{\spc'}]\) for \(\sst\) with its path condition replaced by \(\spc'\), and similarly for the other fields.
The satisfaction relation for symbolic states, denoted $\subst, (\sto, \mem) \models (\ssto, \smem, \spred, \spc)$, is given by:
\[
\exists \mem_1, \mem_2.~\mem = \mem_1 \uplus \mem_2 \land \subst, \sto \stomodels \ssto \land \subst, \mem_1 \memmodels \smem \land \subst, \mem_2 \predmodels \spred \land \esem{\spc}{\subst} = \true,
\]
where the relations $\subst, \sto \stomodels \ssto$, $\subst, \mem \memmodels \smem$, and $\subst, \mem \predmodels \spred$ are defined as expected; in particular, we have $\subst, \mem \memmodels \{ \sexp_a^1 \mapsto \sexp_e^1, \dots, \sexp_a^n \mapsto \sexp_e^n \} \Leftrightarrow \mem = \biguplus_{i=1}^{n} \{ \esem{\sexp_a^i}{\subst} \mapsto \esem{\sexp_e^i}{\subst} \}$.
We say that a symbolic state \(\sst\) is satisfiable, denoted \(\satpred(\sst)\), when \(\exists \subst, \sto, \mem. \; \subst, (\sto, \mem) \models \sst \). Additionally, we say that a symbolic expression \(\sexp\) is entailed by a symbolic state \(\sst\) when it evaluates to \(\true\) under all models of the symbolic state: \(\sst \models \sexp \Leftrightarrow (\forall \subst, \sto, \mem. \; \subst, (\sto, \mem) \models \sst \Rightarrow \esem{\sexp}{\subst} = \true)\).

\paragraph{Engine judgement}

The CSE engine judgement \(\csesemtransabstractm{\sst}{\cmd}{\sst'}{\fsctx}{m}{\outcome}\) is defined over symbolic states.\footnote{L\"{o}\"{o}w et al.\ use \emph{oracles} in the judgement to model angelic non-determinism; we omit these for ease of presentation.} The mode \(m \in \{\mOX, \mUX\}\) switches the behaviour of the engine depending on the type of analysis required. Symbolic outcomes extend the concrete outcomes with \(\oxabort\), to signal an analysis failure that cannot be guaranteed to correspond to a concrete error, which can occur, for example, when a chosen function specification is not applicable. In the rules, \(\cseeval{\pexp}{\ssto}{}{\sexp}{}\) denotes symbolic evaluation of program expression \(\pexp\), which substitutes the symbolic values in \(\ssto\) for its program variables. The successful and missing-resource symbolic rules for lookup are as follows:
\begin{mathpar}
\small
\inferrule
 {\cseeval{\pexp}{\ssto}{}{\sexp_l}{} \\ \smem(\sexp_l') = \sexp \\ \ssto' = \ssto[\pvar{x} \storearrow \sexp] \\\\ \spc' = (\sexp_l = \sexp_l' \land \spc)}
 {\csesemtransabstractm{(\ssto, \smem, \spred, \spc)}{\pderef{\pvar{x}}{\pexp}}{(\ssto', \smem, \spred, \spc')}{\fsctx}{m}{\oxok}}
\and
\inferrule
 {\cseeval{\pexp}{\ssto}{}{\sexp_l}{} \\ \ssto' = \ssto[\pvar{err} \storearrow [\text{``MissingCell''}, \sexp_l]] \\\\ \spc' = (\sexp_l \in \nats \land \bigwedge_{\sexp_l' \in \dom(\smem)} \sexp_l \neq \sexp_l' \land \spc)}
 {\csesemtransabstractm{(\ssto, \smem, \spred, \spc)}{\pderef{\pvar{x}}{\pexp}}{(\ssto', \smem, \spred, \spc')}{\fsctx}{m}{\oxm}}
\end{mathpar}

\paragraph{Engine soundness}
Recall that the main results of L\"{o}\"{o}w et al.\ are the OX and UX soundness results, presented in simplified form in \Cref{sect:overview}. The full statement of OX soundness uses an abort-miss side condition on symbolic executions of \(\cmd\) from \(\sst\). First, every satisfiable path must be non-aborting, since abort is an analysis failure rather than a concrete outcome. Second, if a path reports a missing resource, the final state must contain no remaining predicate resources; otherwise the allegedly missing resource could still be hidden inside one of them. OX soundness is then:
\begin{theorem}[OX Soundness]\label{thm:ox-sound}
Assume $\models (\fictx, \fsctx)$. 
\[
\begin{array}{ll}
\text{If } & \csesemtransabstract{\st}{\cmd}{\st'}{\fictx}{\outcome} \text{ and } \subst, \st \models \sst \text{ and the abort-miss condition holds for } \cmd \text{ from } \sst \\
\text{then } & \exsts{\sst', \subst'}
\csesemtransabstractm{\sst}{\cmd}{\sst'}{\fsctx}{\mOX}{\outcome} \text{ and }
\subst'|_{\lv{\sst}} = \subst|_{\lv{\sst}} \text{ and }
\subst', \st' \models \sst'
\end{array}
\]
\end{theorem}

The consume and produce operations of the engine come with their own soundness results, used to prove soundness of, for example, function call. We omit the details. 

\subsection{Function Verification}%
\label{sect:background:function-verification}

On top of their engine, L\"{o}\"{o}w et al.\ implement both an OX function verification analysis and a UX bug-finding analysis based on bi-abduction.
We recap the former, which we extend to recursive functions (\Cref{sect:frontend}).
It is given by the judgement \(\verifiesfun{\fsctx}{P}{f}{\pvvar x}{\cmd}{\pexp}{Q}\), which takes a function specification with precondition \(\asrtcolor{P}\) and postcondition \(\asrtcolor{Q}\), a function implementation \(\pfunction{f}{\pvvar x}{\psequence{\cmd}{\preturn \pexp}}\), and a context \(\fsctx\) of known function specifications. It checks the specification against the implementation by producing \(\asrtcolor{P}\), symbolically executing the function, and consuming \(\asrtcolor{Q}\). On every satisfiable branch, to rule out memory leaks, no memory may remain after the consumption of \(\asrtcolor{Q}\). The formal definition is in \appref{app:background-defs}{A}.
Their result is limited to non-recursive functions because a recursive function's own specification must be added to \(\fsctx\) before verification, which their proof does not cover.


\section{Extended Frontend: New User-facing Features}%
\label{sect:frontend}

Most frontend extensions follow designs established in CSE tools and Hoare logic. The difficulty lies in proving them sound in one unified formalism, which takes substantial Rocq engineering, and in encoding them into SMT-LIB, which constrains their design. We present them from the perspective of users of tools based on our extended theory, building up to the~verification~task~of~\Cref{fig:example-spec}.

The extensions introduce new contexts, such as an ADT declaration context, like \(\fictx\) and \(\fsctx\) from \Cref{sect:background}. As many judgements depend on them indirectly, we write context indices only when relevant.

\subsection{Verification of Recursive Functions}%
\label{sect:frontend:recursive-functions}

The first limitation with the verification of \(\mathtt{reverse'}\) (\Cref{fig:example-spec:spec}) is that \citet{Loow2025} did not verify the function verification procedure for recursive functions. We address this by adapting \citet{Nipkow2002}'s Hoare logic soundness proof to CSE, as we now explain.
The result we have proved is:
\begin{theorem}[Recursive Function Verification Correctness]%
\label{lem:recursive-verification-correct-idx}
Let \(\fsctx' \defeq \fsctx \cup \{ \triple{P}{\pvar f(\pvvar{x})}{Q} \}\):
  \[
    \begin{array}{ll}
    \text{If } \quad   & \models (\fictx, \fsctx) \text{ and } \pfunction{f}{\pvvar x}{\psequence{\cmd}{\preturn \pexp}} \in \fictx \text{ and } 
                       \verifiesfun{\grayboxed{\fsctx'}}{P}{f}{\pvvar x}{\cmd}{\pexp}{Q} \\
    \text{then} \quad & \models (\fictx, \fsctx')
    \end{array}
  \]
\end{theorem}
The difference from L\"{o}\"{o}w et al.'s result is the highlighted \(\grayboxed{\fsctx'}\): recursive calls in \(\pvar f\)'s body are symbolically executed using \(\pvar f\)'s own specification.\footnote{We consider direct recursion only; mutual recursion, which Nipkow's argument also handles, adds no essential complexity.} Justifying these calls requires \(\models(\fictx,\fsctx')\), which is the theorem's conclusion; their proof has no mechanism to handle this circularity. Nipkow breaks the corresponding circularity in his mechanisation of Hoare logic by indexing the language semantics with function-call depth, leaving the program logic unindexed: the induction hypothesis then covers recursive calls, which happen at smaller depth.

Following Nipkow, we index only the \emph{concrete} semantics, not the \emph{symbolic} one. The depth-indexed concrete semantics \(\baction_\fictx^n\) is identical to \(\baction_\fictx\) except for a depth index \(n \in \nats\), which bounds the function-call depth required to execute a command. All rules except those for function calls thread the index through unchanged; the function-call rules increase the index by~1. We illustrate with one sequencing rule and the successful function-call rule, with the indices highlighted, which are the only difference from the non-indexed rules:
\begin{mathpar}
\inferrule{
    \csesemtransm{\sto, \mem}{\cmd_1}{\sto', \mem'}{\fictx}{\grayboxedscript{n}}{\oxok} \\\\
    \csesemtransm{\sto', \mem'}{\cmd_2}{\sto'', \mem''}{\fictx}{\grayboxedscript{n}}{\outcome}
  }{
    \csesemtransm{\sto, \mem}{\psequence{\cmd_1}{\cmd_2}}{\sto'', \mem''}{\fictx}{\grayboxedscript{n}}{\outcome}
  }
\and
\inferrule{
    \pfunction{f}{\vec{\pvar{x}}}{\cmd; \preturn{\pexp'}} \in \fictx \\
    \esem{\vec{\pexp}}{\sto} = \vec{v} \\\\
    \vec{\pvar{z}} = \pv{\cmd} \setminus \{ \vec{\pvar{x}} \} \\
    \sto_p = \emptyset[ \vec{\pvar{x}} \storearrow \vec{v} ][ \vec{\pvar{z}} \storearrow \nil ] \\\\
    \csesemtransm{\sto_p, \mem}{\cmd}{\sto_q, \mem'}{\fictx}{\grayboxedscript{n}}{\oxok} \\
    \esem{\pexp'}{\sto_q} = v' \\
    \sto' = \sto[\pvar{y} \storearrow v']
  }{
    \csesemtransm{\sto, \mem}{\pfuncall{\pvar{y}}{\pvar{f}}{\vec{\pexp}}}{\sto', \mem'}{\fictx}{\grayboxedscript{n+1}}{\oxok}
  }
\end{mathpar}
 On top of this semantics, we define depth-indexed function-specification validity, \(\models_n(\fictx,\fsctx)\), like \(\models(\fictx,\fsctx)\) but using \(\baction_\fictx^n\). For \(n \neq 0\), all specifications must be valid for executions with function-call depth up to, but not including, \(n\), i.e. depth index \(n-1\); \(\models_0(\fictx,\fsctx)\) holds vacuously.

With this setup, we can now prove \Cref{lem:recursive-verification-correct-idx}. First, we prove a depth-indexed version of OX soundness (\Cref{thm:ox-sound}), with \(\models_n(\fictx,\fsctx)\) in place of \(\models(\fictx,\fsctx)\) and \(\baction_\fictx^n\) in place of \(\baction_\fictx\): if the function context is valid up to but excluding depth \(n\), then symbolic execution is over-approximately sound up to and including depth \(n\). Now, \Cref{lem:recursive-verification-correct-idx} does not mention \(n\), so to induct on \(n\) we use the facts that the indexed and non-indexed concrete semantics coincide, \(\st, \cmd \baction_{\fictx} \outcome: \st' \Leftrightarrow \exists n.\; \st, \cmd \baction_{\fictx}^{n} \outcome: \st'\), as well as \(\models(\fictx,\fsctx) \Leftrightarrow \forall n.\ \models_n(\fictx,\fsctx)\). It therefore suffices to prove \(\forall n.\ \models_n(\fictx,\fsctx')\) by induction on \(n\). The base case is vacuous. In the inductive case, the induction hypothesis \(\models_n(\fictx,\fsctx')\), depth-indexed OX soundness, and the soundness of \(\produce\) and \(\consume\) show that every depth-\(n\) concrete execution of the function body respects its specification; hence \(\models_{n+1}(\fictx,\fsctx')\).

\subsection{Algebraic Data Types}\label{sect:frontend:adts}

Next we add algebraic data types (ADTs), so that users can define, e.g., \(\pname{List}\) (\Cref{fig:example-spec:adt}). The extension is conceptually simple but required substantial binder engineering: our mechanisation, which is locally nameless~\cite{Chargueraud12}, needed machinery to open several binders at once, and we had to thread local-closure side conditions throughout. We focus here on the outward-facing extensions.

We extend program fragments with ADT declarations \(\prgs ::= \ldots \mid (\delta = \ctor_1(\vec\type_1) \mid \cdots \mid \ctor_n(\vec\type_n))\), where \(\delta \in \strings\) is the ADT name and each \(\ctor_i(\vec\type_i)\) is a constructor. These declarations induce an \emph{ADT declaration context} \(\adtctx : \strings \pfmap \types^* \times \strings\), mapping constructors to their argument types and constructed ADT. We write \(\ctor(\vec\type) : \delta \in \adtctx\) when \(\adtctx(\ctor) = (\vec\type,\delta)\), and \(\delta \in \adtctx\) when some constructor builds \(\delta\). We extend types with ADT names, \(\types ::= \ldots \mid \delta\), and define \(\cwf(\type)\) by: (1) \(\type\) is neither a list nor an ADT type; (2) \(\type = \listsof{\type'}\) and \(\cwf(\type')\); or (3) \(\type=\delta\) and \(\delta\in\adtctx\). We assume that the argument types of all constructors in \(\adtctx\) are well formed. Values gain constructor values \(\ctor(\vec\gv)\), typed at \(\delta\) when \(\ctor(\vec\type):\delta\in\adtctx\) and \(\vec\gv\in\vec\type\). Malformed types are uninhabited. In Rocq, values now depend on typing, and typing on values, which Rocq cannot define as a mutual inductive-inductive type. So, we define values as dependent pairs of a \emph{pre-value} (potentially malformed values) and a well-formedness proof, assuming proof irrelevance (see \appref{app:frontend:rocq}{B.3}).

\begin{wrapfigure}[10]{r}{0.42\textwidth}
\vspace{-\baselineskip}
\begin{minipage}{\linewidth}
\begin{mathpar}
\small
\infer{
    \csesemtransabstractm{\sst}{\pmatch{\lexp}{\delta}{ \pcase{p_i}{\gcmd_i}|_{i=1}^{n}}}{\sst''}{\fsctx}{m}{\outcome}
}{
    \begin{array}{c}
    \esem{\lexp}{\access{\sst}{sto}} \baction \sexp \\
    p_j = \ctor(\lvvar x)\ \text{for}\ 1 \leq j \leq n \\
    p_k\ \text{not a variable binder for}\ 1 \le k < j \\
    \spc = (\ctor(\lvvar x) = \sexp) \wedge \bigwedge_{k = 1}^{j - 1} \neg (\tester{\sexp}{\ctor_k}) \\
    \sst' = \sst[\sstupdate{pc}{\spc \wedge \access{\sst}{pc}}] \\
    \csesemtransabstractm{\sst'}{\gcmd_j}{\sst''}{\fsctx}{m}{\outcome}
    \end{array}
}
\end{mathpar}
\end{minipage}
\caption{Symbolic semantics for \(\ghostcolor{\pvar{match}}\).}%
\label{fig:match}
\end{wrapfigure}

At the assertion level, ADTs let us write \([]\) and \(a :: \alpha\) in the \(\pname{list}\) predicate (\Cref{fig:example-spec:predicate}). We extend logical expressions \(\lexps\) (but not program expressions) with constructor applications \(\ctor(\vec\lexp)\), testers \(\tester{\lexp}{\ctor}\), and matches \(\match{\lexp}{\delta}{p_i \rightarrow \lexp_i|_{i = 1}^n}\), where each pattern \(p_i\) is either a variable \(x\) or a constructor pattern \(\ctor(\lvvar x)\). Evaluation is extended as expected, returning \(\undefd\) on match typing errors; it now depends on \(\adtctx\), omitted by convention.
Ghost commands gain \(\pmatch{\lexp}{\delta}{ \pcase{p_i}{\gcmd_i}|_{i = 1}^n }\), whose symbolic semantics (\Cref{fig:match}) branches over all patterns. The branch for constructor pattern \(\ctor_j(\lvvar x)\) adds to the path condition that the scrutinee equals \(\ctor_j(\lvvar x)\) and matches no earlier pattern. Other rules are in \appref{app:frontend-defs}{B}.

\subsection{Lemmas}\label{sect:frontend:lemmas}

With ADTs, we are able to formulate \Cref{fig:example-spec:spec}, but SMT solvers lack inductive reasoning over user-defined ADTs. We therefore allow lemmas to be stated and proved inside CSE. For example, the proof in \Cref{fig:example-spec:lemma} branches on \(\alpha = []\) and \(\alpha = a' :: \alpha'\): the base case follows from the path condition; in the inductive case, \(\alpha \in \pname{List}\) gives \(\alpha' \in \pname{List}\), so the lemma can be applied to strengthen the path condition with \(\alpha' \concat (a :: \beta) = (\alpha' \concat [a]) \concat \beta\).

\paragraph{Lemma definitions and applications}

We extend program fragments \(\prgs\) with lemma statements \(\lemDefM{l}{\lvvar x}{P}{\iffstack}{Q}{\gcmd}{M}\), where \(\pvar{l}\) is the lemma name, \(\lvvar x\) the universally quantified logical variables, \(\asrtcolor{P},\asrtcolor{Q}\in\Assert\) the LHS and RHS of the lemma, and \(\iffstack\in\{\Leftarrow,\Rightarrow,\Leftrightarrow\}\) the implication direction. The remaining components are used for verification: \(\gcmd\) is the proof, a ghost command transforming states satisfying the hypothesis into states satisfying the conclusion; and \(M\in\lexps\) is a decreasing measure ensuring that recursive applications of \(\pvar{l}\) in \(\gcmd\) are well founded. We also extend ghost commands \(\gcmd\) with \(\papply{l}{\vec\lexp}\), applying lemma~\(\pvar{l}\)~with~arguments~\(\vec\lexp\).

\paragraph{Lemma context and semantics of lemmas}

Lemmas induce a new context for the CSE engine, the \emph{lemma context} \(\lemctx\), which the engine judgement carries alongside \(\fsctx\), written \(\baction^m_{\fsctx\lemctx}\). It maps each \(\pvar l\) to its statement \(\lemstmt{\lvvar x}{P}{\iffstack}{Q}\), or, while \(\pvar l\) is being verified (discussed below), to its statement with its measure, \(\lemstmtM{\lvvar x}{P}{\iffstack}{Q}{M}\). Once a lemma is verified, neither its proof \(\gcmd\) nor its measure \(M\) is kept in the context, just as \(\fsctx\) carries function specifications without their implementations. A lemma is valid, \(\models \lemstmt{\lvvar x}{P}{\iffstack}{Q}\), iff \(\subst,\st \models \asrtcolor{P} \iffstack \subst,\st \models \asrtcolor{Q}\) for all \(\subst,\st\) with \(\{\lvvar x\}\subseteq\dom(\subst)\), irrespective of any measure. A context is valid iff all its statements are valid.

\paragraph{Lemma application rule}

The new \(\ghostcolor{\pvar{apply}}\) command mirrors function calls (the formal rule is in~\appref{app:frontend-defs}{B}): it consumes \(\asrtcolor{P}\) and produces \(\asrtcolor{Q}\). We have extended both the OX and UX soundness proofs of the engine to cover this rule: previous lemma-supporting tools only handle OX reasoning; our rule also supports UX, with \(\Rightarrow\) lemmas applicable in OX, \(\Leftarrow\) in UX, and \(\Leftrightarrow\) in both, matching the OX/UX symmetry first formulated in incorrectness logic~\cite{OHearn2019}.

\paragraph{Lemma verification}

Lemma verification checks the inductive proof represented by the ghost command \(\gcmd\). It is given by the judgement \(\verifies{\lemctx}{\lemDefM{l}{\lvvar x}{P}{\iffstack}{Q}{\gcmd}{M}}\), which takes a lemma context \(\lemctx\) and a lemma definition. (The formal definition is again in~\appref{app:frontend-defs}{B}.) In short, verification follows the same three stages, with the same success conditions, as function verification: produce \(\asrtcolor{P}\), symbolically execute \(\gcmd\), and consume \(\asrtcolor{Q}\). Since a lemma must hold for every state satisfying its hypothesis, which only OX reasoning guarantees, every lemma, whatever its direction, is verified in OX mode as a \(\Rightarrow\) lemma: a \(\Leftarrow\) lemma by swapping its two sides, and a \(\Leftrightarrow\) lemma as two \(\Rightarrow\) lemmas, with one proof per direction.

To rule out invalid self-justifying proofs such as
$
\mathtt{absurd} : \asrtcolor{\true} \Rightarrow \asrtcolor{\false}~\{~\papply{absurd}{}~\},
$
we again borrow from \citet{Nipkow2002} and adapt his total-correctness proof for recursive procedures in Hoare logic, taking a measure as the well-founded relation. While a lemma is being verified, its entry in the lemma context carries its measure, \(\lemctx(\pvar{l}) = \lemstmtM{\lvvar x}{P}{\Rightarrow}{Q}{M}\), and the \(\ghostcolor{\pvar{apply}}\) rule checks that the measure decreases at each application; in \Cref{fig:example-spec:lemma}, it checks \(|\alpha'| < |\alpha|\) under \(\alpha = a' :: \alpha'\). Unlike Nipkow's well-founded relations, which are relations on states in the meta-logic, measures are logical expressions, so the CSE engine can check the decrease via SMT. Once verified, the lemma is added to the context without its measure: it is then valid for every value of the measure, so later applications need not check for a decrease.

We prove lemma verification correctness (\Cref{thm:lemma-verif-correct}, below) using two definitions and an auxiliary theorem. First, unlike Nipkow's shallowly embedded relations, measures can fail to evaluate ($\undefd$) or yield values that are not natural numbers; we normalise such outcomes to the smallest possible measure: \(\numify{\esem{M}{\subst, \sto}} = n\) if \(\esem{M}{\subst, \sto} = n\) for some \(n \in \nats\), and \(0\) otherwise. Second, \emph{measured validity} \(\models_n \lemctx\) requires each measured \(\Rightarrow\) lemma to be valid below \(n\): $ \forall \subst, \sto, \mem.\ \{\lvvar x\} \subseteq \dom(\subst) \wedge \numify{\esem{M}{\subst, \sto}} < n \;\Rightarrow\; \subst, (\sto, \mem) \models \asrtcolor{P} \Rightarrow \subst, (\sto, \mem) \models \asrtcolor{Q}$, with all other lemmas valid in the usual sense. Lastly, we need the following measure-indexed version of OX soundness (\Cref{thm:ox-sound}) for ghost commands (cf. depth-indexed soundness, \Cref{sect:frontend:recursive-functions}):
\begin{theorem}[Ghost OX Soundness]\label{thm:ghost-ox-sound}
 Assume $\models_n \lemctx$ and that all lemmas with measures share the same measure expression \(M\).\footnote{Only the lemma being verified carries a measure, so this is not limiting. The measure marks that lemma because the CSE engine, shared with function verification, does not know which lemma it is verifying.}
\[
\begin{array}{ll}
\text{If }
& \subst, (\sto, \mem) \models \sst \text{ and } \numify{\esem{M}{\subst, \sto}} = n \\
\text{and} & \text{the abort-miss condition of \Cref{thm:ox-sound} holds for } \gcmd \text{ from } \sst \\
\text{then } & \exsts{\sst', \subst'}
\csesemtransabstractm{\sst}{\gcmd}{\sst'}{\fsctx\lemctx}{\mOX}{\oxok} \text{ and }
\subst'|_{\lv{\sst}} = \subst|_{\lv{\sst}} \text{ and }
\subst', (\sto, \mem) \models \sst'
\end{array}
\]
\end{theorem}
\begin{proof}[Proof Sketch]
The key case is lemma application. Since \(\models_n \lemctx\), the lemma is valid only for states where the measure is below \(n\). The initial measure \(n\) together with the decreasing-measure check guarantees this holds at each inductive application.
\end{proof}

\begin{theorem}[Lemma Verification Correctness]\label{thm:lemma-verif-correct}
  Let \( \lemctx' \defeq \lemctx[ \pvar{l} \mapsto \lemstmtM{\lvvar x}{P}{\Rightarrow}{Q}{M} ]\):
  \[
    \begin{array}{ll}
    \text{If }  & \models \lemctx \text{, where no lemma in } \lemctx \text{ carries a measure, and } \verifies{\lemctx'}{\lemDefM{l}{\lvvar x}{P}{\Rightarrow}{Q}{\gcmd}{M}} \\
    \text{then} & \models \lemctx[ \pvar{l} \mapsto \lemstmt{\lvvar x}{P}{\Rightarrow}{Q} ]
    \end{array}
  \]
  The \(\Leftarrow\) and \(\Leftrightarrow\) cases follow by applying this to each direction verified.
\end{theorem}
\begin{proof}[Proof Sketch]
  Apply \(\models \lemctx \Leftrightarrow \forall n.\ \models_n \lemctx\) and induct over \(n\): the base case is vacuous; the inductive case follows from the induction hypothesis \(\models_n \lemctx'\), \Cref{thm:ghost-ox-sound}, and  \(\consume\) and \(\produce\) soundness.
\end{proof}

\paragraph{Total function correctness}
The same measure-based technique extends function verification to total correctness. A function specification may carry a measure, which a new engine mode checks decreases at each call, and verification in this mode also proves that the function terminates from every state satisfying its precondition. Details are in \appref{app:frontend-defs}{B}.

\subsection{Logical Functions}%
\label{sect:frontend:logical-functions}

\newcommand{\lpre}{\lexp_\text{pre}}

The final ingredient for \Cref{fig:example-spec} is logical functions, including \(\circ^\dag\) and \(\concat\) from \Cref{fig:example-spec:lfunc}. We extend logical expressions \(\lexps\) with applications \(f(\vec\lexp)\) and program fragments \(\prgs\) with definitions \(\lfunc{f}{\lvvar x : \vec\type}{\type}{\lexp}{\lpre}\), where the parameters \(\vec x\) have types \(\vec\type\), \(\type\) is the return type, \(\lexp\) is the body, and \(\lpre\) is the precondition; the body and precondition contain no program variables, and their logical variables are among \(\vec x\). Logical functions also require a new context, the \emph{logical function context} \(\lfuncctx\), which maps each logical function identifier to its definition. We assume that the parameter and return types of all logical functions in \(\lfuncctx\) are well formed.

\paragraph{Termination and design of preconditions}

Recursive logical functions add a second way for evaluation to fail besides erroring: it may \emph{diverge}, and then \(\esem{\lexp}{\subst, \sto}\) has no result, not even \(\undefd\). We say evaluation \emph{terminates} when it yields a value or \(\undefd\); partiality always refers to erroring, never to divergence. For sound translation to SMT functions, logical functions must terminate. The precondition \(\lpre\) characterises the arguments on which the body terminates without erroring: when it evaluates to \(\true\), the body must evaluate to some \(\gv\in\type\). The encoder also generates side conditions requiring every called function's precondition, preventing partial or underspecified cases from producing spurious \(\sat\) results. For recursive functions, these obligations may be inductive: e.g., a list-of-reciprocals function needs every element to be non-zero. Rather than have the encoder infer such facts, we make preconditions first-class in the logic, dischargeable by lemmas, as in Viper.

\paragraph{Formal definition}
The evaluation \(\esem{f(\vec\lexp)}{\subst, \sto} = \gv\) holds when (1)~\(\lfunc{f}{\lvvar x : \vec\type}{\type}{\lexp}{\lpre} \in \Psi\); (2)~\(\esem{\vec\lexp}{\subst, \sto} = \vec\gv\) with \(\gv_i \in \type_i\) for each \(i\); (3)~\(\subst' \defeq \{\vec x \mapsto \vec\gv\}\); (4)~\(\esem{\lpre}{\subst'} = \true\); and (5)~\(\esem{\lexp}{\subst'} = \gv\) with \(\gv \in \type\). Otherwise, \(f(\vec\lexp)\) evaluates to $\undefd$, unless one of these evaluations diverges. By our convention, we do not add \(\lfuncctx\) as an explicit index to \(\esem{\cdot}{\subst, \sto}\).

SMT soundness relies on preconditions correctly characterising where each function's body terminates without erroring. \(\lfuncctx\) is \emph{valid}, written \(\models \lfuncctx\), iff for all \(\lfunc{f}{\lvvar x: \vec \type}{\type}{\lexp}{\lpre} \in \lfuncctx\):
\[
  \forall \vec\gv \in \vec\type.~ \subst = \{\lvvar x \mapsto \vec\gv\} \Rightarrow
    (\exists r.~ \esem{\lpre}{\subst} = r) \wedge
    (\esem{\lpre}{\subst} = \true \Rightarrow \exists \gv \in \type.~ \esem{\lexp}{\subst} = \gv) .
\]
The first conjunct makes applications outside the precondition evaluate to \(\undefd\) rather than diverge. Together, they ensure that evaluation always terminates for valid \(\lfuncctx\). Moreover, an application errors exactly when its precondition does not hold: if the precondition evaluates to \(\true\), the body cannot error by validity, and otherwise the application errors by definition, even where the body would not. We need this for the SMT-LIB encoding, which guards each application with the callee's precondition: for OX soundness, the guard must not exclude any non-erroring evaluation.

In Rocq, as evaluation may not terminate, it can no longer be defined directly as a function. We define it as a relation, state \(\models \lfuncctx\) in terms of it, and then, assuming \(\models \lfuncctx\), turn it into a function with Hilbert's epsilon operator (see \appref{app:frontend:rocq}{B.3}).

\paragraph{Extending the logical function context} The mechanisation simply assumes \(\models \lfuncctx\). A check would have to show that each new body terminates without erroring under its precondition, which measures alone cannot do: a measure such as \(|\cdot|\) is itself a logical function, whose definition needs a measure. Structural recursion breaks this cycle for termination, but showing that the body does not error may still need induction, and so a lemma with a measure. We leave this to future work.


\section{SMT-LIB and Our New Mechanisation}%
\label{sect:smt}

SMT solvers check satisfiability of first-order formulas modulo a configurable set of background theories. SMT-LIB~\cite{smtlib-website} provides a common solver input language and theory catalogue, but is specified in prose and informal mathematics. We present the first mechanisation of SMT-LIB, a line-by-line Rocq mechanisation of SMT-LIB 2.7 (the release of 9 September 2026, currently the latest) covering its syntax and semantics, theory definitions, and several standard theories. Like the rest of our development, it uses classical axioms.

SMT-LIB has two main artefacts: the \emph{language}, giving syntax and semantics for scripts and formulas; and \emph{theories}, specifying sorts and functions together with their intended meaning. The language is defined by the SMT-LIB standard~\cite{Barrett2026}, roughly 100 pages for version 2.7.\footnote{Our work captures the \emph{semantic layer} of SMT-LIB, which gives the mathematical meaning of queries and defines satisfiability of the logic. The standard also defines S-expression-based lexical and structural layers, which~are~not~required~for~our~work.} Recent versions added ADTs and match expressions (2.6, 2017), then functional sorts and function abstractions for higher-order-style reasoning while retaining first-order quantification (2.7, 2025).\footnote{SMT-LIB 3.0 is expected to build on the Calculus of Inductive Constructions (CIC)~\cite{Paulin1993}, a fully fledged higher-order logic.} Theories, on the other hand, are defined by theory files published on the SMT-LIB website. Because of space, we do not discuss these theory files; the interested reader is instead directed to our mechanisation.

Having finished the mechanisation, we are left with a positive impression of the SMT-LIB standard: we did not find any major flaws during our efforts, possibly because the standard is already relatively formal (compared to, say, the average programming language standard). With that said, in our efforts to prove some of the semantic properties specified in the standard, we found oversights and raised them with the maintainers; some have been acknowledged and will be resolved in future versions. More information can be found with our mechanisation.

\subsection{Syntax}

\Cref{fig:fol-vs-smtlib:smtlib} gives the syntax of SMT-LIB terms, which include variables, named function applications (annotated with their result sort when it is ambiguous due to polymorphism), quantifiers and binders, let expressions, and ADT matches. Note that here and in the rest of our presentation, we use \smtcolor{dark red} to distinguish SMT-LIB's syntactic and semantic constructs. \Cref{fig:fol-vs-smtlib:fol} contrasts the syntax of SMT-LIB with the syntax of traditional first-order logic. Two differences are important. First, SMT-LIB is a \emph{many-sorted} (``typed'') logic, whereas traditional first-order logic is unsorted (``untyped''). Second, SMT-LIB does not syntactically separate \emph{terms} from \emph{formulas}: formulas in SMT-LIB are just terms of sort \(\boolsort\). Thus, even Boolean connectives are syntactically function applications, although we write them infix for readability. We write \(\term\) for arbitrary terms, \(\formula\) for \(\boolsort\) sorted terms, and \(\formulae\) for a set of \(\boolsort\) sorted terms.

\begin{figure}[t]
    \centering
    \begin{subfigure}[b]{0.58\textwidth}
        \centering
        \begin{align*}
          \term \in \termz &::= \termvar x \mid \smtf{f}(\vec{\term}) \mid \termannot{\smtf{f}}{\sort}(\vec{\term}) \\
          & \ \ \ \mid \termforall{\smtvar{x}}{\sort}{\term} \mid \termexists{\smtvar{x}}{\sort}{\term} \mid \termlambda{\smtvar{x}}{\sort}{\term} \\
          & \ \ \ \mid \termlet{x_i = \term_i |_{i=1}^n}{\term} \\
          & \ \ \ \mid \termmatch{\term}{p_i \rightarrow \term_i |_{i=1}^n}
        \end{align*}
        \caption{SMT-LIB term syntax.}%
        \label{fig:fol-vs-smtlib:smtlib}
    \end{subfigure}
    \hfill
    \begin{subfigure}[b]{0.38\textwidth}
        \centering
        \[
        \begin{aligned}
        t &::= x \mid f(\vec t) \\
        \phi &::= P(t_1, \dots, t_n) \mid t_1 = t_2 \\
             & \ \ \ \mid \neg \phi \mid \phi_1 \land \phi_2 \mid \dots \\
             & \ \ \ \mid \forall x.\, \phi \mid \exists x.\, \phi
        \end{aligned}
        \]
        \caption{Traditional first-order syntax.}%
        \label{fig:fol-vs-smtlib:fol}
    \end{subfigure}
    \caption{Comparison between the SMT-LIB syntax and traditional first-order syntax. The former uses explicit sort annotations ($\termvar x : \sort$) and treats all expressions, including quantified formulas, as terms.}%
    \label{fig:fol-vs-smtlib}
    \vspace{-1em}
\end{figure}

SMT-LIB vocabulary is defined by signatures. A \emph{signature} \(\sig\) contains sort symbols \(\sigs\), the arities of these sort symbols, and function symbols \(\sigf\). Non-zero-arity sort symbols form parametric sorts: if \(\sortsymb \in \sigs\) has arity \(n\) and \(\sorti{1}, \dots, \sorti{n}\) are sorts, then \(\sortsymb(\sorti{1},\dots,\sorti{n})\) is a sort. Every signature contains the sort \(\boolsort\) and the arrow sorts \(\sorti{1} \arrowsort \sorti{2}\), the sorts of function abstractions \(\termlambda{\smtvar{x}}{\sort}{\term}\). The set \(\sorts{\sig}\) ranges over the set of well formed sorts based on the sort symbols of \(\sig\) and the arities it assigns them. Each function symbol has one or more \emph{ranks}, written \(\smtf{f} : \smtrank{\sorti{1}, \dots, \sorti{n}}{\sort} \in \sig\) for non-constants and just \(\smtf{f} : \sort \in \sig\) for constants. A signature also includes a \emph{sorting}, a partial map from variables to sorts: we write \(\smtvar{x} : \sort \in \sig\) when \(\sig\) assigns \(\smtvar{x}\) the sort \(\sort\). As of version 2.6, signatures also distinguish function symbols for ADT \emph{constructors}, \emph{selectors}, and \emph{testers}. Each constructor comes with a Boolean tester and a selector for each of its fields, with special semantics given below.

Well-sortedness, \(\sig \vdash \term : \sort\), consults variable sortings and function ranks, updating the sorting under binders. SMT-LIB scripts assert only \(\boolsort\)-sorted terms.

\subsection{Semantics}

A \emph{structure} \(\struct\) gives semantics by defining a universe \(\universe\), interpreting each sort \(\sort\) as a domain \(\sort^\struct \subseteq \universe\), and interpreting each ranked function symbol \(\smtf{f} : \smtrank{\sorti{1}, \dots, \sorti{n}}{\sort}\) as a \emph{total} function \(\smtf{f}^{\struct} : \sorti{1}^\struct \times \dots \times \sorti{n}^\struct \to \sort^\struct\). Function symbols cannot be interpreted partially. The standard also fixes some domains, including \(\boolsort^\struct = \bools = \{\true,\false\}\) and \((\sorti{1} \arrowsort \sorti{2})^\struct\), the set of all total functions \(\sorti{1}^\struct \to \sorti{2}^\struct\).

Representing the universe \(\universe\) as a single Rocq inductive type does not work, because the domain of an arrow sort contains all total functions between other domains, and an inductive type whose constructors take functions over itself is rejected by Rocq's strict positivity check. We instead use the \emph{universe design pattern}~\cite{LeanUniversePattern}, based on universes \emph{à la Tarski}~\cite{MartinLof1984}:
  \begin{minted}{coq}
Fixpoint interpretation (domain : sort -> Type) (args : list sort) (ret : sort) : Type :=
  match args with
  | [] => domain ret
  | arg :: args' => domain arg -> interpretation domain args' ret
  end.

Record structure := {
  domain : sort -> Type;
  domain_bool : domain bool_sort = bool;
  domain_map : forall s1 s2, domain (map_sort s1 s2) = (domain s1 -> domain s2);
  interp : func -> forall (args : list sort) (ret : sort), interpretation domain args ret;
}.
  \end{minted}
  A structure \(\struct\) is represented in Rocq by \texttt{structure}, whose field \texttt{domain} is the decoding function, mapping each code, a syntactic \texttt{sort}, to its carrier type, the counterpart of \(\sort^\struct\). Its field \texttt{interp} gives each function symbol, at each rank, a curried Rocq function between the corresponding carrier types, the counterpart of \(\smtf{f}^\struct\). The fields \texttt{domain\_bool} and \texttt{domain\_map} enforce the constraints above. The universe is never defined explicitly, and is implicitly the union of the sort domains.

How a structure may interpret a function symbol depends on how the symbol entered the signature. Theory-defined symbols get their meaning from the selected background theories. User-\emph{declared} symbols introduced with \smtcolor{\texttt{declare-fun}} are uninterpreted, so structures may interpret them as any total function of the declared rank. User-\emph{defined} symbols introduced with \smtdefinefunsrec{} are constrained by their bodies. Defining \(\smtf{f}\) with parameters \(\vec{\smtvar{x}} : \vec{\sort}\) and body \(\term\), which may mention \(\smtf{f}\) itself, amounts to declaring \(\smtf{f}\) and asserting its universally quantified \emph{defining equation} \mbox{\(\termforall{\vec{\smtvar{x}}}{\vec{\sort}}{\smtf{f}(\vec{\smtvar{x}}) \smteq \term}\)}. Since \(\smtf{f}\) is interpreted as a total function, termination is meaningless here. A definition that would not terminate as a program may therefore leave \(\smtf{f}\) unconstrained. For example, every interpretation of \(\smtf{f}\) satisfies \mbox{\(\termforall{\smtvar{x}}{\sort}{\smtf{f}(\smtvar{x}) \smteq \smtf{f}(\smtvar{x})}\)}. The equation may even have no model, as with the contradictory \mbox{\(\termforall{\smtvar{x}}{\intsort}{\smtf{f}(\smtvar{x}) \smteq \smtf{f}(\smtvar{x}) \smtplus \smtintlit{1}}\)}, making every script containing it unsatisfiable.

For ADTs, the standard requires every structure to interpret constructors \emph{absolutely freely}. The domain \(\delta^\struct\) of each datatype sort \(\delta\) is the set of \emph{ground terms} of sort \(\delta\) given by \(g ::= v \mid \mathtt{c}(g_1, \dots, g_n)\), where \(v\) ranges over values of non-datatype sorts, such as integers, and \(\mathtt{c}\) over constructors. Each constructor's interpretation builds the corresponding ground term, \(\smtf{c}^\struct(g_1, \dots, g_n) = \mathtt{c}(g_1, \dots, g_n)\). For an integer-list datatype with constructors \(\smtf{nil} : \smtcolor{\mathsf{List}}\) and \(\smtf{cons} : \smtrank{\intsort,\smtcolor{\mathsf{List}}}{\smtcolor{\mathsf{List}}}\), the domain \(\smtcolor{\mathsf{List}}^\struct\) is thus the set of ground terms \(\{\mathtt{nil}, \mathtt{cons}(0, \mathtt{nil}), \mathtt{cons}(1, \mathtt{cons}(0, \mathtt{nil})), \dots\}\), and \(\smtf{cons}^\struct(0, \mathtt{nil}) = \mathtt{cons}(0, \mathtt{nil})\). The expected properties of ADTs follow because distinct ground terms are distinct values. Constructors are injective and have disjoint ranges, and since the domain contains nothing else, every value is a finite constructor tree. Our mechanisation requires the domain of each datatype sort to equal a Rocq inductive type of ground terms of that sort.

A tester checks whether a value was built by its constructor, and a selector extracts a field from such a value. As all SMT-LIB functions are interpreted totally, selectors are also defined on values built by other constructors. These cases are \emph{underspecified}, so any value of the result sort is allowed.

A \emph{valuation} \(\valuation\) for a signature \(\sig\) partially maps variable-sort pairs \((\smtvar x,\sort)\) to elements of \(\sort^\struct\), and is defined on \((\smtvar x,\sort)\) whenever \(\smtvar x : \sort \in \sig\). Terms evaluate with respect to a structure-valuation pair \(\esem{\term}{\struct,\valuation}\). We write \(\struct, \valuation \models \formulae\) to mean \(\forall \formula \in \formulae.~ \esem{\formula}{\struct, \valuation} = \true\), omitting \(\valuation\) when \(\formulae\) is closed.

A \emph{theory} \(\theory\) defines a signature \(\sig\) and a set of structures \(\struct \in \theory\) interpreting it. Typical SMT-LIB theories are either a single model, such as the integers, or all structures satisfying some axioms; the axiom set need not be finite or decidable. As an SMT solver works with a \emph{configurable} set of theories, the standard also defines the composition of theories.
A set of \(\boolsort\)-sorted terms \(\formulae\) is \emph{\(\theory\)-satisfiable}, written \(\smtsat_\theory(\formulae)\), iff some \(\struct \in \theory\) and valuation \(\valuation\) for the signature of \(\theory\) satisfy \(\struct, \valuation \models \formulae\).


\section{Extended Backend: SMT-LIB Encoding}%
\label{sect:backend}

We now connect the symbolic expressions \(\sexp \in \sexps\) generated by symbolic execution to the SMT-LIB semantics of \Cref{sect:smt}; in other words, we now build the backend of the CSE engine. We call symbolic expressions, their values, and logical interpretations the \emph{source} side of the encoding, as opposed to the SMT-LIB side.
The encoding must solve four problems: symbolic expressions are dynamically typed, may be partial, may mention user-defined ADTs and logical functions, and may be encoded with SMT-LIB sorts whose domains contain values not representable as source values. We address these challenges as follows. \Cref{sect:backend:background} introduces the background SMT-LIB theory against which our queries are checked, declaring user-defined ADTs, logical function symbols, a distinguished ADT for dynamic values, a second ADT for partial logical interpretations, and representability predicates, which identify exactly those SMT-LIB values that are representable as source values. \Cref{sect:backend:encoding} defines the expression encoding, a type-directed translation that assigns each term a sort and generates side conditions that make the partial evaluation of symbolic expressions faithful inside SMT-LIB's total semantics. \Cref{sect:backend:lfuncs} gives the SMT-LIB definitions that restrict the interpretation of logical function symbols declared in the background theory. \Cref{sect:backend:encoding-correctness} proves the expression encoding correct. Finally, \Cref{sect:backend:expr-sat,sect:backend:satisfiability} define the operational satisfiability check, whose queries use the representability predicates to rule out SMT-LIB models containing non-representable values, and, building on it, the state satisfiability and entailment checks. Detailed definitions are in \appref{app:backend-defs}{C}. As in \Cref{sect:frontend}, we omit context indices (e.g. \(\adtctx\) and \(\lfuncctx\)) unless needed.

\subsection{Symbolic States Extended with Structured Type Information}\label{sect:backend:state-types}

Type information is often only implicit in a symbolic state (e.g.\ if \(\sexp\) is in the domain of the symbolic memory, then \(\sexp \in \nats\)), but the backend's type-based optimisations need it to be explicit. We therefore extend symbolic states to carry a typing environment \(\ste : \lvars \pfmap \types\), writing $\lvar x : \type \in \ste$ when \(\ste\) assigns type \(\type\) to \(x\). We extend state satisfiability $\subst, (\sto, \mem) \models (\ssto, \smem, \spred, \spc, \ste)$ to include typing environment satisfiability, \(\subst \models \ste \Leftrightarrow \forall x,\type.\ \lvar x : \type \in \ste \Rightarrow \exists v.\ \subst(x)=v \wedge v \in \type\).
Adding $\lvar x : \type \in \ste$ is equivalent to adding \(x \in \type\) to the path condition $\spc$, so such entries may be added only when they are already entailed by the state. Thus \(\ste\) only ever records types that the rest of the state already implies, but it need not record all of them. Leaving implied types out is sound but forgoes the type-based optimisations, so they should be propagated into \(\ste\) during execution, similarly to memory consolidation in Viper~\cite{advancing-automated-permissionbased-schwerhoff-2016}.

\subsection{Background Declarations and Dynamic Values}\label{sect:backend:background}

\begin{figure}[hbtp]

  \sidecaption{\(\valsort\) and \(\maybevalsort\) Declarations in \(\sig_{\ste}\)}
  \[
    \begin{gathered}
    \nullval : \smtrank{\nullsort}{\valsort} \qquad
    \boolval : \smtrank{\boolsort}{\valsort} \qquad
    \natval : \smtrank{\intsort}{\valsort} \\
    \ratval : \smtrank{\realsort}{\valsort} \qquad
    \stringval : \smtrank{\stringsort}{\valsort} \qquad
    \listval : \smtrank{\seqsort{\valsort}}{\valsort} \qquad
    \adtval{\delta} : \smtrank{\adtsort}{\valsort}
    \\[0.3em]
    \isnull : \smtrank{\valsort}{\boolsort} \qquad
    \isbool : \smtrank{\valsort}{\boolsort} \qquad
    \isnat : \smtrank{\valsort}{\boolsort} \qquad
    \dots \\
    \getnull : \smtrank{\valsort}{\nullsort} \qquad
    \getbool : \smtrank{\valsort}{\boolsort} \qquad
    \getnat : \smtrank{\valsort}{\intsort} \qquad
    \dots
    \\[0.8em]
    \someval : \smtrank{\valsort}{\maybevalsort} \qquad
    \noval : \maybevalsort
    \\[0.3em]
    \isval : \smtrank{\maybevalsort}{\boolsort} \qquad
    \isnoval : \smtrank{\maybevalsort}{\boolsort} \qquad
    \getval : \smtrank{\maybevalsort}{\valsort}
    \end{gathered}
  \]

  \sidecaption{Representability Declarations in \(\sig_{\ste}\)}
  \[
    \reprnat : \smtrank{\intsort}{\boolsort} \qquad
    \reprrat : \smtrank{\realsort}{\boolsort} \qquad
    \reprval : \smtrank{\valsort}{\boolsort} \qquad
    \repradt{\delta} : \smtrank{\adtsort}{\boolsort}
  \]

  \sidecaption{\(\encsort{\type} = \sort\)}
  \[
    \begin{array}{r@{\;=\;}l@{\qquad}r@{\;=\;}l@{\qquad}r@{\;=\;}l@{\qquad}r@{\;=\;}l}
    \encsort{\vals} & \valsort &
    \encsort{\nils} & \nullsort &
    \encsort{\bools} & \boolsort &
    \encsort{\nats} & \intsort \\
    \encsort{\rationals} & \realsort &
    \encsort{\strings} & \stringsort &
    \encsort{\listsof{\type}} & \seqsort{\encsort{\type}} &
    \encsort{\delta} & \adtsort
    \end{array}
  \]

  \sidecaption{\(\encj{\ste}{\sexp}{\term}{\sort}{\formulae}\) \quad \(\encjc{\ste}{\sexp}{\type}{\term}{\formulae}\)}
  \def\MathparLineskip{\lineskiplimit=0.2cm\lineskip=0.2cm plus 0.1cm}
  \begin{mathpar}
    \infer{\encjc{\ste}{\sexp}{\type}{\term'}{\formulae'}}
          {\encj{\ste}{\sexp}{\term}{\sort}{\formulae}
           \quad
           \totype{\type}(\term, \sort, \formulae) = (\term', \encsort{\type}, \formulae')}

    \infer{\encj{\ste}{x}{\smtvar x}{\encsort{\type}}{\emptyset}}
          {x : \type \in \ste}

    \infer{\encj{\ste}{x}{\smtvar x}{\maybevalsort}{\emptyset}}
          {x \notin \dom{\ste}}

    \infer{\encj{\ste}{\sexp_1 / \sexp_2}
                {\term_1 \smtdiv \term_2}{\intsort}
                {\formulae_1 \cup \formulae_2 \cup \{ \term_2 \smtneq \smtintlit{0} \}}}
          {\encjc{\ste}{\sexp_i}{\nats}{\term_i}{\formulae_i}}

    \infer{\encj{\ste}{\ctor(\sexp_1,\dots,\sexp_n)}
                {\smtconstr{\ctor}(\term_1, \dots, \term_n)}{\adtsort}
                {\textstyle\bigcup_i \formulae_i}}
          {\ctor(\type_1, \dots, \type_n):\delta \in \adtctx
           \quad
           \encjc{\ste}{\sexp_i}{\type_i}{\term_i}{\formulae_i}}

    \infer{\encj{\ste}{f(\sexp_1,\dots,\sexp_n)}
                {\smtlfunc{f}(\term_1, \dots, \term_n)}{\encsort{\type}}
                {\{ \smtlfuncpre{f}(\term_1, \dots, \term_n) \} \cup \textstyle\bigcup_i \formulae_i}}
          {\lfunc{f}{\vec{x} : \vec\type}{\type}{\lexp}{\lexp_{\mathit{pre}}} \in \lfuncctx
           \quad
           \encjc{\ste}{\sexp_i}{\type_i}{\term_i}{\formulae_i}}

    \infer{\encj{\ste}{\sexp \in \type}{\smtfalse}{\boolsort}{\emptyset}}
          {\neg\cwf(\type)}

    \infer{\encj{\ste}{\sexp \in \type}
                {\smtands\formulae \smtand \istype{\type}(\term)}
                {\boolsort}{\emptyset}}
          {\encjc{\ste}{\sexp}{\vals}{\term}{\formulae}
           \quad
           \cwf(\type)}
  \end{mathpar}

  \sidecaption{\(\tonat(\term, \sort, \formulae) = (\term', \sort', \formulae')\) \quad \(\totype{\type}(\term, \sort, \formulae) = (\term', \sort', \formulae')\)}
  \[
    \begin{gathered}
    \tonat(\term, \sort, \formulae) =
    \begin{cases}
      (\term, \intsort, \formulae)
      & \text{if } \sort = \intsort
      \\
      (\getnat(\term), \intsort,
      \formulae \cup \{ \isnat(\term) \})
      & \text{if } \sort = \valsort
      \\
      (\getnat(\getval(\term)), \intsort,
      \formulae \cup \{ \isval(\term), \isnat(\getval(\term)) \})
      & \text{if } \sort = \maybevalsort
      \\
      \text{undefined}
      & \text{otherwise}
    \end{cases}
    \\[0.3em]
    \begin{array}{r@{\;=\;}l@{\qquad}r@{\;=\;}l}
    \totype{\vals}(\term, \sort, \formulae) & \toval(\term, \sort, \formulae) &
    \totype{\bools}(\term, \sort, \formulae) & \tobool(\term, \sort, \formulae) \\
    \totype{\nats}(\term, \sort, \formulae) & \tonat(\term, \sort, \formulae) &
    \totype{\delta}(\term, \sort, \formulae) & \toadt{\delta}(\term, \sort, \formulae) \quad \dots
    \end{array}
    \end{gathered}
  \]
\caption{Background declarations, and the encoding.}%
\label{fig:encoding}
\end{figure}

We first define the SMT-LIB theory against which our encoded queries are checked. This \emph{background theory} extends the SMT-LIB theories we have mechanised with ADT sorts for the declarations in \(\adtctx\), function symbols for the logical functions in \(\lfuncctx\), and, crucially, a wrapper ADT sort for dynamically typed values. Since the type information in \(\ste\) determines the sorts assigned to logical variables, we write \(\theory_{\ste}\) for the resulting background theory, and \(\sig_{\ste}\) for its signature.

The \(\valsort\) ADT allows us to encode dynamically typed symbolic expressions. It has one constructor for each type a dynamic value could inhabit, given in \Cref{fig:encoding}, so a variable declared at sort \(\valsort\) can represent a boolean (constructed with \(\boolval\)), a natural (constructed with \(\natval\)), etc. Each constructor has the usual tester and selector, e.g.\ \(\isnat\) and \(\getnat\) for \(\natval\). For a \(\valsort\)-sorted term \(\term\), \(\istype{\type}(\term)\) checks that \(\term\) has the wrapper shape for \(\type\). It is defined by cases on \(\type\) (see \appref{app:backend:encoder}{C.3}), e.g.\ \(\istype{\nats}(\term) = \isnat(\term)\).

For each user-defined ADT \(\delta\), \(\sig_{\ste}\) declares an SMT-LIB sort \(\adtsort\), with one constructor \(\smtconstr{\ctor} : \smtrank{\encsort{\vec\type}}{\adtsort}\) per \(\ctor(\vec\type):\delta \in \adtctx\), and a dynamic wrapper \(\adtval{\delta}\) into \(\valsort\) (\Cref{fig:encoding}). SMT-LIB requires every datatype to have at least one finite value, so we assume every ADT in \(\adtctx\) is well founded, having a constructor whose argument types are all well founded.

Here, \(\encsort{\type}\) is the SMT-LIB sort encoding \(\type\) (\Cref{fig:encoding}). Some of these sorts admit more values than their source types: e.g., \(\nats\) is encoded as \(\intsort\), which includes negative integers. We address this with \emph{representability predicates}, Boolean-valued function symbols declared in \(\sig_{\ste}\) (\Cref{fig:encoding}) that identify exactly those SMT-LIB values that are representable as source values. The backend defines them with \smtdefinefunsrec{} (\Cref{sect:smt}). For example, the defining equation \mbox{\(\termforall{\smtvar{x}}{\intsort}{\reprnat(\smtvar{x}) \smteq (\smtcolor{x \geq 0})}\)} for \(\reprnat\) asserts non-negativity.
Similarly, \(\reprrat\) holds for the positive reals that are quotients of integers, and \(\reprval\) and each \(\repradt{\delta}\) recursively check constructor payloads, including list elements. We collect these defining equations in \(\smtreprdefs\), and give them in full in \appref{app:backend:lfunc-repr-defs}{C.2}. Analogously to \(\istype{\type}\), we write \(\smtrepr{\type}(\term)\) for the formula, built from these symbols, saying that an \(\encsort{\type}\)-sorted term \(\term\) is representable as a source value of type \(\type\), e.g., \(\smtrepr{\nats}(\term)\) is \(\reprnat(\term)\).

Logical interpretations \(\subst\) are partial (\Cref{sect:background}), but SMT-LIB valuations are total over the variables declared in the signature. To represent variables outside the domain of \(\subst\), the backend uses a second ADT \(\maybevalsort\) with constructors \(\someval\) and \(\noval\), tester \(\isval\), and selector \(\getval\) (\Cref{fig:encoding}).

The optimisation enabled by \(\ste\) is to avoid dynamic wrappers for variables of known type. If \(\lvar x : \type \in \ste\), then \(\subst(x)\) is defined and has type \(\type\) for every logical interpretation \(\subst\) with \(\subst \models \ste\). So \(\smtvar{x}\) can be declared directly at its native sort, as \(\smtvar{x} : \encsort{\type} \in \sig_{\ste}\), without ruling out any such \(\subst\). Otherwise, nothing guarantees that \(\subst(x)\) is defined, so \(\smtvar{x}\) is declared as \(\smtvar{x} : \maybevalsort \in \sig_{\ste}\). Native sorts greatly simplify the encoding, and improve solver performance by avoiding ADT reasoning about nestings of \(\valsort\) and \(\maybevalsort\). The example in \Cref{sect:backend:encoding} shows the difference in the generated query.

For each logical function
\(
  \lfunc{f}{\vec{x} : \vec\type}{\type}{\lexp}{\lexp_{\mathit{pre}}} \in \lfuncctx
\),
the signature contains a function symbol \(\smtlfunc{f}\) with rank
\(\smtrank{\encsort{\type_1}, \dots, \encsort{\type_n}}{\encsort{\type}}\),
a precondition symbol \(\smtlfuncpre{f}\) with rank
\(\smtrank{\encsort{\type_1}, \dots, \encsort{\type_n}}{\boolsort}\),
and an uninterpreted symbol \(\smtlfuncbot{f}\) with the same rank as \(\smtlfunc{f}\). It gives \(\smtlfunc{f}\) an unconstrained value outside its precondition, keeping the definition of \(\smtlfunc{f}\) in \Cref{sect:backend:lfuncs} satisfiable.

\subsection{Encoding Dynamically Typed and Partial Expressions}\label{sect:backend:encoding}

Given a typing environment \(\ste\) and a symbolic expression \(\sexp\), the expression encoder returns an SMT-LIB term \(\term\), its sort \(\sort\), and side conditions \(\formulae\), written \(\encj{\ste}{\sexp}{\term}{\sort}{\formulae}\) (\Cref{fig:encoding}). It ensures that \(\term\) faithfully represents \(\sexp\) when the side conditions \(\formulae\) hold. The sort \(\sort\) is synthesised bottom-up. Values are encoded as native literals, e.g.\ \(\nv\) at \(\intsort\), and a variable \(x\) has the sort it is declared at in \(\sig_{\ste}\), namely \(\encsort{\type}\) if \(x : \type \in \ste\), and \(\maybevalsort\) otherwise. Each operator then fixes the sort of its result, e.g.\ division produces an \(\intsort\)-sorted term, and its operands are coerced to the sorts it expects (again \(\intsort\)). The encoder is thus a \emph{type-directed translation} following a gradual type system~\cite{SiekTaha2006}.

In a gradual type system, as in our setting, type information is partially known, and a distinguished dynamic type (\(\vals\) in our case) represents the lack of type information. The type system allows dynamically typed subexpressions at statically typed use sites, and vice versa, while disallowing subexpressions whose static type is \emph{inconsistent} with their use site. Consistency \(\type \sim \type'\) holds in our gradual type system when the types are equal or one is \(\vals\), and we write \(\ste \vdash \sexp : \type\) for the typing judgement (see \appref{app:backend:typing}{C.6}). \citet{SiekTaha2006} define a cast insertion translation for their gradually typed calculus, elaborating source expressions into an intermediate language in which casts to and from the dynamic type are explicit. Our encoder performs a similar elaboration, and our \emph{coercions} are the analogue of these casts, converting between the dynamic \(\valsort\) sort and native sorts where necessary. Whereas an inserted cast can fail at run time, with type safety guaranteeing that the failure is signalled rather than silently producing a wrong answer, our coercions generate side conditions ruling out the SMT-LIB models in which they would fail.

For example, \(\tonat\) (\Cref{fig:encoding}) coerces a \(\valsort\)-sorted term \(\term\) to \(\getnat(\term)\) with the side condition \(\isnat(\term)\). Like all SMT-LIB function symbols, the selector \(\getnat\) is interpreted totally, and so is underspecified on arguments not built by \(\natval\), e.g.\ \(\getnat(\boolval(\smttrue))\) could be interpreted as any integer. The side condition \(\isnat(\term)\) rules out such underspecified evaluations, which correspond to type-erroring evaluations of the source expression. Coercions out of \(\valsort\) and \(\maybevalsort\) likewise apply selectors guarded by the corresponding testers. \(\totype{\type}\) coerces an encoding to the native sort \(\encsort{\type}\), dispatching on \(\type\) to the corresponding coercion (\Cref{fig:encoding}). We write \(\encjc{\ste}{\sexp}{\type}{\term}{\formulae}\) for the encoding of \(\sexp\) at type \(\type\), which encodes \(\sexp\) under \(\ste\) and then coerces the result with \(\totype{\type}\) (the first rule in \Cref{fig:encoding}). Conversely, \(\toval\) coerces into \(\valsort\). It wraps natively sorted terms in the matching constructor (e.g.\ \(\natval\) for \(\intsort\)) without side conditions, and unwraps \(\maybevalsort\) terms with \(\getval\) under the side condition \(\isval(\term)\). Both are defined in full in \appref{app:backend:encoder}{C.3}.

The encoder is a partial function, and where an expression form has several rules, their premises are disjoint. Apart from undeclared constructors, ADTs, or logical functions, it is undefined only where a coercion is. For example, \(1 / \true\) has no encoding, since \(\true\) is encoded at sort \(\boolsort\), from which \(\tonat\) is undefined. The encoder thus type-checks its input as it translates it. Like cast insertion, it is defined on every expression accepted by the gradual type system.

\begin{theorem}[Expression Encoder Totality]\label{thm:encode-total}
  If \(\; \ste \vdash \sexp : \type\), then \(\; \exists \term, \formulae. \, \encjc{\ste}{\sexp}{\type}{\term}{\formulae}\).
\end{theorem}

Side conditions also account for partiality beyond type errors, as the representative cases of the encoder in \Cref{fig:encoding} show. In the division case, the side condition \(\term_2 \smtneq \smtcolor{0}\) rules out the underspecified SMT-LIB evaluation of division by zero, which corresponds to an erroring evaluation of the source expression. The application case does the same with the side condition \(\smtlfuncpre{f}(\term_1, \dots, \term_n)\), where \(\smtlfuncpre{f}\) is the symbol for \(f\)'s precondition. This is exact because an application errors precisely when its precondition does not hold (\cref{sect:frontend:logical-functions}). Neither the body nor the precondition of \(f\) is encoded here: both are encoded once, in the SMT-LIB definitions of \(\smtlfunc{f}\) and \(\smtlfuncpre{f}\) (\Cref{sect:backend:lfuncs}). The type-membership case illustrates why expression side-conditions cannot all be asserted at the top level: \(\sexp \in \type\) should evaluate to \(\false\) when \(\sexp\) errors, rather than eliminating the whole  model, and so the side conditions are folded into the main term here.

To ground this all in a concrete example, consider \(\sexp \defeq (x / y = 2)\). If \(x\) and \(y\) are \(\maybevalsort\)-sorted, the encoder produces the term \(\getnat(\getval(\smtvar{x})) \smtdiv \getnat(\getval(\smtvar{y})) \smteq \smtintlit{2}\) with side conditions
\(
  \isval(\smtvar{x}) \smtand \isnat(\getval(\smtvar{x})) \smtand
  \isval(\smtvar{y}) \smtand \isnat(\getval(\smtvar{y})) \smtand
  \getnat(\getval(\smtvar{y})) \smtneq \smtintlit{0}
\). If \(x : \nats, y : \nats \in \ste\), the expression is encoded natively as \(\smtvar{x} \smtdiv \smtvar{y} \smteq \smtintlit{2}\), with the side condition \(\smtvar{y} \smtneq \smtintlit{0}\). Thus \(\ste\) removes dynamic wrappers in the encoding without changing the semantic contract.

\subsection{Logical Function Definitions}\label{sect:backend:lfuncs}

Logical functions are again defined with \smtdefinefunsrec{}. Given a logical function \(\lfunc{f}{\lvvar x : \vec\type}{\type}{\lexp}{\lexp_{\mathit{pre}}}\) in \(\lfuncctx\), its SMT-LIB definitions generate the following defining equations:
\[
  \begin{array}{cl}
  \termforall{\vec{\smtvar{x}}}{\encsort{\vec\type}}
    {\smtlfunc{f}(\vec{\smtvar{x}}) \smtcolor{=}
      \smtite(\smtlfuncpre{f}(\vec{\smtvar{x}}), \term_f, \smtlfuncbot{f}(\vec{\smtvar{x}}))},
  & \text{where } \encjc{\ste_f}{\lexp}{\type}{\term_f}{\formulae_f} \\
  \termforall{\vec{\smtvar{x}}}{\encsort{\vec\type}}
    {\smtlfuncpre{f}(\vec{\smtvar{x}}) \smtcolor{=}
     (\smtands_i \smtrepr{\type_i}(\smtvar{x}_i) \smtand
      \smtands \formulae_{\mathit{pre}} \smtand \formula_{\mathit{pre}})},
  & \text{where } \encjc{\ste_f}{\lexp_{\mathit{pre}}}{\bools}{\formula_{\mathit{pre}}}{\formulae_{\mathit{pre}}}
  \end{array}
\]
and \(\ste_f \defeq \{ x_i\mapsto\type_i \}_{i=1}^n\). In the definition of \(\smtlfunc{f}\), whose \(\smtite\) is SMT-LIB's if-then-else, the side conditions \(\formulae_f\) are discarded: body partiality is handled by the precondition. For \(\smtlfuncpre{f}\), the side conditions \(\formulae_{\mathit{pre}}\) of the precondition's encoding are kept instead, conjoined with the encoded precondition \(\formula_{\mathit{pre}}\) and the representability guards \(\smtrepr{\type_i}(\smtvar{x}_i)\) on the parameters. To guarantee that these definitions exist, we assume throughout, besides \(\models \lfuncctx\), that \(\lfuncctx\) is consistently typed: \(\ste_f \vdash \lexp : \type\) and \(\ste_f \vdash \lexp_{\mathit{pre}} : \bools\); by \Cref{thm:encode-total}, both encodings are then defined. We collect the generated formulae in \(\smtlfuncs\).

Validity of \(\lfuncctx\) guarantees that the function body terminates without erroring whenever its precondition holds. Elsewhere, a defining equation not guarded by the precondition may have no model (\Cref{sect:smt}); for example, \(\lfunc{f}{\lvar{n} : \nats}{\nats}{f(\lvar{n}) + 1}{\false}\) would generate the contradictory \mbox{\(\termforall{\smtvar{n}}{\intsort}{\smtlfunc{f}(\smtvar{n}) \smteq \smtlfunc{f}(\smtvar{n}) \smtplus \smtintlit{1}}\)}, which has no model, so every query would (unsoundly) be unsatisfiable. The definition therefore sets \(\smtlfunc{f}\) to the uninterpreted \(\smtlfuncbot{f}\) outside the precondition, and the application encoding in \Cref{sect:backend:encoding} guards calls with \(\smtlfuncpre{f}\). The parameter representability guards in \(\smtlfuncpre{f}\) are needed too: with the precondition \(\lvar{n} < 0\) in place of \(\false\), which no natural satisfies, \(\smtlfuncpre{f}(\smtintlit{-1})\) would otherwise hold over the SMT-LIB integers, reviving the contradiction.

\subsection{Correctness of Expression Encoding}\label{sect:backend:encoding-correctness}

We prove two facts about the expression encoder. First, the encoder emits well-sorted SMT-LIB terms and side conditions in the generated signature.

\begin{theorem}[Expression Encoding Well-Sortedness]\label{thm:encode-well-sorted}
  Suppose that \(\cwf(\type)\) for every \(x : \type \in \ste\), and that \(\encj{\ste}{\sexp}{\term}{\sort}{\formulae}\). Then, \(\sig_{\ste} \vdash \term : \sort \), and for all \(\formula \in \formulae\), \(\sig_{\ste} \vdash \formula : \boolsort\).
\end{theorem}

Second, the encoding is sound. To formulate our soundness theorem, we relate source values to SMT-LIB values by \emph{semantic equivalence}: a relation \(v \simeq \smtval{u}\) between a source value \(v\), or absence \(\undefd\), and an SMT-LIB value \(\smtval{u}\) (defined in full in \appref{app:backend:equiv}{C.5}). At native sorts, \(v \simeq \smtval{u}\) is direct: it relates equal boolean values, naturals to equal integers, and rationals to equal reals, recursing pointwise through lists and ADTs. At \(\valsort\), it unwraps the dynamic constructor. At \(\maybevalsort\), it likewise unwraps \(\someval\), and relates \(\undefd\) to \(\noval\). We write \(\subst \simeq \valuation\) for the pointwise lifting; that is, \(\subst(x)\simeq\valuation(x)\) for each variable \(x\), reading an undefined \(\subst(x)\) as \(\undefd\).

In \Cref{thm:encode-sound}, the assumptions fix the encoding and the evaluation of \(\sexp\): \(\lfuncctx\) is valid, \(\sexp\) evaluates under \(\subst\) to a value or \(\undefd\), and \(\subst \models \ste\). They relate the SMT-LIB side to it via \(\subst \simeq \valuation\), \(\struct \in \theory_{\ste}\), and \(\struct \models \smtlfuncs \cup \smtreprdefs\), so that \(\struct\) interprets logical functions and representability predicates as defined.

\begin{theorem}[Expression Encoding Soundness]\label{thm:encode-sound}
Assume \(\models \lfuncctx\), and let \(r, r' \in \vals \cup \{\undefd\}\).
\[
\begin{array}{@{}ll@{}}
\text{If} & \encj{\ste}{\sexp}{\term}{\sort}{\formulae},\quad
            \esem{\sexp}{\subst}=r,\quad
            \subst\models\ste,\\
          & \subst\simeq\valuation,\quad
            \struct\in\theory_{\ste},\quad
            \struct\models\smtlfuncs\cup\smtreprdefs,\\
\text{then} & \text{\normalfont (OX)}~~
              r \neq \undefd \Rightarrow
              \struct,\valuation\models\formulae \wedge
              \exists \smtval{u}.~\esem{\term}{\struct,\valuation}=\smtval{u} \wedge
              r\simeq\smtval{u},
              \text{ and}\\
          & \text{\normalfont (UX)}~~
              \forall r'.~\bigl(\struct,\valuation\models\formulae \wedge
              \exists \smtval{u}.~\esem{\term}{\struct,\valuation}=\smtval{u} \wedge
              r'\simeq\smtval{u}\bigr)
              \Rightarrow r'=r .
\end{array}
\]
\end{theorem}
\begin{proof}[Proof Sketch]
  Induct on expression evaluation, using functional induction for logical function calls to obtain hypotheses for the body and precondition. The two
  directions are proved simultaneously: compound cases sometimes need the opposite direction, notably type membership.
\end{proof}

The OX direction says that if the source evaluation does not error, the side conditions hold and the encoded term evaluates to an equivalent value. The UX direction says that an SMT-LIB value obtained while the side conditions hold corresponds to the same source result, even when it is \(\undefd\), so an SMT-LIB model cannot give an erroneous expression a value. Thus the side conditions are exact, admitting every evaluation of \(\sexp\) and excluding witnesses due only to underspecified selectors or unguarded partial operations. The theorem assumes a terminating evaluation \(\esem{\sexp}{\subst}=r\), which only context validity guarantees, requiring preconditions, and bodies under them, to terminate.

\subsection{Expression Satisfiability Check}\label{sect:backend:expr-sat}

The operational satisfiability check $\issat_{\ste}(\sexp)$ needs representability guards, making satisfying SMT-LIB models correspond to logical interpretations, and ill-typing, enabling a type optimisation.

Consider UX soundness of the operational satisfiability check; given a \(\sat\) answer, we must conclude that \(\sexp\) is satisfiable. To apply \Cref{thm:encode-sound} to a satisfying model, it must construct, from its SMT-LIB valuation \(\valuation\), a logical interpretation \(\subst\) with \(\subst \simeq \valuation\). This is impossible in general, as \(\valuation\) might assign non-representable values. For example, nothing in the encoding of \(x/y=2\) from \Cref{sect:backend:encoding} rules out negative integer assignments to \(\smtvar{x}\) and \(\smtvar{y}\), so the query also needs the guards \(\smtrepr{\nats}(\smtvar{x})\) and \(\smtrepr{\nats}(\smtvar{y})\). The OX direction goes the other way: it starts from a logical interpretation \(\subst\) and builds an equivalent valuation, so it needs only that the guards admit this valuation.

In general, the query asserts one guard for each variable in \(\lv{\sexp}\cup\dom{\ste}\), collected in
\[
  \reprguards_{\ste}(\sexp) \defeq
  \{ \smtrepr{\type}(\smtvar{x}) \mid x : \type \in \ste \}
  \cup
  \{ \isval(\smtvar{x}) \smtimplies \smtrepr{\vals}(\getval(\smtvar{x})) \mid x \in \lv{\sexp} \setminus \dom{\ste} \} .
\]
A variable outside \(\dom{\ste}\) is declared at \(\maybevalsort\), so its guard constrains only the defined case: \(\noval\) remains available, but every value it wraps is representable. A valuation thus satisfies the guards exactly when every value it assigns to these variables is representable. For UX, such a valuation determines a logical interpretation \(\subst\) with \(\subst \simeq \valuation\), reading \(\noval\) as \(\undefd\). For OX, a valuation built from a logical interpretation assigns only representable values, so it satisfies the guards.

The representability guards must be at the top level of the query to immediately rule out any unrepresentable models. They are not expression side conditions, because expression side conditions may need to be folded into encoded boolean terms, as in the type-membership case of \Cref{sect:backend:encoding}.

To rule out type-based \(\unsat\) cases without calling the underlying SMT solver, we define a negative counterpart to the gradual judgement \(\ste \vdash \sexp : \type\) which statically identifies expressions which \emph{must} result in a typing error. Following two-sided ill-typedness reasoning~\cite{Ramsay2024}, \(\ste \dashv \sexp : \type\) says that \(\sexp\) cannot evaluate to a value of type \(\type\) (which, note, is not just the negation of \(\ste \vdash \sexp : \type\)). Thus, if \(\ste \dashv \sexp : \bools\), a satisfiability check can return \(\unsat\) without calling the solver.

\begin{theorem}[Completeness of \(\dashv\)]\label{thm:type-completeness}
$\subst \models \ste \wedge \ste \dashv \sexp : \type \wedge \esem{\sexp}{\subst} = \gv \Rightarrow \gv \notin \type$
\end{theorem}

We are now ready to define the operational check, $\issat_{\ste}(\sexp)$, which builds a boolean query and invokes an SMT-LIB compliant solver:
\[
  \issat_{\ste}(\sexp) =
  \begin{cases}
    \unsat
      & \text{if } \ste \dashv \sexp : \bools \text{ or } \exists x : \type \in \ste.\ \neg\cwf(\type)
    \\[0.5ex]
    \begin{aligned}[t]
      &\smtcolor{\mathsf{solve}}(\theory_{\ste}, \{\formula\} \cup \formulae \cup \smtlfuncs
      \\
      &\qquad {} \cup \smtreprdefs \cup \reprguards_{\ste}(\sexp))
    \end{aligned}
      & \text{if } \encjc{\ste}{\sexp}{\bools}{\formula}{\formulae}
    \\[0.5ex]
    \unknown
      & \text{otherwise.}
  \end{cases}
\]
The \(\dashv\) condition of the first branch is the optimisation enabled by $\dashv$: it avoids a solver call when the type system proves that \(\sexp\) cannot evaluate to a boolean. Its well-formedness condition rejects ill-formed typing environments: malformed types are uninhabited (\Cref{sect:frontend:adts}), so no \(\subst\) satisfies \(\subst \models \ste\). It also ensures that the second branch meets the premise of \Cref{thm:encode-well-sorted}, so that the solver receives a well-sorted query. The second branch asserts the encoded boolean formula, side conditions, logical function definitions, representability-predicate definitions, and representability guards. \(\solvercheck{\theory}{\formulae}\) may return \(\sat\), \(\unsat\), or \(\unknown\). We assume only standard compliance: for well-sorted boolean queries, a \(\sat\) answer implies \(\smtsat_\theory(\formulae)\) and an \(\unsat\) answer implies \(\neg\smtsat_\theory(\formulae)\) (\Cref{sect:smt}).\footnote{A CSE tool following our theory inherits this assumption, but need not trust it blindly. A \(\sat\) answer's model can be lifted to a logical interpretation and checked against the source expression, and an \(\unsat\) answer can come with a proof, e.g.\ a cvc5~\cite{Barbosa2022} Alethe proof~\cite{Schurr21}, checked by Carcara~\cite{Andreotti23}.}

\begin{lemma}[Model Existence]\label{lem:model-existence}
  If \(\models \lfuncctx\), then some \(\struct \in \theory_{\ste}\) satisfies \(\struct \models \smtlfuncs \cup \smtreprdefs\).
\end{lemma}

\begin{theorem}[Satisfiability Check Soundness]\label{thm:smt-sat-sound}
\[
  \begin{array}{rr@{\;\;}c@{\;\;}l}
    \text{\normalfont (OX)}
    &
      \esem{\sexp}{\subst}=\true \wedge \subst \models \ste
    &\Rightarrow&
      \issat_{\ste}(\sexp) \in \{ \sat, \unknown\}
    \\
    \text{\normalfont (UX)}
    &
      \issat_{\ste}(\sexp) =\sat
    &\Rightarrow&
      \exists \subst.~ \esem{\sexp}{\subst}=\true \wedge \subst \models \ste .
  \end{array}
\]
\end{theorem}
\begin{proof}[Proof Sketch]
  Use \Cref{thm:type-completeness,thm:encode-well-sorted,thm:encode-sound} and solver compliance. For OX, take the model of Lemma~\ref{lem:model-existence}. A logical interpretation \(\subst\) witnessing satisfiability induces an SMT-LIB valuation \(\valuation\) with \(\subst \simeq \valuation\) that satisfies the representability guards; \Cref{thm:encode-sound} then gives the encoded formula and side conditions, hence satisfiability. For UX, the guards for \(\lv{\sexp}\cup\dom{\ste}\) rule out non-representable values in the model, so any satisfying model determines a \(\subst\) with \(\subst\models\ste\) and \(\subst\simeq\valuation\); applying \Cref{thm:encode-sound} transfers the boolean result back to the source side.
\end{proof}

\subsection{State Satisfiability and Entailment}\label{sect:backend:satisfiability}

State satisfiability and entailment are both built on the expression check \(\issat_{\ste}(\sexp)\).

We overload \(\issat\) for state satisfiability: the state check \(\issat(m, \sst)\) takes a mode \(m\) and a symbolic state \(\sst = (\ssto, \smem, \spred, \spc, \ste)\). It invokes the expression check \(\issat_{\ste}\) on the path condition \(\spc\) conjoined with the constraints implicit in the store and memory. The typing environment itself needs no such constraint, since \(\issat_{\ste}\) already enforces it: by \Cref{thm:smt-sat-sound}, a \(\sat\) answer yields a \(\subst\) with \(\subst \models \ste\). The store constraint \(\satof{\ssto}\) holds exactly when the symbolic store is satisfiable, that is, \(\esem{\satof{\ssto}}{\subst} = \true \Leftrightarrow \exists \sto.\ \subst, \sto \stomodels \ssto\), and is defined by \(\satof{\ssto} \defeq \bigwedge_{\sexp \in \rng(\ssto)} \sexp \in \vals\). The memory constraint \(\satof{\smem}\) is specific to the memory model, which must define it to satisfy the analogous contract \(\esem{\satof{\smem}}{\subst} = \true \Leftrightarrow \exists \mem.\ \subst, \mem \memmodels \smem\). For the linear heap of \Cref{sect:background}, \(\smem = \{\sexp_a^1 \mapsto \sexp_c^1, \dots, \sexp_a^n \mapsto \sexp_c^n\}\) represents a disjoint~union~of~singleton~heaps, so its constraint is
\[
    \satof{\smem} \defeq
      \bigwedge \{ \sexp_a^i \neq \sexp_a^j \mid 1 \leq i < j \leq n \}
      \wedge
      \bigwedge \{ \sexp_a^i \in \nats \mid 1 \leq i \leq n \}
    \wedge
      \bigwedge \{ \sexp_c^i \in \vals
        \mid 1 \leq i \leq n \}.
\]

The remaining approximation point is folded predicates. In OX mode, predicates may be ignored, since OX soundness only requires not pruning feasible executions: ignoring a predicate can only miss a contradiction in the state, never introduce one. Conversely, in UX mode, \(\sat\) must witness a real state, so the check returns \(\unknown\) while folded predicates remain, as their bodies may hold constraints, invisible to the check, that make the whole state unsatisfiable:
\[
  \issat(m, \sst) =
    \begin{cases}
      \unknown
        & \text{if } m = \mUX \text{ and } \spred \neq \emptyset
      \\
      \issat_{\ste}(\spc \wedge \satof{\ssto} \wedge \satof{\smem})
        & \text{otherwise.}
    \end{cases}
\]
Folded predicates can still contribute by unfolding before the state check, e.g. unfolding a list predicate into empty and non-empty cases until a branch exposes enough pure constraints.

A simple implementation of state entailment $\isentails(\sst, \sexp)$ checks, under the typing environment \(\ste\) of \(\sst\), that \(\issat_{\ste}(\spc \wedge \neg (\sexp \in \bools)) = \unsat\), so \(\sexp\) is necessarily boolean, and that \(\issat_{\ste}(\spc \wedge \neg \sexp) = \unsat\), so it cannot be false.

\begin{theorem}[State Check Soundness]\label{thm:state-check-sound}
Given memory-model support for each analysis:
\[
  \begin{array}{rl}
    \text{\normalfont (OX)}
    & \satpred(\sst)
      \Rightarrow
      \issat(\mOX, \sst) \in \{ \sat, \unknown \}
    \\
    \text{\normalfont (UX)}
    & \issat(\mUX, \sst) = \sat
      \Rightarrow
      \satpred(\sst)
    \\
    \text{\normalfont (Entailment)}
    &
      \isentails(\sst, \sexp)
      \Rightarrow
      \sst \models \sexp .
  \end{array}
\]
\end{theorem}

As discussed in \Cref{sect:overview}, state satisfiability checks happen throughout symbolic execution. To prove such checks sound, we model that they can happen at any point by encoding them as ghost commands: $\gcmds ::= \ldots \mid \psatpc \mid \psatstate$. The ghost command \(\psatpc\) checks only \(\issat_{\ste}(\spc)\); \(\psatstate\) invokes the full state check \(\issat(m, \sst)\).
\begin{theorem}[Ghost Check Soundness]
The ghost commands \(\psatpc\) and \(\psatstate\) are both OX and UX sound, that is, they both concretely correspond to no-ops.
\end{theorem}


\section{Empirical Evaluation}%
\label{sect:eval}

So far, our evaluation has been mathematical. Mathematical proof can answer the question \emph{``Is the CSE engine our theory models sound?''} It cannot answer the question \emph{``How does the modelled CSE engine relate to real-world CSE tools?''} We address that question, necessarily, by empirical~evaluation.

\subsection{Frontend Evaluation}\label{sect:eval:frontend}

We have designed the frontend of our theory to map directly and cleanly to important features of existing CSE tools. Together, algebraic data types, logical functions, lemmas, and user-defined predicates relating heap structures to their mathematical abstractions take us, perhaps surprisingly, far. We present three pieces of evidence that our frontend captures a useful fragment of real-tool frontends. This is not a full frontend model: constructs used by tools, such as sets, are still omitted.

\paragraph{Evidence 1: Our CSE theory to VeriFast} Examples written for our theory translate easily to real CSE tools: \Cref{fig:example-spec}, for instance, maps directly to VeriFast (see \appref{app:eval}{D}).

\begin{figure}[t]
\footnotesize
\begin{verbatim}
pred CRawEncryptionContext(+buffer : List, ECKs : List) :
(0i i<# |buffer|) * (#rawKC == l-sub(buffer, 0i, 2i)) * rawToUInt16(#rawKC, false, #keyCount) *
(0i i<# #keyCount) * Elements("Complete", buffer, 2i, #keyCount, 2i, ECKs, #ECKsLength) *
toUtf8PairMap(ECKs, #utf8ECKs) * FirstProj(ECKs, #ECKeys) *
Unique(#ECKeys) * ((2i i+ #ECKsLength) == |buffer|);
\end{verbatim}
\vspace{-1em}
\caption{A predicate for encryption contexts from the Gillian repository, lightly reformatted. See the file \texttt{Gillian-C/examples/amazon/logic/EncryptionHeaderLogic.gil}.}%
\label{fig:aws-string-pred}
\end{figure}

\paragraph{Evidence 2: VeriFast to our CSE theory} With ADTs, our theory can verify many data structures and their algorithms, as VeriFast's case studies illustrate. For example, \texttt{sorted\_bintree.c} verifies sorted binary trees and their operations. We have translated the parts of this example needed to verify its insert function into our theory (see \appref{app:eval}{D}).

\paragraph{Evidence 3: Our CSE theory vs. real-world verification case studies} Our frontend also scales to real-world case studies, as we illustrate using the AWS Encryption SDK message header deserialisation module case study of Gillian~\cite{Maksimovic2021}, the largest Gillian case study, covering a JavaScript~(${\approx}200$~LoC) and a C~(${\approx}1$~KLoC) implementation with ${\approx}3.5$~KLoC specifications, predicates, lemmas, etc. Our theory has all the frontend features the study needs except sets. As is common at this size, the study builds its abstractions layer by layer, each using our frontend's features. For example, the predicate of~\Cref{fig:aws-string-pred}, defining encryption contexts for the C study, builds on lower-level predicates such as \texttt{Unique}, stating that its list argument has no duplicates, itself built on lower-level ones.

\subsection{Backend Evaluation}

To evaluate the backend, we use Gillian, a production symbolic execution engine, as a source of solver queries. We extended Gillian to send each of its SMT queries to two backends: its own SMT-LIB encoder and our verified encoder, extracted to OCaml. Gillian's analysis continues on its own backend's answer, so the verified backend only observes. A bridge translates Gillian's symbolic expressions into those of our theory. Most of Gillian's expressions have a direct counterpart in ours. Only the encoder is extracted: the bridge, and the OCaml code that sends the encoded expressions to the solver, are unverified. This code omits the representability guards of \cref{sect:backend:expr-sat}, since Gillian's floats include zero and the negatives, and with the guards the comparison would measure that difference rather than the encodings (see \appref{app:eval:backend}{D.4}).

We ran 32 Gillian examples and case studies, covering JavaScript, C, and WISL (Gillian's demonstrator language), including the AWS case studies (\Cref{sect:eval:frontend}). We recorded both backends' answers and solver times, using Z3 4.15.4 and timing only its \texttt{check-sat} call.

\begin{figure}[t]
\centering
\begin{minipage}[c]{0.43\linewidth}
\centering
\begin{tikzpicture}
\begin{axis}[
  width=\linewidth,
  height=0.8\linewidth,
  xmode=log,
  ymode=log,
  xmin=0.015, xmax=10,
  ymin=0.015, ymax=10000,
  xlabel={Gillian Solver Time (ms)},
  ylabel={Verified Solver Time (ms)},
  xtick={0.1,1,10},
  ytick={0.1,1,10,100,1000,10000},
  xticklabels={0.1,1,10},
  yticklabels={0.1,1,10,$10^2$,$10^3$,$10^4$},
  grid=major,
  xminorticks=false,
  yminorticks=false,
  major grid style={gray!20},
  tick label style={font=\scriptsize},
  label style={font=\small},
  legend image post style={mark size=1.6pt},
  legend cell align=left,
  legend style={
    at={(0.02,0.98)},
    anchor=north west,
    draw=none,
    fill=white,
    fill opacity=0.85,
    text opacity=1,
    font=\scriptsize,
  },
]
\addplot[only marks, mark=*, mark size=0.8pt, mark options={draw=blue!60, fill=blue!20, draw opacity=0.8, fill opacity=0.6}]
  table[x=gillian_ms,y=certified_ms] {data/backend-solver-times-js.tsv};
\addlegendentry{JavaScript}
\addplot[only marks, mark=square*, mark size=0.8pt, mark options={draw=DarkGreen!70, fill=DarkGreen!25, draw opacity=0.8, fill opacity=0.6}]
  table[x=gillian_ms,y=certified_ms] {data/backend-solver-times-c.tsv};
\addlegendentry{C}
\addplot[only marks, mark=triangle*, mark size=1pt, mark options={draw=red!60, fill=red!20, draw opacity=0.8, fill opacity=0.6}]
  table[x=gillian_ms,y=certified_ms] {data/backend-solver-times-wisl.tsv};
\addlegendentry{WISL}
\addplot+[black, dashed, domain=0.015:10, samples=2, mark=none] {x};
\addlegendentry{Equal Time}
\end{axis}
\end{tikzpicture}
\label{fig:backend-solver-scatter}
\end{minipage}\hfill
\begin{minipage}[c]{0.55\linewidth}
\centering
\footnotesize
\setlength{\tabcolsep}{5pt}
\begin{tabular}{lrrrrr}
\toprule
\textbf{Lang.} & \textbf{Runs} & \textbf{Qs} & \textbf{Tr.} & \textbf{Cov.} & \textbf{Agree} \\
\midrule
JS    & 16 & 1324 &  985 & 74\% &  923/937 \\
C     &  7 & 4073 & 2989 & 73\% & 2988/2988 \\
WISL  &  9 &  423 &  416 & 98\% &  416/416 \\
\midrule
\textbf{Total} & 32 & 5820 & 4390 & 75\% & 4327/4341 \\
\bottomrule
\end{tabular}

\vspace{0.4em}
\footnotesize
``Tr.'' counts queries the bridge translated; ``Agree'' counts agreeing answers among queries both backends answered.
\label{fig:backend-coverage-table}
\end{minipage}
\vspace{-0.75em}
\caption{Data from Gillian backend comparison. The scatter plot shows solver time for all 4,341 queries answered by both backends; both axes are logarithmic and points near the dashed line have equal solver time under the two encodings. The table gives the corresponding query coverage and agreement counts.}%
\label{fig:backend-eval-summary}
\vspace{-1em}
\end{figure}

\Cref{fig:backend-eval-summary} summarises the results of this experiment. Gillian generated 5,820 SMT queries, of which the bridge translated 4,390 (75.4\%), covering all three languages. Most of the 1,430 untranslated queries contain negative integer literals, because Gillian normalises a subtraction \(a - b\) to \(a + ({-1}) \times b\), whereas the integers of our theory are the naturals; the others use quantifiers, sets, or list operations that our theory lacks. The verified encoder declines 48 of the translated queries, all from the JavaScript AWS case study, which apply a float comparison to integers derived from list lengths; this is not consistently typed in our theory, so lies outside the encoder's domain (\Cref{thm:encode-total}).

Both backends answered 4,341 of the remaining 4,342 queries, and agreed on 4,327 of them: 2,767 were answered \sat{} and 1,560 \unsat{} by both. The 14 disagreements, all \sat{} from Gillian and \unsat{} from the verified backend, reflect a difference between the expression languages rather than between the encodings: each casts the integer 0 to a float, but the rationals of our theory are positive, so the corresponding cast from naturals to rationals is undefined at 0, and its encoding has a side condition requiring a positive argument. Without it, every disagreeing query is \sat{}.

Solver times are comparable on most queries: the median is 0.09~ms under Gillian's encoding and 0.16~ms under the verified one. The exception is a small tail of 67 queries (1.5\%), all \sat{} and all from the AWS case studies, that take over 100~ms (up to 7.3~s) under the verified encoding but at most 1.3~ms under Gillian's. The verified encoder encodes list lengths exactly, so to answer \sat{} the solver must construct lists, whereas Gillian's encoder encodes some lengths as an uninterpreted function, which is sound only under invariants that Gillian's engine maintains. With lengths encoded as Gillian's encoder does, these queries take as long as the rest (see \appref{app:eval:backend}{D.4}).

Thus, the extracted encoder is compatible with Gillian's solver loop: it handles three quarters of real verification queries, agrees with Gillian's backend except where the expression languages differ in meaning, and needs comparable solver time apart from a small tail with a known cause.


\section{Related Work}

\paragraph{Frontend work}

Our frontend covers many, but not all, CSE features from existing nonverified tools. Notable omissions include Viper's magic wands~\cite{Dardinier22} and iterated separating conjunctions~\cite{Muller16}. We believe both features could be implemented on top of our theory. Extending the theory to concurrency would require more substantial changes. Existing approaches either reduce concurrent verification to sequential verification, as mechanised by~\citet{Dardinier2025}, or handle concurrency natively, as in the not-yet-mechanised~Raven~verifier~\cite{Gupta25}.
Unlike stand-alone tools such as VeriFast, Viper, and Gillian, some tools live inside an ITP: Steel~\cite{Fromherz21} and Pulse~\cite{Ebner25} in F$^\star$, RefinedC~\cite{Sammler21} and RefinedRust~\cite{Gaeher24} in Rocq, and, outside SL, Loom~\cite{Gladshtein26} in Lean. These tools can reuse the host ITP's mathematics, including ADTs and logical functions, rather than reconstructing it as we do here.

\paragraph{SMT-LIB work} We are not aware of a previous line-by-line mechanisation of the full SMT-LIB standard. The only dedicated SMT-LIB mechanisation we know is~\citet{Lotz23}, specialised to \texttt{Strings}. Other SMT-LIB mechanisations are by-products of projects that build only the fragment they need.
The most substantial is SMTCoq~\cite{Armand11}, a Rocq plugin for SMT-backed proof automation, whose SMT-LIB mechanisation is described only briefly and hard-codes a fixed set of theories rather than supporting general theory loading. A smaller example is CoqQFBV~\cite{Shi21}, covering only \texttt{FixedSizeBitVectors}.
We are not aware of any verified general-purpose SMT solver; verified SAT solvers, with a much simpler input format, are well established (e.g. IsaSAT~\cite{Blanchette2018,Fleury19,Fleury23}). Proof-producing SMT solvers, such as Z3 and cvc5, are common, with proposed proof formats such as LFSC~\cite{Stump12}, Alethe~\cite{Schurr21}, and eDRAT~\cite{Codel24}, plus independent checkers such as Carcara~\cite{Andreotti23}; none has a mechanised meta-theory with respect to SMT-LIB. SMT reasoning is also integrated into ITPs via Isabelle's Sledgehammer~\cite{Blanchette13}, Rocq's SMTCoq~\cite{Armand11}, and Lean's lean-smt~\cite{Abdalrhman25}. Most use proof reconstruction in the ITP logic rather than mechanised SMT-LIB semantics; SMTCoq instead uses computational reflection, with a checker verified against its own SMT-LIB mechanisation.

\paragraph{Backend work} Given the lack of an SMT-LIB mechanisation large enough for CSE tools, no previous work has mechanised a CSE tool's SMT encoding. Beyond CSE, the most comprehensive description of a nonverified SMT-LIB encoding is F$^\star$'s~\cite{fstar_smt}. ``Destabilising'' Iris~\cite{Spies25} lays foundations for connecting Iris to SMT solvers but does not model SMT-LIB itself, while Loom~\cite{Gladshtein26} uses Lean automation, some SMT-powered. Verification-condition generators have been mechanised without SMT-LIB connections: \citet{Parthasarathy2024} give a proof-producing generator from Viper to Boogie, and \citet{Nezamabadi26} a verified generator from Dafny to Dafny expressions.

\section{Conclusion}

We have presented mechanised progress towards a CSE theory that is simultaneously foundational, realistic, and end-to-end. Building on the CSE theory of \citet{Loow2025}, we extended the frontend of its CSE engine with expressive features used in real-world program analysis, and connected its backend to standard SMT solvers through a proven-correct encoding into SMT-LIB, built on our line-by-line mechanisation of the SMT-LIB standard.
We see two immediate avenues for future work. The first is executability: only our SMT-LIB encoding is executable, whereas the CSE engine is specified by inference rules, and validity of logical-function contexts is assumed rather than checked. The second is concurrency, which will require further frontend features such as invariants.

\section*{Disclosure of Use of AI in Research}

We used Claude (Anthropic) to develop proofs in the Rocq mechanisation and to write evaluation code. The authors wrote or closely reviewed all definitions and top-level theorem statements, and an author reviewed the evaluation code. Rocq checks all proofs, and we have audited their axioms: they rely only on the classical axioms listed in \Cref{sect:introduction}.

\section*{Data-availability Statement}

Our artefact~\cite{TamilKumaran2026Artefact} is archived on Figshare. It contains the Rocq mechanisation and, for the evaluation of \Cref{sect:eval}, the VeriFast translations and a fork of Gillian with the verified backend built in. Our SMT-LIB mechanisation is also published separately, as the package \texttt{rocq-smt-lib} on Rocq Packages, \url{https://rocq-prover.org/p/rocq-smt-lib}.

\begin{acks}
  We thank Shing Hin Ho and Alyssa Renata for pointing us to Tarski universes, used in our SMT-LIB mechanisation, and the anonymous POPL reviewer for helping us understand our expression encoder as a type-directed translation.
  This work was supported by Gardner's faculty gift from Meta and an Imperial College London Department of Computing PhD Scholarship for Tamil Kumaran.
\end{acks}


\bibliography{cse4.bib}

\ifarxiv
\newpage
\appendix
\crefalias{section}{appendix}
\crefalias{subsection}{subappendix}
\clearpage\section{Additional Background Definitions}%
\label{app:background-defs}

\subsection{Semantics of Assertions and Program Logics}

We now define the assertion satisfaction relation \(\subst, (\sto, \mem) \models \asrtcolor{P}\), which states that assertion \(\asrtcolor{P}\) is satisfied by logical interpretation \(\subst\) and concrete state \((\sto, \mem)\). Predicate definitions \(\pred(\vec x) = \asrtcolor{P}\) come from the program fragment \(\prg\) under analysis (\Cref{fig:syntax}); we require the parameters \(\vec x\) to be distinct logical variables and the body to satisfy \(\pv{\asrtcolor{P}} = \emptyset\) and \(\lv{\asrtcolor{P}} \subseteq \vec x\). The relation is defined inductively as follows:
\[
\begin{array}{lcl}
\subst, (\sto, \mem) \models \asrtcolor{\lexp} &\Leftrightarrow& \esem{\lexp}{\subst, \sto} = \true \ \land\ \mem = \emptyset \\
\subst, (\sto, \mem) \models \asrtcolor{\cell{\lexp_1}{\lexp_2}} &\Leftrightarrow& \mem = \{ \esem{\lexp_1}{\subst, \sto} \mapsto \esem{\lexp_2}{\subst, \sto} \} \\
\subst, (\sto, \mem) \models \asrtcolor{\pred(\vec\lexp)} &\Leftrightarrow& \esem{\vec\lexp}{\subst, \sto} = \vec\gv \ \land\ \{ \vec x \mapsto \vec\gv \}, (\emptyset, \mem) \models \asrtcolor{P} \text{ for } \pred(\vec x) = \asrtcolor{P} \in \prg \\
\subst, (\sto, \mem) \models \asrtcolor{\emp} &\Leftrightarrow& \mem = \emptyset \\
\subst, (\sto, \mem) \models \asrtcolor{\AssTrue} &\Leftrightarrow& \mathrm{true} \\
\subst, (\sto, \mem) \models \asrtcolor{P \lstar Q} &\Leftrightarrow& \exists \mem_1, \mem_2 \ldotp \mem = \mem_1 \uplus \mem_2 \ \land\ \subst, (\sto, \mem_1) \models \asrtcolor{P} \ \land\ \subst, (\sto, \mem_2) \models \asrtcolor{Q} \\
\subst, (\sto, \mem) \models \asrtcolor{P \vee Q} &\Leftrightarrow& \subst, (\sto, \mem) \models \asrtcolor{P} \ \lor\ \subst, (\sto, \mem) \models \asrtcolor{Q} \\
\subst, (\sto, \mem) \models \asrtcolor{P \Rightarrow Q} &\Leftrightarrow& \subst, (\sto, \mem) \models \asrtcolor{P} \ \Rightarrow\ \subst, (\sto, \mem) \models \asrtcolor{Q} \\
\subst, (\sto, \mem) \models \asrtcolor{\exists x \ldotp P} &\Leftrightarrow& \exists \gv \ldotp \subst[x \mapsto \gv], (\sto, \mem) \models \asrtcolor{P}
\end{array}
\]

We write \(\genquadruple{P}{\cmd}{\Qok}{\Qerr}\) for a quadruple of either kind, SL or ISL. The OX and UX judgements for SL and ISL quadruples (in terms of a function implementation context \(\fictx\)) are defined in terms of assertion satisfaction:
\[
  \begin{array}{ccc}
    \fictx \models \quadruple{P}{\cmd}{\Qok}{\Qerr} & \Leftrightarrow & \begin{array}{c}
      \forall \subst, \st, \st', \outcome. \;
      \subst, \st \models \asrtcolor{P} \, \wedge \,
      \csesemtransabstract{\st}{\cmd}{\st'}{\fictx}{\outcome} \; \Rightarrow \\
      \outcome \neq \oxm \, \wedge \, \subst, \st' \models \asrtcolor{Q_{\outcome}}
    \end{array}\\[2mm]
    \fictx \models \islquadruple{P}{\cmd}{\Qok}{\Qerr} & \Leftrightarrow & \begin{array}{c}
      \forall \subst, \st', \outcome. \;
      \subst, \st' \models \asrtcolor{Q_\outcome} \, \wedge \, \outcome \neq \oxm \; \Rightarrow \\
      \exists \st. \; \subst, \st \models \asrtcolor{P} \, \wedge \,
      \csesemtransabstract{\st}{\cmd}{\st'}{\fictx}{\outcome}
    \end{array}
  \end{array}
\]

A function quadruple
\[
S = \genquadruple{\pvvar x = \lvvar x \lstar P}{\pfuncspec{f}{\pvvar x}}{\Qok}{\Qerr}
\]
has program parameters \(\pvvar x\) and logical parameters \(\lvvar x\). It is \emph{well formed} iff the logical parameters are distinct and as many as the program parameters, \(\asrtcolor{P}\) contains no program variables, and \(\asrtcolor{\Qok}\) and \(\asrtcolor{\Qerr}\) contain no program variables other than the distinguished \(\pvar{ret}\) and \(\pvar{err}\), respectively, which stand for the return value and the error description. It is \emph{valid} under a function implementation context \(\fictx\) (denoted \(\fictx \models S\)) iff it is well formed, \(\pfunction{f}{\pvvar x}{\psequence{\cmd}{\preturn{\pexp}}} \in \fictx\), and
\[
\fictx \models \genquadruple{\pvvar x = \lvvar x \lstar P'}{\cmd}{\Qok'}{\Qerr'}
\]
for some internal assertions \(\asrtcolor{P'}\), \(\asrtcolor{\Qok'}\), and \(\asrtcolor{\Qerr'}\), where \(\asrtcolor{P'}\) extends \(\asrtcolor{P}\) with the initialisation of the local variables of \(\cmd\), and \(\asrtcolor{\Qok'}\) and \(\asrtcolor{\Qerr'}\), with the variables of \(\cmd\) existentially closed and \(\pvar{ret}\) bound to the return expression \(\pexp\), imply \(\asrtcolor{\Qok}\) and \(\asrtcolor{\Qerr}\) for OX specifications and are implied by them for UX specifications. We defer the specifics to \citet{Loow2025}. A function specification context \(\fsctx \in 2^{\specs}\) collects such function quadruples, and is valid under an implementation context \(\fictx\) iff for all \(S \in \fsctx\), \(\fictx \models S\). We write \(\models (\fictx, \fsctx)\) when this is the case.

\subsection{Symbolic Engine}

We now give the \(\consume\) and \(\produce\) operations of the engine in more detail. A symbolic substitution \(\ssubst : \lvars \pfmap \sexps\) instantiates free logical variables (e.g. \(a\) and \(\alpha\) in the precondition of \(\mathtt{prepend}\) in \Cref{fig:example-spec:append}) when matching assertions in \(\consume\). The judgements of \(\consume\) and \(\produce\) are:
\[
\consume(m, \asrtcolor{P}, \ssubst, \sst) \rightsquigarrow (\outcome, \ssubst', \sst') \qquad \text{ and } \qquad \produce(\asrtcolor{Q}, \ssubst, \sst) \rightsquigarrow \sst'
\]
where the judgement for \(\consume\) states that the consumption of assertion \(\asrtcolor{P}\) from state \(\sst\), in mode \(m\) and with initial symbolic substitution \(\ssubst\), results in outcome \(\outcome\), frame state \(\sst'\), and a learned symbolic substitution \(\ssubst'\) which now contains bindings for all free logical variables in \(\asrtcolor{P}\); and the judgement for \(\produce\) adds the resources and constraints imposed by \(\asrtcolor{Q}\) (with free variables bound by \(\ssubst\)) to the state \(\sst\) resulting in \(\sst'\).

\subsection{Function Verification}

The function verification procedure is defined by the following rule:
\[
\infer{
  \verifiesfun{\fsctx}{P}{f}{\pvvar x}{\cmd}{\pexp}{Q}
}{
  \begin{array}{@{}l@{}}
  \forall \sst, \outcome, \sst'.~ \produce(\asrtcolor{P}, \ssubst_{\mathrm{init}}, \sst_{\mathrm{init}}) \rightsquigarrow \sst \text{ and } \csesemtransabstractm{\sst}{\cmd}{\sst'}{\fsctx}{\mOX}{\outcome} \text{ and } \issat(\mOX, \sst') \neq \unsat \Rightarrow {} \\
  \quad\quad \outcome = \oxok \text{ and } {} \\
  \quad\quad (\forall \sexp_{\mathit{ret}}.~ \cseeval{\pexp}{\access{\sst'}{sto}}{}{\sexp_{\mathit{ret}}}{} \Rightarrow \isentails(\sst', \sexp_{\mathit{ret}} \in \vals) \text{ and } {} \\
  \quad\quad\quad\quad (\forall \outcome', \ssubst_1, \sst''.~ \consume(\mOX, \asrtcolor{Q[r/\pvar{ret}]}, \ssubst_{\mathrm{init}}[r \mapsto \sexp_{\mathit{ret}}], \sst') \rightsquigarrow (\outcome', \ssubst_1, \sst'') \text{ and } {} \\
  \quad\quad\quad\quad\quad\quad \issat(\mOX, \sst'') \neq \unsat \Rightarrow \outcome' \neq \oxabort \text{ and } \access{\sst''}{mem} = \emptyset \text{ and } \access{\sst''}{preds} = \emptyset))
  \end{array}
}
\]
The specification being verified is \(\triple{\pvvar x = \lvvar x \lstar P}{\pfuncspec{f}{\pvvar x}}{Q}\), where the logical parameters \(\lvvar x\) are the logical counterparts of the program parameters \(\pvvar x\) (in the mechanisation, the same names). The precondition is produced using the initial symbolic substitution \(\ssubst_{\mathrm{init}} \defeq \{\, \lvar{x} \mapsto \lvar{x} \mid \lvar{x} \in \lvvar{x} \cup \lv{\asrtcolor{P}} \,\}\), the identity substitution over the logical parameters and the logical variables of \(\asrtcolor{P}\), into the initial symbolic state \(\sst_{\mathrm{init}} \defeq (\ssto_{\mathrm{init}},\ \emptyset,\ \emptyset,\ \true)\), where the initial store \(\ssto_{\mathrm{init}} \defeq \{\, \pvar{x} \mapsto \lvar{x} \mid \pvar{x} \in \pvvar{x} \,\} \cup \{\, \pvar{y} \mapsto \nil \mid \pvar{y} \in \pv{\cmd} \setminus \pvvar{x} \,\}\) maps each parameter to its logical counterpart and every other variable of the body to \(\nil\), the memory and predicates are empty, and the path condition is \(\true\). After symbolic execution, the return expression \(\pexp\) is evaluated in the final store, and its value \(\sexp_{\mathit{ret}}\) must be entailed to be a proper value. The postcondition is then consumed with the distinguished variable \(\pvar{ret}\) replaced by a logical variable \(r \notin \lv{\sst'}\), which the substitution binds to \(\sexp_{\mathit{ret}}\). The final step is a leak check: the frame state \(\sst''\) left after consuming the postcondition must have empty memory and empty predicates, ensuring that the postcondition accounts for all remaining memory and predicates so that nothing is leaked.

\clearpage
\clearpage\section{Additional Extended Frontend Definitions}%
\label{app:frontend-defs}

\subsection{Algebraic Data Types}%

\paragraph{Logical expression evaluation}%

We give the formal definition of the logical expression evaluation function \(\esem{\cdot}{\subst, \sto}\) for the three constructs added for ADTs.

\paragraph{Constructor application.} \(\ctor(\vec\lexp)\) evaluates its arguments and returns the resulting constructor value, provided that value is well-typed:
\[
\esem{\ctor(\vec\lexp)}{\subst, \sto} =
\begin{cases}
\ctor(\vec\gv) & \text{if } \esem{\vec\lexp}{\subst, \sto} = \vec\gv,\ \ctor(\vec\type) : \delta \in \adtctx, \text{ and } \vec\gv \in \vec\type \\
\undefd & \text{otherwise.}
\end{cases}
\]

\paragraph{Testers.} \(\tester{\lexp}{\ctor}\) always evaluates to a Boolean, and in particular never to \(\undefd\), even when the scrutinee does:
\[
\esem{\tester{\lexp}{\ctor}}{\subst, \sto} =
\begin{cases}
\true & \text{if } \esem{\lexp}{\subst, \sto} = \ctor(\vec\gv) \text{ for some } \vec\gv \\
\false & \text{otherwise.}
\end{cases}
\]

\paragraph{Match expressions.} \(\match{\lexp}{\delta}{p_i \rightarrow \lexp_i|_{i = 1}^n}\), where each pattern \(p_i\) is either a variable \(x\) or a constructor pattern \(\ctor(\vec x)\), first evaluates the scrutinee. Unless it is a constructor value \(\ctor(\vec\gv)\) of the ADT \(\delta\) (that is, \(\ctor(\vec\type) : \delta \in \adtctx\)), the result is \(\undefd\). Otherwise, let \(j\) be the least index such that \(p_j\) matches \(\ctor(\vec\gv)\): a variable pattern \(p_j = x\) matches any value, whereas a constructor pattern \(p_j = \ctor'(\vec x)\) matches iff \(\ctor' = \ctor\). Evaluation then continues with the corresponding branch under an extended interpretation:
\[
\esem{\match{\lexp}{\delta}{p_i \rightarrow \lexp_i|_{i = 1}^n}}{\subst, \sto} =
\begin{cases}
\esem{\lexp_j}{\subst', \sto} & \text{if } \esem{\lexp}{\subst, \sto} = \ctor(\vec\gv),\ \ctor(\vec\type) : \delta \in \adtctx, \text{ and} \\
& \quad p_j \text{ is the first pattern matching } \ctor(\vec\gv) \\
\undefd & \text{otherwise,}
\end{cases}
\]
where \(\subst' \defeq \subst[x \mapsto \ctor(\vec\gv)]\) for a variable pattern \(p_j = x\), and \(\subst' \defeq \subst[\vec x \mapsto \vec\gv]\) for a constructor pattern \(p_j = \ctor(\vec x)\).

\paragraph{Command-level match}%

The main text gives the symbolic-execution rule for \(\ghostcolor{\pvar{match}}\) that handles a constructor pattern. We give here the remaining rules. The variable-binder rule mirrors the constructor rule, but binds a fresh logical variable \(x\) to the scrutinee and additionally records that the scrutinee is a value of the matched ADT \(\delta\) (which, unlike for a constructor pattern, is not already implied by the branch's equation):
\begin{mathpar}
\infer{
    \csesemtransabstractm{\sst}{\pmatch{\lexp}{\delta}{ \pcase{p_i}{\gcmd_i}|_{i=1}^{n}}}{\sst''}{\fsctx}{m}{\outcome}
}{
    \begin{array}{c}
    \esem{\lexp}{\access{\sst}{sto}} \baction \sexp \\
    p_j = x\ \text{a variable binder, for}\ 1 \leq j \leq n \\
    p_k\ \text{not a variable binder for}\ 1 \le k < j \\
    \spc = (x = \sexp) \wedge (\sexp \in \delta) \wedge \bigwedge_{k = 1}^{j - 1} \neg (\tester{\sexp}{\ctor_k}) \\
    \sst' = \sst[\sstupdate{pc}{\spc \wedge \access{\sst}{pc}}] \\
    \csesemtransabstractm{\sst'}{\gcmd_j}{\sst''}{\fsctx}{m}{\outcome}
    \end{array}
}
\end{mathpar}
The three error rules each evaluate the scrutinee and then abort, strengthening the path condition with the (mutually exclusive) condition characterising the error and recording the error kind in the \(\pvar{err}\) store variable: the scrutinee fails to evaluate to a value; it is a value but not of the ADT \(\delta\); or it is an ADT value matched by no pattern (which requires every pattern to be a constructor pattern, as a variable binder matches any value).
\begin{mathpar}
\infer{
    \csesemtransabstractm{\sst}{\pmatch{\lexp}{\delta}{ \pcase{p_i}{\gcmd_i}|_{i=1}^{n}}}{\sst'}{\fsctx}{m}{\oxabort}
}{
    \begin{array}{c}
    \esem{\lexp}{\access{\sst}{sto}} \baction \sexp \\
    \sst' = \sst[\sstupdate{sto}{\access{\sst}{sto}[\pvar{err} \storearrow \text{``ExprEval''}]}][\sstupdate{pc}{\neg(\sexp \in \vals) \wedge \access{\sst}{pc}}]
    \end{array}
}
\and
\infer{
    \csesemtransabstractm{\sst}{\pmatch{\lexp}{\delta}{ \pcase{p_i}{\gcmd_i}|_{i=1}^{n}}}{\sst'}{\fsctx}{m}{\oxabort}
}{
    \begin{array}{c}
    \esem{\lexp}{\access{\sst}{sto}} \baction \sexp \\
    \spc = (\sexp \in \vals) \wedge \neg(\sexp \in \delta) \\
    \sst' = \sst[\sstupdate{sto}{\access{\sst}{sto}[\pvar{err} \storearrow \text{``Type''}]}][\sstupdate{pc}{\spc \wedge \access{\sst}{pc}}]
    \end{array}
}
\and
\infer{
    \csesemtransabstractm{\sst}{\pmatch{\lexp}{\delta}{ \pcase{p_i}{\gcmd_i}|_{i=1}^{n}}}{\sst'}{\fsctx}{m}{\oxabort}
}{
    \begin{array}{c}
    \esem{\lexp}{\access{\sst}{sto}} \baction \sexp \\
    p_i = \ctor_i(\lvvar x_i)\ \text{for}\ 1 \le i \le n \\
    \spc = (\sexp \in \delta) \wedge \bigwedge_{i = 1}^{n} \neg (\tester{\sexp}{\ctor_i}) \\
    \sst' = \sst[\sstupdate{sto}{\access{\sst}{sto}[\pvar{err} \storearrow \text{``NoMatch''}]}][\sstupdate{pc}{\spc \wedge \access{\sst}{pc}}]
    \end{array}
}
\end{mathpar}

\subsection{Lemmas and Total Function Correctness}

\paragraph{Apply rule} We give the successful rule for \(\ghostcolor{\pvar{apply}}\), omitting the uninteresting error rules, which all abort:
\begin{mathpar}
\infer{
    \csesemtransabstractm{\sst}{\papply{l}{\vec\lexp}}{\sst'}{\fsctx \lemctx}{m}{\oxok}
}{
    \begin{array}{c}
    \lv{\vec\lexp} \subseteq \lv{\sst} \\
    (\lem{l}{\lvvar x}{P}{\iffstack}{Q}) \in \lemctx \\
    \iffstack\ \text{is sound in mode}\ m \\
    \esem{\vec\lexp}{\access{\sst}{sto}} \baction \vec\sexp\ \text{and}\  \ssubst = \{ \lvvar x \mapsto \vec\sexp \} \\
    \sst_{1} = \sst[\sstupdate{pc}{(\vec\sexp \in \vals \wedge \access{\sst}{pc})}] \\
    \consume(m, \asrtcolor{P}, \ssubst, \sst_{1}) \rightsquigarrow (\oxok, \ssubst', \sst_2) \\
    \produce(\asrtcolor{Q}, \ssubst', \sst_{2}) \rightsquigarrow \sst' \\
    \end{array}
}
\end{mathpar}
When the applied lemma specifies a measure \(M\), the \(\ghostcolor{\pvar{apply}}\) rule gains an additional premise, \(\isentails(\sst_2, \ssubst'(M) < M)\), ensuring the measure decreases on each inductive application.

\paragraph{Lemma verification}

The lemma verification judgement is defined by the following rule, which mirrors function verification (\Cref{app:background-defs}) with the proof \(\gcmd\) in place of the function body:
\[
\infer{
  \verifies{\lemctx}{\lemDefM{l}{\lvvar x}{P}{\Rightarrow}{Q}{\gcmd}{M}}
}{
  \begin{array}{@{}l@{}}
  \forall \sst, \outcome, \sst'.~ \produce(\asrtcolor{P}, \ssubst_{\mathrm{init}}, \sst_{\mathrm{init}}) \rightsquigarrow \sst \text{ and } \csesemtransabstractm{\sst}{\gcmd}{\sst'}{\fsctx\lemctx}{\mOX}{\outcome} \text{ and } \issat(\mOX, \sst') \neq \unsat \Rightarrow {} \\
  \quad\quad \outcome = \oxok \text{ and } {} \\
  \quad\quad (\forall \outcome', \ssubst_1, \sst''.~ \consume(\mOX, \asrtcolor{Q}, \ssubst_{\mathrm{init}}, \sst') \rightsquigarrow (\outcome', \ssubst_1, \sst'') \text{ and } \issat(\mOX, \sst'') \neq \unsat \Rightarrow {} \\
  \quad\quad\quad\quad \outcome' \neq \oxabort \text{ and } \access{\sst''}{mem} = \emptyset \text{ and } \access{\sst''}{preds} = \emptyset)
  \end{array}
}
\]
where \(\ssubst_{\mathrm{init}} \defeq \{\, x \mapsto x \mid x \in \lvvar x \cup \lv{\asrtcolor{P}} \,\}\) and \(\sst_{\mathrm{init}} \defeq (\emptyset, \emptyset, \emptyset, \true, \emptyset)\), with \(\ste'\) and \(\ste''\) the typing environments of \(\sst'\) and \(\sst''\): unlike a function body, a proof contains no program variables, so the initial store is empty. The rule verifies a lemma in the \(\Rightarrow\) direction; a \(\Leftarrow\) lemma is verified as the \(\Rightarrow\) lemma with \(\asrtcolor{P}\) and \(\asrtcolor{Q}\) swapped, and a \(\Leftrightarrow\) lemma in both directions. In each case, \(\lemctx\) contains the lemma being verified together with its measure, so that every application of the lemma within its own proof is subject to the measure check above.

\paragraph{Total function correctness}

The measure-based approach for lemma verification extends naturally to function verification, enabling sound verification of total correctness. The construction is almost identical:
\begin{enumerate}
  \item A termination judgement \(\st, \cmd \terminates_{\fictx}\) asserts that \(\cmd\) terminates from state \(\st\) under function implementation context \(\fictx\).
  \item Total OX specifications \(\totalquadrupleM{P}{\pfuncspec{f}{\pvvar x}}{\Qok}{\Qerr}{M}\) strengthen the usual validity condition to additionally require, for \(\pfunction{f}{\pvvar x}{\cmd; \preturn \pexp} \in \fictx\), that \(\forall \subst, \st.~\subst, \st \models \asrtcolor{P} \Rightarrow \st, \cmd \terminates_{\fictx}\).
  \item A new engine mode \(\mOXT\) operates identically to \(\mOX\), but consults only total specifications. A call to a function whose specification carries a measure checks that the measure decreases, analogously to the \(\ghostcolor{\pvar{apply}}\) rule above.
  \item \emph{Measured termination} of a function specification context requires the body of each measured total specification to terminate from every state that satisfies its (internalised) precondition and in which the measure is below \(n\).
\end{enumerate}
With these definitions, we prove that verification in mode \(\mOXT\) establishes total correctness:
\begin{theorem}[Total Function Verification Correctness]
Let
\[
\fsctx' \defeq \fsctx \cup \{ \totaltriple{P}{\pfuncspec{f}{\pvvar x}}{Q} \; (M\ \mathrm{\mathbf{decreases}}) \}
\quad\text{and}\quad
\fsctx'' \defeq \fsctx \cup \{ \totaltriple{P}{\pfuncspec{f}{\pvvar x}}{Q} \},
\]
and assume that no specification in \(\fsctx\) carries a measure, \(\pv{M} = \emptyset\), and \(\lv{M} \subseteq \lvvar x \cup \lv{\asrtcolor{P}}\):
\[
\begin{array}{ll}
\text{If } \quad & \models (\fictx, \fsctx) \text{ and } \pfunction{f}{\pvvar x}{\psequence{\cmd}{\preturn \pexp}} \in \fictx \text{ and } \\
                 & \verifiesfun{\fsctx'}{P}{f}{\pvvar x}{\cmd}{\pexp}{Q} \text{ in mode } \mOXT \\
\text{then} \quad & \models (\fictx, \fsctx'')
\end{array}
\]
where verification in mode \(\mOXT\) is the procedure of \Cref{app:background-defs} with \(\mOXT\) in place of \(\mOX\).
\end{theorem}
\begin{proof}[Proof Sketch]
The proof follows the same structure as that of \Cref{thm:lemma-verif-correct}, except that the measure now bounds termination rather than validity. Under measured termination, and validity of the context in the usual, partial sense, a command whose symbolic executions in mode \(\mOXT\) never abort on satisfiable states terminates from every concrete state that satisfies the initial symbolic state and in which the measure equals \(n\). Total correctness then follows by induction over the measure, with partial correctness as in \Cref{lem:recursive-verification-correct-idx}.
\end{proof}

\subsection{Mechanisation Notes}%
\label{app:frontend:rocq}

\paragraph{Values}
In the unextended CSE theory, values were simple to define:
\begin{minted}{coq}
Inductive val := VNull | VBool (b : bool) | VNat (n : nat) | (* ... *).
\end{minted}
With ADTs, however, value well-formedness depends on typing and typing depends on values. Since Rocq does not support such mutual inductive-inductive definitions, we define values as dependent pairs, assuming proof irrelevance to recover standard equality:
\begin{minted}{coq}
(* Pre-values, the old val type extended with ADT values *)
Inductive preval := pVADT (adt ctor : string) (args : list preval) | pVNull | (* ... *).
(* Well-typed values *)
Inductive preval_wf : preval -> Prop := (* ... *).
Definition val := { v : preval | preval_wf v }.
\end{minted}

\paragraph{Evaluation of logical expressions}
Before our extension, $\esem{\lexp}{\subst, \sto}$ was defined, straightforwardly, as follows:
\begin{minted}{coq}
Fixpoint lexp_eval (t:theta) (s:cstore) (e:lexp) : option val :=
 match e with
 | LEVal v => Some v
 (*...*).
\end{minted}
With logical functions, \texttt{lexp\_eval} is a function only when all functions in \(\lfuncctx\) terminate, but termination is most naturally stated using evaluation. We break this circularity in two stages:
\begin{minted}{coq}
Notation lfun_impl_ctx := (*...*).
Inductive lexp_eval_rel :
  lfun_impl_ctx -> theta -> cstore -> lexp -> option val -> Prop := (*...*).
Definition lfun_ctx_terminates (ctx : lfun_impl_ctx) := (*...*).
Notation lfun_ctx := { ctx : lfun_impl_ctx | lfun_ctx_terminates ctx }.
Definition lexp_eval (ctx : lfun_ctx) t s e : option val :=
 epsilon inhabited_option_val (lexp_eval_rel (`ctx) t s e).
\end{minted}
First, \texttt{lexp\_eval\_rel} states evaluation relationally; this lets \texttt{lfun\_ctx\_terminates} express termination. Then \texttt{lfun\_ctx} packages an implementation context with its termination proof, and Hilbert's epsilon operator turns the relation into \texttt{lexp\_eval}.

Because of Hilbert's epsilon operator, we cannot compute with \texttt{lexp\_eval} (e.g., \texttt{simpl} does nothing in proofs). We can, however, manually compute by rewriting with~the~following~theorem:
\begin{minted}{coq}
Theorem lexp_eval_eval : forall t s e,
 lexp_eval t s e =
 match e with
 | LEVal v => Some v
 (*...*).
\end{minted}

\clearpage
\clearpage\section{Backend SMT-LIB Encoding Definitions}%
\label{app:backend-defs}

\subsection{Background Theory}%
\label{app:backend:background}

The background theory is parameterised by the symbolic typing environment \(\ste : \lvars \pfmap \types\). It is the composition of the \texttt{Core}, \texttt{Reals\_Ints}, \texttt{Seq}, \texttt{Strings} standard SMT-LIB theories, extended with the ADT and function symbols introduced in \Cref{sect:backend:background}.

\subsubsection{Sort Symbols}
Sort symbols are extended with
\[
  \nullsort,\quad \valsort,\quad \maybevalsort,\quad
  \adtsort \; \text{ for each } \delta \in \adtctx.
\]
The translation \(\encsort{\tau}\), mapping source types \(\tau\) to sorts in this background theory, is the one used in the main text.

\subsubsection{Function Symbols}
Function symbols are extended to include logical functions, precondition
functions, uninterpreted counterparts of logical functions, representability predicates, ADT constructors, selectors, and
testers.

For each logical function \(f(\vec\type):\type \in \lfuncctx\),%
\footnote{We write \(f(\vec\type):\type \in \lfuncctx\) when
\(\lfunc{f}{\vec x : \vec\type}{\type}{\lexp}{\lexp_{\mathit{pre}}} \in \lfuncctx\) for some \(\vec x\), \(\lexp\),
and \(\lexp_{\mathit{pre}}\).}
the
background theory declares the function symbol, its precondition symbol, and
the uninterpreted symbol that gives its value outside the precondition:
\[
\begin{array}{@{}rcl@{}}
\smtlfunc{f}
  &:& \smtrank{\encsort{\type_1},\dots,\encsort{\type_n}}{\encsort{\type}}
      \quad
      \text{for } f(\vec\type):\type \in \lfuncctx
\\
\smtlfuncpre{f}
  &:& \smtrank{\encsort{\type_1},\dots,\encsort{\type_n}}{\boolsort}
\\
\smtlfuncbot{f}
  &:& \smtrank{\encsort{\type_1},\dots,\encsort{\type_n}}{\encsort{\type}}.
\end{array}
\]

The generated representability predicate symbols have ranks:
\[
\begin{array}{@{}rcl@{}}
\reprnat
  &:& \smtrank{\intsort}{\boolsort}
\\
\reprrat
  &:& \smtrank{\realsort}{\boolsort}
\\
\reprval
  &:& \smtrank{\valsort}{\boolsort}
\\
\repradt{\delta}
  &:& \smtrank{\adtsort}{\boolsort}.
\end{array}
\]

The constructors of the \(\nullsort\), \(\valsort\), and \(\maybevalsort\) ADTs are:
\[
\begin{array}{rcl@{\quad}rcl}
\nullconst &:& \nullsort &
\nullval &:& \smtrank{\nullsort}{\valsort} \\
\boolval &:& \smtrank{\boolsort}{\valsort} &
\natval &:& \smtrank{\intsort}{\valsort} \\
\ratval &:& \smtrank{\realsort}{\valsort} &
\stringval &:& \smtrank{\stringsort}{\valsort} \\
\listval &:& \smtrank{\seqsort{\valsort}}{\valsort} &
\adtval{\delta} &:& \smtrank{\adtsort}{\valsort} \\
\someval &:& \smtrank{\valsort}{\maybevalsort} &
\noval &:& \maybevalsort .
\end{array}
\]
Each non-nullary constructor has the corresponding selector: \(\getnull\), \(\getbool\),
\(\getnat\), \(\getrat\), \(\getstr\), \(\getlist\), \(\getadt{\delta}\), and
\(\getval\).  The tester function symbols are denoted similarly, for example,
\(\isnat\) and \(\isval\).

For every source constructor declaration
\(\ctor(\type_1,\dots,\type_n):\delta \in \adtctx\), the generated ADT sort
\(\adtsort\) contains constructor \(\smtconstr{\ctor}\)
with rank
\[
  \smtrank{\encsort{\type_1},\dots,\encsort{\type_n}}{\adtsort} .
\]
Its \(i\)-th selector \(\getctor{\ctor}{i}\) has rank
\(\smtrank{\adtsort}{\encsort{\type_i}}\), and its tester \(\isctor{\ctor}\) has rank
\(\smtrank{\adtsort}{\boolsort}\).

\subsection{Logical Function and Representability Predicate Definitions}
\label{app:backend:lfunc-repr-defs}

For a logical function
\(\lfunc{f}{\vec x: \vec\type}{\type}{\lexp}{\lexp_{\mathit{pre}}}\),
let \(\ste \defeq \{ \vec x \mapsto \vec\type \} \).
Recall from \Cref{sect:smt} that defining a function is semantically equivalent to asserting its defining equation; in the case of our logical function, the generated defining equation is:
\[
  \termforall{\vec{\smtvar{x}}}{\encsort{\vec\type}}
    {\smtlfunc{f}(\vec{\smtvar{x}}) \smtcolor{=}
      \smtite(\smtlfuncpre{f}(\vec{\smtvar{x}}), \term, \smtlfuncbot{f}(\vec{\smtvar{x}}))}
  \quad
  \text{where }
  \encjc{\ste}{\lexp}{\type}{\term}{\formulae}.
\]
The side conditions of the body encoding are not asserted in this equation; soundness
uses the semantic validity of logical functions under their preconditions.
The precondition function does include those side conditions, together with
representability guards \(\smtrepr{\type_i}\) (defined below) on the parameters:
\[
  \termforall{\vec{\smtvar{x}}}{\encsort{\vec\type}}
    {\smtlfuncpre{f}(\vec{\smtvar{x}}) \smtcolor{=}
     (\smtands_i \smtrepr{\type_i}(\smtvar{x}_i) \smtand
      \smtands \formulae \smtand \formula)},
  \quad
  \text{where }
  \encjc{\ste}{\lexp_{\mathit{pre}}}{\bools}{\formula}{\formulae}.
\]
The set of all such equations is \(\smtlfuncs\).

The representability predicates are generated by the same mechanism: each
displayed clause below abbreviates the defining equation emitted
by \(\smtdefinefunsrec{}\). We write
\[
  \smtcolor{p}(\smtvar{x} : \sort) \smtdefeq \formula
  \quad\text{for}\quad
  \termforall{\smtvar{x}}{\sort}
    {\smtcolor{p}(\smtvar{x}) \smtcolor{=} \formula}.
\]
For the generated predicates:
\[
\begin{array}{rcl}
\reprnat(\smtvar{x} : \intsort)
  &\smtdefeq& \smtcolor{x \geq 0}
\\[0.3em]
\reprrat(\smtvar{x} : \realsort)
  &\smtdefeq&
    \termexists{\smtvar{n}}{\intsort}{
      \termexists{\smtvar{d}}{\intsort}{
      \smtcolor{
        d \neq 0} \smtand
      \smtcolor{
        x = \smtf{to\_real}(n) /
            \smtf{to\_real}(d)}
        \smtand \smtcolor{x > 0}}}
\\[0.3em]
\reprval(\smtvar{x} : \valsort)
  &\smtdefeq&
    \isnull(\smtvar{x})
\\
&&{} \smtor
    \isbool(\smtvar{x})
\\
&&{} \smtor
    (\isnat(\smtvar{x}) \smtand
      \reprnat(\getnat(\smtvar{x})))
\\
&&{} \smtor
    (\israt(\smtvar{x}) \smtand
      \reprrat(\getrat(\smtvar{x})))
\\
&&{} \smtor
    \isstring(\smtvar{x})
\\
&&{} \smtor
    (\islist(\smtvar{x}) \smtand
      \termforall{\smtvar{i}}{\intsort}{
        \smtcolor{0} \smtleq \smtvar{i}
        \smtlt \seqlen(\getlist(\smtvar{x}))
        \smtimplies
\\
&&\qquad\quad
        \reprval(
          \seqnth(\getlist(\smtvar{x}), \smtvar{i}))})
\\
&&{} \smtor
    \smtors_{\delta \in \adtctx}
      (\isadt{\delta}(\smtvar{x}) \smtand
       \repradt{\delta}
        (\getadt{\delta}(\smtvar{x}))).
\end{array}
\]
For each ADT \(\delta\),
\[
  \repradt{\delta}(\smtvar{a} : \adtsort)
  \smtdefeq
  \smtors_{\ctor(\vec\type):\delta \in \adtctx}
    \isctor{\ctor}(\smtvar{a})
    \smtand
    \smtands_i
      \smtrepr{\type_i}(\getctor{\ctor}{i}(\smtvar{a})).
\]
The generated defining equations for \(\reprnat\), \(\reprrat\),
\(\reprval\), and each \(\repradt{\delta}\) are collected in
\(\smtreprdefs\).

The type-indexed guard \(\smtrepr{\type}\), used in the \(\repradt{\delta}\)
equation, is a meta-level abbreviation:
\[
\begin{gathered}
\begin{array}{rcl@{\qquad}rcl}
\smtrepr{\vals}(\term) &=& \reprval(\term) &
\smtrepr{\nils}(\term) &=& \smttrue
\\
\smtrepr{\bools}(\term) &=& \smttrue &
\smtrepr{\nats}(\term) &=& \reprnat(\term)
\\
\smtrepr{\rationals}(\term) &=& \reprrat(\term) &
\smtrepr{\strings}(\term) &=& \smttrue
\\
\smtrepr{\delta}(\term) &=& \repradt{\delta}(\term)
\end{array}
\\[0.4em]
\smtrepr{\listsof{\type}}(\term)
  = \termforall{\smtvar{i}}{\intsort}{
      \smtcolor{0} \smtleq \smtvar{i} \smtlt \seqlen(\term) \smtimplies
      \smtrepr{\type}(\seqnth(\term,\smtvar{i}))}.
\end{gathered}
\]

\subsection{Expression Encoder and Coercions}
\label{app:backend:encoder}

An encoding state is a triple \((\term,\sort,\formulae)\) of an SMT-LIB term, its sort,
and a finite set of boolean side conditions.

The value coercion \(\toval\) turns a term into a \(\valsort\)-sorted
term:
\[
\begin{array}{rcl}
\toval(\term,\valsort,\formulae) &=& (\term,\valsort,\formulae)\\
\toval(\term,\maybevalsort,\formulae)
  &=& (\getval(\term),\valsort,\formulae \cup \{\isval(\term)\})\\
\toval(\term,\nullsort,\formulae)
  &=& (\nullval(\term),\valsort,\formulae)\\
\toval(\term,\boolsort,\formulae)
  &=& (\boolval(\term),\valsort,\formulae)\\
\toval(\term,\intsort,\formulae)
  &=& (\natval(\term),\valsort,\formulae)\\
\toval(\term,\realsort,\formulae)
  &=& (\ratval(\term),\valsort,\formulae)\\
\toval(\term,\stringsort,\formulae)
  &=& (\stringval(\term),\valsort,\formulae)\\
\toval(\term,\encsort{\delta},\formulae)
  &=& (\adtval{\delta}(\term),\valsort,\formulae).
\end{array}
\]
For a sequence sort \(\seqsort{\sort}\), the mechanisation maps each element
through \(\toval\):
\[
  \toval(\term,\seqsort{\sort},\formulae)
  =
  (\listval(\seqmap(\termlambda{\smtvar{x}}{\sort}{\term_x},\term)),
    \valsort,\formulae),
\]
where \(\toval(\smtvar{x},\sort,\emptyset)=(\term_x,\valsort,\emptyset)\).

The primitive coercions \(\tonull\), \(\tobool\),
\(\tonat\), \(\torat\), \(\tostring\), and
\(\toadt{\delta}\) are the inverse projections at native sort, at
\(\valsort\), and at \(\maybevalsort\).  For example:
\[
\begin{array}{rcl}
\tonat(\term,\intsort,\formulae) &=& (\term,\intsort,\formulae)\\
\tonat(\term,\valsort,\formulae)
  &=& (\getnat(\term),\intsort,\formulae \cup \{\isnat(\term)\})\\
\tonat(\term,\maybevalsort,\formulae)
  &=& (\getnat(\getval(\term)),\intsort,
       \formulae \cup \{\isval(\term),\isnat(\getval(\term))\})\\
\tonat(\term,\sort,\formulae)
  &=& \text{undefined, for any other } \sort.
\end{array}
\]
The type-directed coercion \(\totype{\type}\) dispatches to the corresponding
primitive coercion. The list case is special.  If the input already has sort
\(\seqsort{\encsort{\type}}\), it is returned unchanged.  If it has sort
\(\seqsort{\sort}\) for some different element sort, the output is
\(\seqmap(\termlambda{\smtvar{x}}{\sort}{\term_x},\term)\), where
\(\totype{\type}(\smtvar{x},\sort,\emptyset)=(\term_x,\encsort{\type},\formulae_x)\), together with the
side condition
\[
  \termforall{\smtvar{i}}{\intsort}{
    \smtcolor{0} \smtleq \smtvar{i} \smtlt \seqlen(\term)
    \smtimplies
    (\smtands\formulae_x)[\seqnth(\term,\smtvar{i})/\smtvar{x}]},
\]
where \(\formula[\term'/\smtvar{x}]\) denotes the substitution of \(\term'\) for
\(\smtvar{x}\) in \(\formula\): each element of \(\term\) must satisfy the side
conditions of its coercion.
The \(\valsort\) and \(\maybevalsort\) list cases first select the wrapped list
payload and add the corresponding tester side conditions.

The encoding judgement \(\encj{\ste}{\sexp}{\term}{\sort}{\formulae}\) is defined by the following syntax-directed rules, which extend those of \Cref{fig:encoding}. As there, rules with an index \(i\) have their premises for each \(i\), and where an expression form has several rules, their side conditions are disjoint. Encoding at a type, \(\encjc{\ste}{\sexp}{\type}{\term}{\formulae}\), coerces the encoding with \(\totype{\type}\), where \(\totype{\vals}\) is \(\toval\):
\begin{mathpar}
  \infer{\encjc{\ste}{\sexp}{\type}{\term'}{\formulae'}}
        {\encj{\ste}{\sexp}{\term}{\sort}{\formulae}
         \quad
         \totype{\type}(\term, \sort, \formulae) = (\term', \encsort{\type}, \formulae')}
\end{mathpar}
Primitive values become native SMT-LIB literals with no side conditions:
\begin{mathpar}
  \infer{\encj{\ste}{\nil}{\nullconst}{\nullsort}{\emptyset}}{}

  \infer{\encj{\ste}{\bv}{\smtboollit{b}}{\boolsort}{\emptyset}}{}

  \infer{\encj{\ste}{\nv}{\smtintlit{n}}{\intsort}{\emptyset}}{}

  \infer{\encj{\ste}{q}{\smtreallit{q}}{\realsort}{\emptyset}}{}

  \infer{\encj{\ste}{\sv}{\smtstrlit{s}}{\stringsort}{\emptyset}}{}
\end{mathpar}
Variables, constructors, lists, and logical-function calls are encoded by:
\begin{mathpar}
  \infer{\encj{\ste}{x}{\smtvar x}{\encsort{\type}}{\emptyset}}
        {x : \type \in \ste}

  \infer{\encj{\ste}{x}{\smtvar x}{\maybevalsort}{\emptyset}}
        {x \notin \dom{\ste}}

  \infer{\encj{\ste}{\ctor(\sexp_1,\dots,\sexp_n)}
              {\smtconstr{\ctor}(\term_1, \dots, \term_n)}{\adtsort}
              {\textstyle\bigcup_i \formulae_i}}
        {\ctor(\type_1, \dots, \type_n):\delta \in \adtctx
         \quad
         \encjc{\ste}{\sexp_i}{\type_i}{\term_i}{\formulae_i}}

  \infer{\encj{\ste}{[\sexp_1,\dots,\sexp_n]}
              {\term_\ell}{\seqsort{\valsort}}
              {\textstyle\bigcup_i \formulae_i}}
        {\encjc{\ste}{\sexp_i}{\vals}{\term_i}{\formulae_i}}

  \infer{\encj{\ste}{f(\sexp_1,\dots,\sexp_n)}
              {\smtlfunc{f}(\term_1, \dots, \term_n)}{\encsort{\type}}
              {\{ \smtlfuncpre{f}(\term_1, \dots, \term_n) \} \cup \textstyle\bigcup_i \formulae_i}}
        {\lfunc{f}{\vec{x} : \vec\type}{\type}{\lexp}{\lexp_{\mathit{pre}}} \in \lfuncctx
         \quad
         \encjc{\ste}{\sexp_i}{\type_i}{\term_i}{\formulae_i}}
\end{mathpar}
where, in the list rule, \(\term_\ell \defeq \sequnit(\term_1) \seqconcat \cdots \seqconcat \sequnit(\term_n) \seqconcat \seqempty{\seqsort{\valsort}}\).

Representative operation cases follow. Negation and length encode their operand at \(\bools\) and at \(\listsof{\vals}\), and the natural-number operations encode both operands at \(\nats\):
\begin{mathpar}
  \infer{\encj{\ste}{\neg \sexp}{\smtcolor{\neg}\term}{\boolsort}{\formulae}}
        {\encjc{\ste}{\sexp}{\bools}{\term}{\formulae}}

  \infer{\encj{\ste}{|\sexp|}{\seqlen(\term)}{\intsort}{\formulae}}
        {\encjc{\ste}{\sexp}{\listsof{\vals}}{\term}{\formulae}}

  \infer{\encj{\ste}{\sexp_1 + \sexp_2}{\term_1 \smtplus \term_2}{\intsort}{\formulae_1 \cup \formulae_2}}
        {\encjc{\ste}{\sexp_i}{\nats}{\term_i}{\formulae_i}}

  \infer{\encj{\ste}{\sexp_1 - \sexp_2}
              {\smtite(\term_1 \smtminus \term_2 \smtlt \smtcolor{0}, \smtcolor{0}, \term_1 \smtminus \term_2)}
              {\intsort}{\formulae_1 \cup \formulae_2}}
        {\encjc{\ste}{\sexp_i}{\nats}{\term_i}{\formulae_i}}

  \infer{\encj{\ste}{\sexp_1 / \sexp_2}
              {\term_1 \smtdiv \term_2}{\intsort}
              {\formulae_1 \cup \formulae_2 \cup \{ \term_2 \smtneq \smtcolor{0} \}}}
        {\encjc{\ste}{\sexp_i}{\nats}{\term_i}{\formulae_i}}

  \infer{\encj{\ste}{\sexp_1 \bmod \sexp_2}
              {\term_1 \smtmod \term_2}{\intsort}
              {\formulae_1 \cup \formulae_2 \cup \{ \term_2 \smtneq \smtcolor{0} \}}}
        {\encjc{\ste}{\sexp_i}{\nats}{\term_i}{\formulae_i}}
\end{mathpar}
Equality compares the encoded terms directly when both have the same sort other than \(\maybevalsort\), and otherwise coerces both with \(\toval\). The coercion is necessary, since terms of different sorts can still denote equal values, so differing sorts do not make an equality false. For example, if \(x : \listsof{\vals} \in \ste\) and \(y : \listsof{\nats} \in \ste\), then \(x\) and \(y\) are encoded at the different sorts \(\seqsort{\valsort}\) and \(\seqsort{\intsort}\), yet \(x = y\) evaluates to \(\true\) whenever both denote the same list of naturals. Both sides are coerced even when both have sort \(\maybevalsort\). Compared directly, two variables outside the domain of \(\subst\) would both be \(\noval\), making the equality true, whereas in the source it evaluates to \(\undefd\); the side conditions \(\isval(\term_i)\) added by \(\toval\) rule this out:
\begin{mathpar}
  \infer{\encj{\ste}{\sexp_1 = \sexp_2}{\term_1 \smteq \term_2}{\boolsort}{\formulae_1 \cup \formulae_2}}
        {\encj{\ste}{\sexp_i}{\term_i}{\sort_i}{\formulae_i}
         \quad
         \sort_1 = \sort_2 \neq \maybevalsort}

  \infer{\encj{\ste}{\sexp_1 = \sexp_2}{\term'_1 \smteq \term'_2}{\boolsort}{\formulae'_1 \cup \formulae'_2}}
        {\encj{\ste}{\sexp_i}{\term_i}{\sort_i}{\formulae_i}
         \quad
         \sort_1 \neq \sort_2 \text{ or } \sort_1 = \maybevalsort
         \quad
         \toval(\term_i, \sort_i, \formulae_i) = (\term'_i, \valsort, \formulae'_i)}
\end{mathpar}
The list operations encode their element at \(\vals\) and their list at \(\listsof{\vals}\):
\begin{mathpar}
  \infer{\encj{\ste}{\sexp_1 :: \sexp_2}
              {\sequnit(\term_1) \seqconcat \term_2}{\seqsort{\valsort}}
              {\formulae_1 \cup \formulae_2}}
        {\encjc{\ste}{\sexp_1}{\vals}{\term_1}{\formulae_1}
         \quad
         \encjc{\ste}{\sexp_2}{\listsof{\vals}}{\term_2}{\formulae_2}}

  \infer{\encj{\ste}{\sexp_1 \in \sexp_2}
              {\seqcontains(\term_2, \sequnit(\term_1))}{\boolsort}
              {\formulae_1 \cup \formulae_2}}
        {\encjc{\ste}{\sexp_1}{\vals}{\term_1}{\formulae_1}
         \quad
         \encjc{\ste}{\sexp_2}{\listsof{\vals}}{\term_2}{\formulae_2}}
\end{mathpar}
Rational arithmetic is analogous to natural arithmetic, encoding operands at \(\rationals\) and using \(\realsort\) and its operators.
Subtraction is the exception: \(\rationals\) contains only positive rationals, so rational subtraction is undefined
unless its result is positive, rather than saturating like natural subtraction.
Its encoding is \(\term_1 \smtminus \term_2\) with the side condition
\(\term_2 \smtlt \term_1\).

The type-test case deliberately folds side conditions into the boolean term:
\begin{mathpar}
  \infer{\encj{\ste}{\sexp \in \type}{\smtfalse}{\boolsort}{\emptyset}}
        {\neg\cwf(\type)}

  \infer{\encj{\ste}{\sexp \in \type}
              {\smtands\formulae \smtand \istype{\type}(\term)}
              {\boolsort}{\emptyset}}
        {\encjc{\ste}{\sexp}{\vals}{\term}{\formulae}
         \quad
         \cwf(\type)}
\end{mathpar}
Here \(\istype{\type}\) is the wrapper-shape predicate:
\[
\begin{gathered}
\begin{array}{rcl@{\qquad}rcl}
\istype{\vals}(\term) &=& \smttrue &
\istype{\nils}(\term) &=& \isnull(\term)
\\
\istype{\bools}(\term) &=& \isbool(\term) &
\istype{\nats}(\term) &=& \isnat(\term)
\\
\istype{\rationals}(\term) &=& \israt(\term) &
\istype{\strings}(\term) &=& \isstring(\term)
\\
\istype{\delta}(\term) &=& \isadt{\delta}(\term)
\end{array}
\\[0.4em]
\istype{\listsof{\type}}(\term)
  = \islist(\term)
    \smtand
    \termforall{\smtvar{i}}{\intsort}{
      \smtcolor{0} \smtleq \smtvar{i} \smtlt \seqlen(\getlist(\term))
      \smtimplies
      \istype{\type}(\seqnth(\getlist(\term),\smtvar{i}))}.
\end{gathered}
\]

The constructor-test case is similar. An undeclared constructor gives \(\smtfalse\) without encoding \(\sexp\):
\begin{mathpar}
  \infer{\encj{\ste}{\tester{\sexp}{\ctor}}{\smtfalse}{\boolsort}{\emptyset}}
        {\ctor \notin \dom{\adtctx}}

  \infer{\encj{\ste}{\tester{\sexp}{\ctor}}{\smtfalse}{\boolsort}{\emptyset}}
        {\ctor(\vec\type):\delta \in \adtctx
         \quad
         \encj{\ste}{\sexp}{\term}{\sort}{\formulae}
         \quad
         \toadt{\delta}(\term, \sort, \formulae) \text{ undefined}}

  \infer{\encj{\ste}{\tester{\sexp}{\ctor}}
              {\smtands\formulae \smtand \isctor{\ctor}(\term)}
              {\boolsort}{\emptyset}}
        {\ctor(\vec\type):\delta \in \adtctx
         \quad
         \encjc{\ste}{\sexp}{\delta}{\term}{\formulae}}
\end{mathpar}

Match branches are encoded under the typing environment their pattern induces.
SMT-LIB matches are total but evaluating a source match can fail, so a second match
records the side conditions of the branch taken:
\begin{mathpar}
  \infer{\encj{\ste}{\match{\sexp}{\delta}{p_i \rightarrow \sexp_i|_{i=1}^n}}
              {\term_m}{\valsort}{\formulae \cup \{\formula_m\}}}
        {\delta \in \adtctx
         \quad
         n \geq 1
         \quad
         \encjc{\ste}{\sexp}{\delta}{\term}{\formulae}
         \quad
         \encjc{\ste_i}{\sexp_i}{\vals}{\term_i}{\formulae_i}}
\end{mathpar}
where \(y\) is fresh,
\[
\begin{array}{@{}l@{}}
  \term_m = \termmatch{\term}{\left( p_i \rightarrow \term_i \right)|_{i=1}^n
    \quad y \rightarrow \nullval(\nullconst)},
  \\
  \formula_m = \termmatch{\term}{\left( p_i \rightarrow \smtands\formulae_i \right)|_{i=1}^n
    \quad y \rightarrow \smtfalse},
\end{array}
\]
and \(\ste_i \defeq \ste[x \mapsto \delta]\) if \(p_i = x\), and \(\ste_i \defeq \ste[\vec x \mapsto \vec\type]\) if \(p_i = \ctor(\vec x)\) with \(\ctor(\vec\type):\delta \in \adtctx\).
The catch-all branch keeps \(\term_m\) total. Its counterpart in \(\formula_m\)
is \(\smtfalse\), so the side conditions also fail when no pattern matches, as
evaluating the source match does.

\subsection{Representability Guards}
\label{app:backend:reprguards}

The top-level query asserts:
\[
  \reprguards_{\ste}(\sexp)
  \defeq
  \{
    G_{\ste}(x)
    \mid x \in \lv{\sexp}\cup\dom{\ste}
  \},
\]
where
\[
  G_{\ste}(x)=
  \begin{cases}
    \smtrepr{\type}(x) & x : \type \in \ste\\
    \isval(x) \smtimplies \reprval(\getval(x))
      & x \notin \dom{\ste}.
  \end{cases}
\]

\subsection{Semantic Equivalence}
\label{app:backend:equiv}

Fix an SMT-LIB structure \(\struct\) for the generated background theory.  As in
the main text, we write \(v \simeq \smtval{u}\), read \emph{\(v\) is semantically
equivalent to \(\smtval{u}\)}, when the source result \(v\) corresponds to the
SMT-LIB value \(\smtval{u}\), interpreted in \(\struct\).  The surrounding sort
determines which case of the relation is used.  At native sorts, the
representative cases are:
\[
\begin{array}{rcll}
\nil
  &\simeq& \nullconst^\struct\\
b
  &\simeq& \smtval{u}
  & b\in\bools,\ \smtval{u}=b\\
n
  &\simeq& \smtval{u}
  & n\in\nats,\ \smtval{u}=n,\ \smtval{u}\in\ints\\
q
  &\simeq& \smtval{u}
  & q\in\rationals,\ \smtval{u}=q,\ \smtval{u}\in\reals\\
s
  &\simeq& \smtval{u}
  & s\in\strings,\ \smtval{u}=s\\
{[v_1,\dots,v_n]}
  &\simeq& [\smtval{u_1},\dots,\smtval{u_n}]
  & \text{if }v_i\simeq\smtval{u_i}\\
\ctor(v_1,\dots,v_n)
  &\simeq& \smtconstr{\ctor}^{\struct}(\smtval{u_1},\dots,\smtval{u_n})
  & \text{if }v_i\simeq\smtval{u_i}.
\end{array}
\]
Since the left-hand sides range over source values, native SMT-LIB values that
are not representable, such as negative integers, are related to none.

At \(\valsort\), the relation unwraps the dynamic constructor and then uses the
same relation on the payload:
\[
\begin{array}{rcl@{\qquad}l}
\nil &\simeq& \nullval^\struct(\smtval{u})
  &\text{if }\nil\simeq\smtval{u}\\
b &\simeq& \boolval^\struct(\smtval{u})
  &\text{if }b\simeq\smtval{u}\\
n &\simeq& \natval^\struct(\smtval{u})
  &\text{if }n\simeq\smtval{u}\\
q &\simeq& \ratval^\struct(\smtval{u})
  &\text{if }q\simeq\smtval{u}\\
s &\simeq& \stringval^\struct(\smtval{u})
  &\text{if }s\simeq\smtval{u}\\
{[v_1,\dots,v_n]} &\simeq& \listval^\struct(\smtval{u})
  &\text{if }[v_1,\dots,v_n]\simeq\smtval{u}\\
\ctor(v_1,\dots,v_n) &\simeq& \adtval{\delta}^{\struct}(\smtval{u})
  &\text{if }\ctor(v_1,\dots,v_n)\simeq\smtval{u}.
\end{array}
\]
At \(\maybevalsort\), \(v\simeq\someval^\struct(\smtval{u})\) when \(v\simeq\smtval{u}\),
while \(\undefd\simeq\noval^\struct\).

\subsection{Gradual and Ill Typing}
\label{app:backend:typing}

The gradual typing judgement uses type consistency:
\[
  \type \sim \type'
  \quad\Leftrightarrow\quad
  \type=\type' \text{ or } \type=\vals \text{ or } \type'=\vals.
\]
The key rules are:
\begin{mathpar}
\infer{\ste \vdash \gv : \type}
      {\gv \in \type \quad \type\neq\vals}

\infer{\ste \vdash x : \type'}
      {x : \type \in \ste \quad
       \cwf(\type') \quad
       (\type' = \type \text{ or } \type = \vals)}

\infer{\ste \vdash x : \type}
      {x \notin \dom{\ste} \quad \cwf(\type)}

\infer{\ste \vdash \ctor(\vec\sexp) : \delta}
      {\ctor(\vec\type):\delta \in \adtctx \quad
       \ste \vdash \vec\sexp : \vec\type' \quad
       \vec\type \sim \vec\type'}

\infer{\ste \vdash f(\vec\sexp) : \type}
      {f(\vec\type):\type \in \lfuncctx \quad
       \ste \vdash \vec\sexp : \vec\type' \quad
       \vec\type \sim \vec\type'}

\infer{\ste \vdash \odot\sexp : \type_{\mathit{out}}}
      {\odot : \type_{\mathit{in}}\to\type_{\mathit{out}} \quad
       \cwf(\type_{\mathit{in}}) \quad
       \ste \vdash \sexp : \type'_{\mathit{in}} \quad
       \type_{\mathit{in}}\sim\type'_{\mathit{in}}}

\infer{\ste \vdash \sexp_1 \oplus \sexp_2 : \type}
      {\oplus : \type_1\times\type_2\to\type \quad
       \cwf(\type_i) \quad
       \ste \vdash \sexp_i : \type_i' \quad
       \type_i\sim\type_i'}

\infer{\ste \vdash \sexp \in \type : \bools}
      {\ste \vdash \sexp : \type'}

\infer{\ste \vdash \tester{\sexp}{\ctor} : \bools}
      {\ste \vdash \sexp : \type} .
\end{mathpar}
Lists require all elements to have types consistent with the chosen element type. A match expression over \(\delta\) requires \(\delta \in \adtctx\), a scrutinee consistently typed with \(\delta\), a well-formed result type, consistently typed branch bodies under the pattern-bound variables, and at least one branch whose pattern is a variable or a constructor of \(\delta\).

The negative judgement \(\ste \dashv \sexp : \type\) has the property that whenever \(\subst\models\ste\), \(\ste\dashv\sexp:\type\), and \(\esem{\sexp}{\subst}=v\), then \(v\notin\type\).  We write \(\type'\leq\type\) for the subtype relation on types: it is reflexive, every type is below the top type \(\vals\), and list types are covariant, so \(\listsof{\type_1}\leq\listsof{\type_2}\) whenever \(\type_1\leq\type_2\).  We write \(\type\#\type'\) when the two types are disjoint, i.e. no value has both types.  Its rules include:
\begin{mathpar}
\infer{\ste\dashv\sexp:\type'}
      {\ste\dashv\sexp:\type \quad \type'\leq\type}

\infer{\ste\dashv\gv:\type'}
      {\gv \in \type \quad \type\#\type'}

\infer{\ste\dashv x:\type'}
      {x : \type \in \ste \quad \type\#\type'}

\infer{\ste\dashv\ctor(\vec\sexp):\vals}
      {\ctor\notin\dom{\adtctx}}

\infer{\ste\dashv\ctor(\vec\sexp):\vals}
      {\ctor(\vec\type):\delta \in \adtctx \quad |\vec\sexp|\neq|\vec\type|}

\infer{\ste\dashv\ctor(\vec\sexp):\vals}
      {\ctor(\vec\type):\delta \in \adtctx \quad \ste\dashv\sexp_i:\type_i}

\infer{\ste\dashv\ctor(\vec\sexp):\type}
      {\ctor(\vec\type'):\delta \in \adtctx \quad \delta\#\type}

\infer{\ste\dashv[\vec\sexp]:\listsof{\type}}
      {\sexp\in\vec\sexp \quad \ste\dashv\sexp:\type}

\infer{\ste\dashv f(\vec\sexp):\vals}
      {f\notin\dom{\lfuncctx}}

\infer{\ste\dashv f(\vec\sexp):\vals}
      {f(\vec\type):\type \in \lfuncctx \quad |\vec\sexp|\neq|\vec\type|}

\infer{\ste\dashv f(\vec\sexp):\vals}
      {f(\vec\type):\type \in \lfuncctx \quad \ste\dashv\sexp_i:\type_i}

\infer{\ste\dashv f(\vec\sexp):\type'}
      {f(\vec\type):\type \in \lfuncctx \quad \type\#\type'} .
\end{mathpar}

\subsection{Satisfiability Queries and Symbolic Commands}
\label{app:backend:sat-commands}

The symbolic semantics extends the ghost-command relation with the backend satisfiability commands \(\psatpc\) and \(\psatstate\). For \(\sst=(\ssto,\smem,\spred,\spc,\ste)\), the rules continue exactly when the corresponding query does not return \(\unsat\):
\begin{mathpar}
\infer{\csesemtransabstractm{\sst}{\psatpc}{\sst}{\fsctx}{m}{\oxok}}
      {\issat_{\ste}(\spc)\in\{\sat,\unknown\}}

\infer{\csesemtransabstractm{\sst}{\psatstate}{\sst}{\fsctx}{m}{\oxok}}
      {\issat(m,\sst)\in\{\sat,\unknown\}}
\end{mathpar}
There is no successful rule when the result is \(\unsat\); such branches are pruned. Thus \(\psatpc\) checks only the path condition, while \(\psatstate\) invokes the full state check \(\issat(m,\sst)\) from \Cref{sect:backend:satisfiability}.

\clearpage
\clearpage\section{Evaluation Examples}%
\label{app:eval}

We used VeriFast 26.10 for all VeriFast experiments. Neither of the VeriFast files below has a \texttt{main} function, so VeriFast requires the \texttt{-shared} flag to verify them: e.g., \texttt{verifast -shared reverse\_list.c}.

\subsection{VeriFast: Non-destructive List Reversal}%

\begin{minted}{c}
// Included for malloc (used by prepend below). stdlib.h transitively
// includes VeriFast's list<t> library, but we do not use that list
// library here, to make this file self-contained.
#include "stdlib.h"

/*
 * Non-destructive list reversal from Fig. 2.
 *
 * This file intentionally does not reuse VeriFast's
 * standard library (prelude) list<t> type or its append/reverse fixpoints.
 * Instead, it defines its own List ADT (Fig. 2b) and its own ++ and dagger
 * logical functions (Fig. 2c) from scratch, so the translation is
 * self-contained and mirrors the figure directly.
 *
 * Naming: the paper's [], ::, ++, and dagger are named LNil, LCons, concat,
 * and dagger here, since C/VeriFast identifiers cannot contain these
 * symbols. Likewise reverse' is named reverseP (and alpha', a' are
 * named alphaP, aP), since C identifiers cannot contain '.
 *
 * NOTE: the paper's Val is translated into int here, since VeriFast has no
 * type corresponding to Val.
 */

struct node {
    int value;
    struct node *next;
};

// C requires prepend to be declared before reverseP calls it.
struct node *prepend(int a, struct node *x);

// (a) Function and its specification
// {list(x, alpha) * list(y, beta)}
// reverse'(x, y)
// {list(x, alpha) * list(ret, alpha^dagger ++ beta)}
struct node *reverseP(struct node *x, struct node *y)
    //@ requires list(x, ?alpha) &*& list(y, ?beta);
    //@ ensures list(x, alpha) &*& list(result, concat(dagger(alpha), beta));
{
    struct node *r;
    //@ open list(x, alpha);
    if (x == 0) {
        r = y;
        //@ close list(x, alpha);
    } else {
        int h = x->value;
        struct node *t = x->next;
        y = prepend(h, y);
        r = reverseP(t, y);
        /*@
        switch (alpha) {
            case LNil:
            case LCons(a, alphaP):
                concatr(dagger(alphaP), a, beta);
        }
        @*/
        //@ close list(x, alpha);
    }
    return r;
}

/*@

// (b) ADT definition: List ::= [] | Val :: List
inductive List = LNil | LCons(int, List);

// (c) Logical function definition: (alpha : List) ++ (beta : List) : List
fixpoint List concat(List alpha, List beta) {
    switch (alpha) {
        case LNil: return beta;
        case LCons(a, alphaP): return LCons(a, concat(alphaP, beta));
    }
}

// (c) Logical function definition: (alpha : List)^dagger : List
fixpoint List dagger(List alpha) {
    switch (alpha) {
        case LNil: return LNil;
        case LCons(a, alphaP): return concat(dagger(alphaP), LCons(a, LNil));
    }
}

// (d) List predicate definition
predicate list(struct node *x, List alpha) =
    x == 0
    ? alpha == LNil
    : x->value |-> ?a &*& x->next |-> ?y &*& malloc_block_node(x) &*&
      list(y, ?alphaP) &*& alpha == LCons(a, alphaP);

// (e) Lemma statement and proof
// concatr : forall alpha, a, beta.
//   alpha in List * beta in List ==>
//   alpha ++ (a :: beta) = (alpha ++ [a]) ++ beta
// (|alpha| decreases)
//
// VeriFast is statically typed, so the paper's type hypotheses on alpha and
// beta hold by construction and the precondition below is just true.
//
// VeriFast has no decreases/terminates clause for lemmas (unlike for
// fixpoints, loops, and terminates-annotated C functions). Lemma
// recursion is instead checked automatically, because it is structural:
// each recursive call is on alphaP, a strict sub-term of alpha obtained
// by pattern matching.
lemma void concatr(List alpha, int a, List beta)
    requires true;
    ensures concat(alpha, LCons(a, beta)) == concat(concat(alpha, LCons(a, LNil)), beta);
{
    switch (alpha) {
        case LNil:
        case LCons(aP, alphaP):
            concatr(alphaP, a, beta);
    }
}

@*/

// (f) Prepend's specification
// {list(x, alpha)}
// prepend(a, x)
// {list(ret, a :: alpha)}
struct node *prepend(int a, struct node *x)
    //@ requires list(x, ?alpha);
    //@ ensures list(result, LCons(a, alpha));
{
    // Unlike the linear memory model we work with in this paper, where
    // allocation always succeeds, malloc can fail (return 0) in VeriFast/C,
    // so we must null-check.
    struct node *n = malloc(sizeof(struct node));
    if (n == 0) { abort(); }
    n->value = a;
    n->next = x;
    //@ close list(n, LCons(a, alpha));
    return n;
}

/*@

// Only needed for reverse(x) below. With the standard library, the
// analogous fact is available for free via the automatic append_nil lemma.
lemma void concat_nil(List alpha)
    requires true;
    ensures concat(alpha, LNil) == alpha;
{
    switch (alpha) {
        case LNil:
        case LCons(a, alphaP): concat_nil(alphaP);
    }
}

@*/

// Top-level function we are verifying (not shown in Fig. 2)
// {list(x, alpha)}
// reverse(x)
// {list(x, alpha) * list(ret, alpha^dagger)}
struct node *reverse(struct node *x)
    //@ requires list(x, ?alpha);
    //@ ensures list(x, alpha) &*& list(result, dagger(alpha));
{
    struct node *r;
    //@ close list(0, LNil);
    r = reverseP(x, 0);
    //@ concat_nil(dagger(alpha));
    return r;
}
\end{minted}

\subsection{VeriFast: Sorted Binary Tree Insertion}%
\label{app:verifast:bintree}

The following is the \texttt{add} function from VeriFast's \texttt{sorted\_bintree.c} example, including all required definitions. We include only the definitions needed to verify \texttt{add} itself, omitting all other functions and their supporting lemmas.

\begin{minted}{c}
#include "stdlib.h"

/*
 * Sorted binary tree insertion, from VeriFast's sorted_bintree.c example.
 *
 * We keep exactly what is needed to verify add() and omit all other
 * functions and their supporting lemmas. Otherwise, no identifiers,
 * formatting, or proof hints are changed, except that the "Needed by Redux"
 * hints are now line comments after each line, not block comments before
 * it. We moved them because block comments do not nest in C: inside the
 * ghost annotation block (itself a block comment), the original hints ended
 * that block early under the simple C syntax highlighter we use here,
 * garbling the highlighting.
 */

struct tree{
  int value;
  struct tree *left;
  struct tree *right;
};

/*@
predicate tree(struct tree *t,bintree b) =
  switch(b){
    case empty: return t==0;
    case tree(a,bl,br): return t->value |-> ?v &*& t->left |-> ?l &*& t->right |-> ?r 
			&*& malloc_block_tree(t) &*& tree(l,bl) &*& tree(r,br) &*& t!=0 &*& a==v ;
  }&*& inorder(b)==true;

inductive bintree = |empty |tree(int,bintree,bintree);

fixpoint bool t_contains(bintree b, int v) {
  switch (b) {
    case empty: return false;
    case tree(a,l,r): return a==v || (v < a? t_contains(l,v):t_contains(r,v));
  }
}
fixpoint bintree tree_add(bintree b, int x) {
  switch (b) {
    case empty: return tree(x,empty,empty);
    case tree(v,l,r): return x < v? tree(v,tree_add(l,x),r):
			(x==v? tree(v,l,r):tree(v,l,tree_add(r,x)));
  }
}
fixpoint int max(bintree b){
  switch(b){
    case tree(a,bl,br): return (br==empty? a:max(br));
    case empty: return 0;
  }
}
fixpoint int min(bintree b){
  switch(b){
    case tree(a,bl,br): return (bl==empty? a:min(bl));
    case empty: return 0;
  }
}
fixpoint bool inorder(bintree b){
  switch(b){
	case empty: return true;
	case tree(a,bl,br): return (bl==empty? true:max(bl)<a)&& (br==empty? true: a < min(br))
		&& inorder(bl) && inorder(br);
  }
}
lemma void max_conj_add(bintree l,int v,int x)
  requires x < v &*& (max(l)<v||l==empty) &*& inorder(l)==true;
  ensures max(tree_add(l,x))<v &*& inorder(l)==true;
{
  switch(l){
	case empty:
	case tree(a,b,c):if(x < a){
                          if (b == empty) {} else {} // Needed by Redux.
			  max_conj_add(b,a,x);
			 }
			 if(a < x){
                          if (c == empty) {} else {} // Needed by Redux.
			  max_conj_add(c,v,x);
			 }
                         if ((x < a) == true) {} else { if (x == a) {} else {} } // Needed by Redux.
                         if (c == empty) {} else {} // Needed by Redux.
                         if (x == a) {} else {} // Needed by Redux.
                         if (tree_add(c, x) == empty) {} else {} // Needed by Redux.
  }
}
lemma void min_conj_add(bintree r,int v,int x)
  requires v < x &*& (v < min(r)||r==empty) &*& inorder(r)==true;
  ensures v < min(tree_add(r,x)) &*& inorder(r)==true;
{
  switch(r){
	case empty:
	case tree(a,b,c):if(a < x){
			  min_conj_add(c,a,x);
			 }
			 if(x < a){
			  min_conj_add(b,v,x);
			 }
  }
}

lemma void tree_add_inorder(bintree b, int x)
    requires inorder(b)==true &*& t_contains(b,x)==false;
    ensures inorder(tree_add(b,x))==true &*& t_contains(tree_add(b,x),x)==true;
{
    switch (b) {
        case empty:
        case tree(v,l,r):if(x < v){
			  max_conj_add(l,v,x);
			  tree_add_inorder(l,x);
			 }
			 if(v < x){
			  min_conj_add(r,v,x);
			  tree_add_inorder(r,x);
		  	 }
    }
}
@*/
struct tree *init_tree(int x)
  //@ requires true;
  //@ ensures tree(result,tree(x,empty,empty));
{
  struct tree *t = malloc(sizeof(struct tree));
  if(t!=0){
    t->value=x;
    t->left=0;
    t->right=0;
    //@ struct tree *l = t->left;
    //@ close tree(l,empty);
    //@ struct tree *r = t->right;
    //@ close tree(r,empty);
    //@ int v = t->value;
    //@ close tree(t,tree(x,empty,empty));
    return t;
  }else{
	abort();
  }
}

void add(struct tree *t, int x)
  //@ requires tree(t,?b) &*& b!=empty &*& false==t_contains(b,x) &*& inorder(b)==true;
  //@ ensures tree(t,tree_add(b,x)) &*& inorder(tree_add(b,x))==true;
 {
  //@ open tree(t,b);
  int v=t->value;
  struct tree *l=t->left;
  //@ open tree(l,?bl);
  //@ close tree(l,bl);
  struct tree *r=t->right;
  //@ open tree(r,?br);
  //@ close tree(r,br);
  if(x < v){
    if(l!=0){
      add(l,x);
      //@ tree_add_inorder(b,x);
      //@ close tree(t,tree(v,tree_add(bl,x),br));
    }else{
      struct tree *temp=init_tree(x);
      t->left=temp;
      //@ open tree(l,bl);
      //@ close tree(t,tree(v,tree(x,empty,empty),br));
      //@ tree_add_inorder(b,x);
    }
  }else{
    if(v < x){
      if(r!=0){
        add(r,x);
        //@ tree_add_inorder(b,x);
        //@ close tree(t,tree(v,bl,tree_add(br,x)));	
      }else{
        struct tree *temp=init_tree(x);
        t->right=temp;
        //@ open tree(r,br);
        //@ close tree(t,tree(v,bl,tree(x,empty,empty)));
      }
    }
  }
}
\end{minted}

\subsection{Our CSE Theory: Sorted Binary Tree Insertion}%
\label{app:verifast:bintree-cse}

\newcommand{\ptree}{\predd{tree}}
\makeatletter
\newenvironment{leftdisplay}{%
  \par\if@noskipsec\leavevmode\par\fi 
  \medskip\noindent\hspace*{\parindent}$\displaystyle}{$\par\medskip}
\makeatother

We render the verification of \texttt{add} (\Cref{app:verifast:bintree}) in our theory's notation, in the style of \Cref{fig:example-spec}. Unlike \Cref{fig:example-spec}, we keep the ghost commands for folding and unfolding the \(\pname{tree}\) predicate, to show how these are used in practice. We give the definitions in the same order as the source file, and translate VeriFast's \texttt{int} to our naturals \(\nats\). Our theory also needs a measure for each recursive lemma (\cref{sect:frontend:lemmas}), so we add the logical function \(\pname{size}\), which is not in the source file, and use it for the measures.

\small

\paragraph{Predicate definition}%
\begin{leftdisplay}
\ptree{x, b} =
\asrtcolor{
\begin{array}[t]{l}
\big( (x = \nil \lstar b = \pname{empty}) \lor {} \\
\quad (\exists a, l, r, bl, br.\ \threecell{x}{a}{l}{r} \lstar b = \pname{tree}(a, bl, br) \lstar x \neq \nil \lstar \ptree{l, bl} \lstar \ptree{r, br}) \big) \\
\lstar \pname{inorder}(b) = \true
\end{array}
}
\end{leftdisplay}

\paragraph{ADT definition}%
\begin{leftdisplay}
\pname{bintree} ::= \pname{empty} \mid \pname{tree}(\nats, \pname{bintree}, \pname{bintree})
\end{leftdisplay}

\paragraph{Logical function definitions}%
\begin{leftdisplay}
\begin{array}{l}
\pname{t\_contains}(b : \pname{bintree}, v : \nats) : \bools = {} \\
\quad \match{b}{\pname{bintree}}{} \\
\quad \begin{array}{l l}
    \pname{empty} & \rightarrow \false \\
    \pname{tree}(a, l, r) & \rightarrow a = v \lor (v < a\ ?\ \pname{t\_contains}(l, v) : \pname{t\_contains}(r, v))
\end{array} \\
\mathrm{\mathbf{when}}\ \true
\end{array}
\end{leftdisplay}
\begin{leftdisplay}
\begin{array}{l}
\pname{tree\_add}(b : \pname{bintree}, x : \nats) : \pname{bintree} = {} \\
\quad \match{b}{\pname{bintree}}{} \\
\quad \begin{array}{l l}
    \pname{empty} & \rightarrow \pname{tree}(x, \pname{empty}, \pname{empty}) \\
    \pname{tree}(v, l, r) & \rightarrow \begin{array}[t]{@{}l@{}}
    x < v\ ?\ \pname{tree}(v, \pname{tree\_add}(l, x), r) \\
    {} : (x = v\ ?\ \pname{tree}(v, l, r) : \pname{tree}(v, l, \pname{tree\_add}(r, x)))
    \end{array}
\end{array} \\
\mathrm{\mathbf{when}}\ \true
\end{array}
\end{leftdisplay}
\begin{leftdisplay}
\begin{array}{l}
\pname{max}(b : \pname{bintree}) : \nats = {} \\
\quad \match{b}{\pname{bintree}}{} \\
\quad \begin{array}{l l}
    \pname{tree}(a, bl, br) & \rightarrow br = \pname{empty}\ ?\ a : \pname{max}(br) \\
    \pname{empty} & \rightarrow 0
\end{array} \\
\mathrm{\mathbf{when}}\ \true
\end{array}
\end{leftdisplay}
\begin{leftdisplay}
\begin{array}{l}
\pname{min}(b : \pname{bintree}) : \nats = {} \\
\quad \match{b}{\pname{bintree}}{} \\
\quad \begin{array}{l l}
    \pname{tree}(a, bl, br) & \rightarrow bl = \pname{empty}\ ?\ a : \pname{min}(bl) \\
    \pname{empty} & \rightarrow 0
\end{array} \\
\mathrm{\mathbf{when}}\ \true
\end{array}
\end{leftdisplay}
\begin{leftdisplay}
\begin{array}{l}
\pname{inorder}(b : \pname{bintree}) : \bools = {} \\
\quad \match{b}{\pname{bintree}}{} \\
\quad \begin{array}{l l}
    \pname{empty} & \rightarrow \true \\
    \pname{tree}(a, bl, br) & \rightarrow \begin{array}[t]{@{}l@{}}
    (bl = \pname{empty}\ ?\ \true : \pname{max}(bl) < a) \\
    {} \land (br = \pname{empty}\ ?\ \true : a < \pname{min}(br)) \\
    {} \land \pname{inorder}(bl) \land \pname{inorder}(br)
    \end{array}
\end{array} \\
\mathrm{\mathbf{when}}\ \true
\end{array}
\end{leftdisplay}
\begin{leftdisplay}
\begin{array}{l}
\pname{size}(b : \pname{bintree}) : \nats = {} \\
\quad \match{b}{\pname{bintree}}{} \\
\quad \begin{array}{l l}
    \pname{empty} & \rightarrow 0 \\
    \pname{tree}(a, l, r) & \rightarrow 1 + \pname{size}(l) + \pname{size}(r)
\end{array} \\
\mathrm{\mathbf{when}}\ \true
\end{array}
\end{leftdisplay}

\paragraph{Lemma statements and proofs}%
\begin{leftdisplay}
\begin{array}{l}
\mathtt{max\_conj\_add} : \forall l, v, x.\\
\quad \asrtcolor{x < v \lstar (\pname{max}(l) < v \lor l = \pname{empty}) \lstar \pname{inorder}(l) = \true} \Rightarrow \\
\quad \asrtcolor{\pname{max}(\pname{tree\_add}(l, x)) < v \lstar \pname{inorder}(l) = \true} \\
\{ \\
\quad \pmatchs{l}{\pname{bintree}} \\
\quad\quad \pcase{\pname{empty}}{\pskip} \\
\quad\quad \ghostcolor{\pname{tree}(a, b, c) \rightarrow} \\
\quad\quad\quad \ghostcolor{\pifelses{x < a}{\papply{max\_conj\_add}{b, a, x}} \pifelsem \pskip \pifelsee;} \\
\quad\quad\quad \ghostcolor{\pifelses{a < x}{\papply{max\_conj\_add}{c, v, x}} \pifelsem \pskip \pifelsee} \\
\quad \pmatche \\
\} \\
(\pname{size}(l)\ \mathrm{\mathbf{decreases}})
\end{array}
\end{leftdisplay}
\begin{leftdisplay}
\begin{array}{l}
\mathtt{min\_conj\_add} : \forall r, v, x.\\
\quad \asrtcolor{v < x \lstar (v < \pname{min}(r) \lor r = \pname{empty}) \lstar \pname{inorder}(r) = \true} \Rightarrow \\
\quad \asrtcolor{v < \pname{min}(\pname{tree\_add}(r, x)) \lstar \pname{inorder}(r) = \true} \\
\{ \\
\quad \pmatchs{r}{\pname{bintree}} \\
\quad\quad \pcase{\pname{empty}}{\pskip} \\
\quad\quad \ghostcolor{\pname{tree}(a, b, c) \rightarrow} \\
\quad\quad\quad \ghostcolor{\pifelses{a < x}{\papply{min\_conj\_add}{c, a, x}} \pifelsem \pskip \pifelsee;} \\
\quad\quad\quad \ghostcolor{\pifelses{x < a}{\papply{min\_conj\_add}{b, v, x}} \pifelsem \pskip \pifelsee} \\
\quad \pmatche \\
\} \\
(\pname{size}(r)\ \mathrm{\mathbf{decreases}})
\end{array}
\end{leftdisplay}
\begin{leftdisplay}
\begin{array}{l}
\mathtt{tree\_add\_inorder} : \forall b, x.\\
\quad \asrtcolor{\pname{inorder}(b) = \true \lstar \pname{t\_contains}(b, x) = \false} \Rightarrow \\
\quad \asrtcolor{\pname{inorder}(\pname{tree\_add}(b, x)) = \true \lstar \pname{t\_contains}(\pname{tree\_add}(b, x), x) = \true} \\
\{ \\
\quad \pmatchs{b}{\pname{bintree}} \\
\quad\quad \pcase{\pname{empty}}{\pskip} \\
\quad\quad \ghostcolor{\pname{tree}(v, l, r) \rightarrow} \\
\quad\quad\quad \ghostcolor{\pifelses{x < v}{\papply{max\_conj\_add}{l, v, x}; \papply{tree\_add\_inorder}{l, x}} \pifelsem \pskip \pifelsee;} \\
\quad\quad\quad \ghostcolor{\pifelses{v < x}{\papply{min\_conj\_add}{r, v, x}; \papply{tree\_add\_inorder}{r, x}} \pifelsem \pskip \pifelsee} \\
\quad \pmatche \\
\} \\
(\pname{size}(b)\ \mathrm{\mathbf{decreases}})
\end{array}
\end{leftdisplay}

\paragraph{Helper function and its specification}%
\begin{leftdisplay}
\begin{array}{l}
\specline{\pvar{x} = x \lstar \true} \\
\pfuncspec{init\_tree}{\pvar x}\ \{ \\
\quad \palloc{\pvar t}{3}; \\
\quad \pmutate{\pvar t}{\pvar x}; \\
\quad \pmutate{\pvar t + 1}{\nil}; \\
\quad \pmutate{\pvar t + 2}{\nil}; \\
\quad \pfold{\pname{tree}}{\nil, \pname{empty}}; \\
\quad \pfold{\pname{tree}}{\nil, \pname{empty}}; \\
\quad \pfold{\pname{tree}}{\pvar t, \pname{tree}(\pvar x, \pname{empty}, \pname{empty})}; \\
\quad \preturn{\pvar t} \\
\} \\
\specline{\ptree{\pvar{ret}, \pname{tree}(x, \pname{empty}, \pname{empty})}}
\end{array}
\end{leftdisplay}

\paragraph{Function and its specification} Strictly speaking, folding and unfolding of predicates in the CSE theory by \citet{Loow2025} are based on so-called in- and out-parameters. This is relevant here because the unfold commands below with question-marked variables are, strictly speaking, not fully translated to the theory. But as we do not cover in- and out-parameters in this paper, we ignore this minor difference between VeriFast and the theory in the below and simply re-use VeriFast's syntax.
\par\medskip
\noindent
\hspace*{\parindent}\(\displaystyle \specline{\pvar{t} = t \lstar \pvar{x} = x \lstar \ptree{t, b} \lstar b \neq \pname{empty} \lstar \pname{t\_contains}(b, x) = \false \lstar \pname{inorder}(b) = \true}\)\\
\hspace*{\parindent}\(\displaystyle \pfuncspec{add}{\pvar t, \pvar x}\ \{\)\\
\hspace*{\parindent}\(\displaystyle \quad \punfold{\pname{tree}}{\pvar t, b};\)\\
\hspace*{\parindent}\(\displaystyle \quad \pderef{\pvar v}{\pvar t};\)\\
\hspace*{\parindent}\(\displaystyle \quad \pderef{\pvar l}{\pvar t + 1};\)\\
\hspace*{\parindent}\(\displaystyle \quad \punfold{\pname{tree}}{\pvar l, ?bl};\)\\
\hspace*{\parindent}\(\displaystyle \quad \pfold{\pname{tree}}{\pvar l, bl};\)\\
\hspace*{\parindent}\(\displaystyle \quad \pderef{\pvar r}{\pvar t + 2};\)\\
\hspace*{\parindent}\(\displaystyle \quad \punfold{\pname{tree}}{\pvar r, ?br};\)\\
\hspace*{\parindent}\(\displaystyle \quad \pfold{\pname{tree}}{\pvar r, br};\)\\
\hspace*{\parindent}\(\displaystyle \quad \pifelses{\pvar x < \pvar v}\)\\
\hspace*{\parindent}\(\displaystyle \quad\quad \pifelses{\pvar l \neq \nil}\)\\
\hspace*{\parindent}\(\displaystyle \quad\quad\quad \pfuncall{\pvar{ignore}}{\mathtt{add}}{\pvar l, \pvar x};\)\\
\hspace*{\parindent}\(\displaystyle \quad\quad\quad \papply{tree\_add\_inorder}{b, \pvar x};\)\\
\hspace*{\parindent}\(\displaystyle \quad\quad\quad \pfold{\pname{tree}}{\pvar t, \pname{tree}(\pvar v, \pname{tree\_add}(bl, \pvar x), br)}\)\\
\hspace*{\parindent}\(\displaystyle \quad\quad \pifelsem\)\\
\hspace*{\parindent}\(\displaystyle \quad\quad\quad \pfuncall{\pvar{temp}}{\mathtt{init\_tree}}{\pvar x};\)\\
\hspace*{\parindent}\(\displaystyle \quad\quad\quad \pmutate{\pvar t + 1}{\pvar{temp}};\)\\
\hspace*{\parindent}\(\displaystyle \quad\quad\quad \punfold{\pname{tree}}{\pvar l, bl};\)\\
\hspace*{\parindent}\(\displaystyle \quad\quad\quad \pfold{\pname{tree}}{\pvar t, \pname{tree}(\pvar v, \pname{tree}(\pvar x, \pname{empty}, \pname{empty}), br)};\)\\
\hspace*{\parindent}\(\displaystyle \quad\quad\quad \papply{tree\_add\_inorder}{b, \pvar x}\)\\
\hspace*{\parindent}\(\displaystyle \quad\quad \pifelsee\)\\
\hspace*{\parindent}\(\displaystyle \quad \pifelsem\)\\
\hspace*{\parindent}\(\displaystyle \quad\quad \pifelses{\pvar v < \pvar x}\)\\
\hspace*{\parindent}\(\displaystyle \quad\quad\quad \pifelses{\pvar r \neq \nil}\)\\
\hspace*{\parindent}\(\displaystyle \quad\quad\quad\quad \pfuncall{\pvar{ignore}}{\mathtt{add}}{\pvar r, \pvar x};\)\\
\hspace*{\parindent}\(\displaystyle \quad\quad\quad\quad \papply{tree\_add\_inorder}{b, \pvar x};\)\\
\hspace*{\parindent}\(\displaystyle \quad\quad\quad\quad \pfold{\pname{tree}}{\pvar t, \pname{tree}(\pvar v, bl, \pname{tree\_add}(br, \pvar x))}\)\\
\hspace*{\parindent}\(\displaystyle \quad\quad\quad \pifelsem\)\\
\hspace*{\parindent}\(\displaystyle \quad\quad\quad\quad \pfuncall{\pvar{temp}}{\mathtt{init\_tree}}{\pvar x};\)\\
\hspace*{\parindent}\(\displaystyle \quad\quad\quad\quad \pmutate{\pvar t + 2}{\pvar{temp}};\)\\
\hspace*{\parindent}\(\displaystyle \quad\quad\quad\quad \punfold{\pname{tree}}{\pvar r, br};\)\\
\hspace*{\parindent}\(\displaystyle \quad\quad\quad\quad \pfold{\pname{tree}}{\pvar t, \pname{tree}(\pvar v, bl, \pname{tree}(\pvar x, \pname{empty}, \pname{empty}))}\)\\
\hspace*{\parindent}\(\displaystyle \quad\quad\quad \pifelsee\)\\
\hspace*{\parindent}\(\displaystyle \quad\quad \pifelsee\)\\
\hspace*{\parindent}\(\displaystyle \quad \pifelsee;\)\\
\hspace*{\parindent}\(\displaystyle \quad \preturn{\nil}\)\\
\hspace*{\parindent}\(\displaystyle \}\)\\
\hspace*{\parindent}\(\displaystyle \specline{\ptree{t, \pname{tree\_add}(b, x)} \lstar \pname{inorder}(\pname{tree\_add}(b, x)) = \true \lstar \pvar{ret} = \nil}\)
\par\medskip

\normalsize

\subsection{Backend Evaluation Details}%
\label{app:eval:backend}

\Cref{tab:backend-eval} gives the language-level breakdown for the backend evaluation. Each run had a timeout of 120~s. One run, of the main function of the C AWS case study, reaches it; since the comparison is query-local, we include the 2,948 queries it produced before then. Likewise, a few runs end with analysis errors that Gillian also reports without the verified backend, and we include the queries they produced.

\begin{table}[ht]
\centering
\small
\caption{Backend comparison on SMT queries generated by Gillian examples. Timing columns report median~/~95th percentile solver time in milliseconds over the queries both backends answered, measuring solver time only.}%
\label{tab:backend-eval}
\begin{tabular}{lrrrrrcc}
\toprule
\textbf{Language} & \textbf{Runs} & \textbf{Queries} & \textbf{Translated} & \textbf{Answered} & \textbf{Agree} & \textbf{Gillian ms} & \textbf{Verified ms} \\
\midrule
JavaScript & 16 & 1324 &  985 &  937 &  923 & 0.25 / 0.77 & 0.47 / 311.59 \\
C          &  7 & 4073 & 2989 & 2988 & 2988 & 0.07 / 0.42 & 0.14 / 2.29 \\
WISL       &  9 &  423 &  416 &  416 &  416 & 0.09 / 0.22 & 0.40 / 0.96 \\
\midrule
\textbf{Total} & 32 & 5820 & 4390 & 4341 & 4327 & 0.09 / 0.59 & 0.16 / 2.68 \\
\bottomrule
\end{tabular}
\end{table}

\paragraph{Unverified code} Only the expression encoder of \cref{sect:backend:encoding} is extracted. The bridge from Gillian's expressions, and the OCaml code that sends the encoded expressions to the SMT solver, are unverified; the latter also emits the background declarations. In particular, the experiment does not run the operational satisfiability check of \cref{sect:backend:expr-sat}: it applies neither the \(\dashv\) check nor the representability guards, and it declares no user-defined ADTs or logical functions, which Gillian's queries do not use. Leaving out the guards only weakens a query, so an \unsat{} answer without them remains \unsat{} with them. For the same reason as the guards, the bridge translates Gillian's non-positive float literals, nearly all of them \texttt{0.}, which the encoder handles as it does any rational. It does not translate negative integer literals, since the integers of our theory are the naturals. The extracted code carries a small patch, consisting of additions only, for three kinds of Gillian values that our theory lacks.

\paragraph{Untranslated queries} Of the 1,430 untranslated queries, 1,111 contain negative integer literals, and for 1,068 of them these are the only obstacle; 988 of the 1,111 come from C. Further, 212 contain quantifiers, 195 sets, 39 list slicing, and 3 list reversal; these categories overlap.

\paragraph{Solver-time tail} An encoder can encode the length of a list in one of two ways: exactly, as the length of the SMT-LIB sequence that encodes the list, or abstractly, as an uninterpreted function of the list, whose value the solver may choose freely without constructing a list. The verified encoder always encodes lengths exactly. Gillian's encoder chooses separately for each application of length in the query, depending on its argument. It encodes exactly the length of a list literal, such as \([1, 2, 3]\), of any other compound list expression, and of a list variable that the query also uses elsewhere. When the list is a variable that occurs in the query only as the argument of length, however, the query constrains the list only through its length, so its contents do not matter, and Gillian's encoder encodes that length abstractly. The abstract encoding never makes a satisfiable query unsatisfiable, since a model of the exact query is one of the abstract query, so its \unsat{} answers are always correct. Its \sat{} answers are correct when the query also requires the length to be non-negative and the list variable is not bound by a quantifier, since any list of the chosen length is then a witness. Both conditions hold for every query of our run, of which 3,170 use the abstract encoding, but Gillian's symbolic engine maintains them, not its encoder. The 67 slowest queries under the verified encoding each require the solver to construct lists: in the slowest, the constraint \(32 \leq |l|\) requires a list of at least 32 elements, and encoding the length of \(l\) alone abstractly makes the query as fast as the others. If we modify the verified encoding to encode these lengths abstractly, as Gillian's encoder does, each of the 67 queries gives the same answer and takes as long as the other queries.

\clearpage
\fi

\end{document}